\documentclass[%
 amsmath,amssymb,
 aps,
]{revtex4-2}

\usepackage{graphicx}
\usepackage{epstopdf, epsfig}
\usepackage{amsfonts}
\usepackage{float}
\usepackage{color}
\usepackage{caption}
\usepackage{subcaption}

\def\log{{\rm{log}}}

\def\v{{\rm{\bf v}}}
\def\e{{\rm{\bf e}}}

\def\n{{\rm \bf n}}
\def\t{{\rm \bf t}}

\usepackage{color} %new package
\newcommand{\ks}{\textcolor{black}}
\begin{document}

%\preprint{APS/123-QED}

\title{Freezing of sessile droplet and frost halo formation}

\author{Sivanandan Kavuri$^1$, George Karapetsas$^2$, Chander Shekhar Sharma$^3$ and Kirti Chandra Sahu$^1$\footnote{ksahu@che.iith.ac.in}}
\affiliation{ 
$^1$Department of Chemical Engineering, Indian Institute of Technology Hyderabad, Sangareddy, 502 284, Telangana, India \\
$^2$Department of Chemical Engineering, Aristotle University of Thessaloniki, Thessaloniki 54124, Greece\\
$^3$Department of Mechanical Engineering, Indian Institute of Technology Ropar, 140001 Rupnagar, India}%

\date{\today}% It is always \today, today,

\begin{abstract}
The freezing of a sessile droplet unveils fascinating physics, characterised by the emergence of a frost halo on the underlying substrate, the progression of the liquid-ice interface, and the formation of a cusp-like morphology at the tip of the droplet. We investigate the freezing of a volatile sessile droplet, focusing on the frost halo formation, which has not been theoretically explored. The formation of the frost halo is associated with the inherent evaporation process in the early freezing stages. We observe a negative evaporation flux enveloping the droplet in the initial stages, which indicates that vapour produced during freezing condenses on the substrate close to the contact line, forming a frost halo. The condensate accumulation triggers re-evaporation, resulting in a temporal shift of the frost halo region away from the contact line. Eventually, it disappears due to the diffusive nature of the water vapour far away from the droplet. We found that increasing the relative humidity increases the lifetime of the frost halo due to a substantial reduction in evaporation that prolonged the presence of net condensate on the substrate. Increasing liquid volatility increases the evaporation flux, and condensation occurs closer to the droplet, as a higher amount of vapour is in the periphery of the droplet. We also found that decreasing the thermal conductivity of the substrate increases the total freezing time. The slower freezing process is accompanied by increased vaporized liquid, resulting in condensation with its concentration reaching supersaturation.
\end{abstract}


\maketitle

%\tableofcontents

\section{Introduction} \label{sec:intro}

Droplet freezing is ubiquitous in many practical applications, such as icing in aircraft and wind turbines, spray, food and pharmaceutical engineering, and natural phenomena, such as raindrop formation \citep{myers1998modeling,cao2015aircraft,mensah2015review,pruppacher1980microphysics}. Thus, several researchers examined the intricate physics underlying the freezing behaviour of a sessile droplet on supercooled substrates \citep{angell1983supercooled,jung2012frost}. However, despite significant advancements, there are still considerable gaps in our fundamental understanding of the freezing phenomenon and frost halo formation around a sessile droplet during the freezing process because of the coupling between the heat and mass transfer processes.

\begin{figure}
\centering
\includegraphics[width=0.95\textwidth]{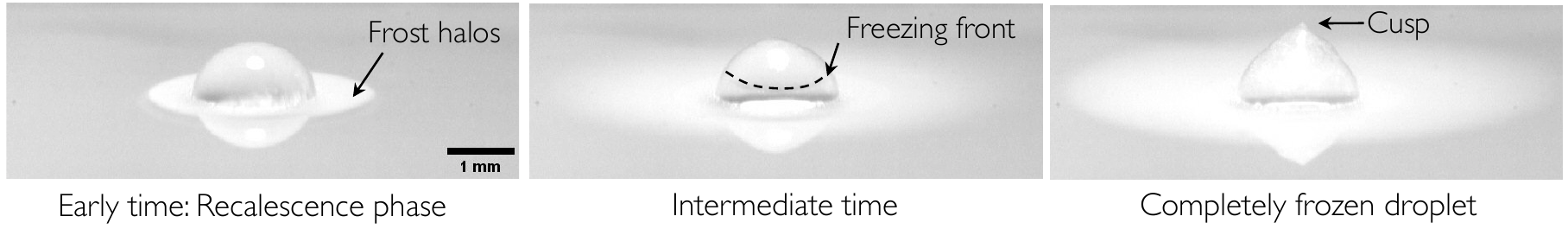}
\caption{Different stages observed during the freezing of a water droplet on a hydrophilic Polymethyl methacrylate (PMMA) substrate. Courtesy: Vineet Kumar - some preliminary experiments conducted in our group.}
\label{stages}
\end{figure}

The freezing of a sessile droplet occurs in two distinct phases. The early recalescence phase involves the rapid development of an ice-crystal scaffold and the temperature rise in the remaining liquid \citep{hu2010icing,jung2012frost}. The energy released as a liquid transforms into ice causes an increase in temperature. Subsequently, the second, much slower phase follows, which causes the remaining liquid to solidify through isothermal freezing. At the same time, more heat is released as the freezing front (liquid-ice interface) propagates towards the apex of the sessile droplet. The condensation of the vapor produced by the evaporation induced by the release of latent heat during this freezing process creates a frost halo around the frozen droplet \citep{jung2012frost}. Eventually, the apex of the frozen droplet exhibits an inflection point and a cusp-like morphology \citep{anderson1996case,zadravzil2006droplet} as a result of the volume expansion \citep{sanz1987influence} caused by the density difference between the ice and water phases. The different stages, namely the early recalescence, the intermediate stage when the freezing front moves towards the apex of the droplet and the final stage associated with the formation of the cusp at the tip of the droplet, during freezing of a water droplet on a hydrophilic Polymethyl methacrylate (PMMA) substrate obtained experimentally, are illustrated in figure \ref{stages}.

The earlier experimental study by \citet{hu2010icing} investigated the icing of a water droplet by examining the temperature field and demonstrated that the drop freezes from its bottom. \citet{enriquez2012freezing,marin2014universality} employed shadowgraphy to investigate the freezing of a water droplet on a copper plate immersed in a cooling bath of ethylene-glycol, ethanol and dry ice. They examined the temporal evolution of the freezing front and cusp formation at the tip of the droplet. \citet{peng2020study,pan2019experimental} studied the effect of wettability on the freezing behaviour of a water droplet. \citet{peng2020study} considered a superhydrophobic aluminum substrate, whereas \citet{pan2019experimental} used micropatterned stainless steel surfaces to regulate wettability. A few researchers also examined the effect of substrate curvature on the freezing behaviour of droplets \citep{liu2021supercooled,jin2017impact,zhang2018experimental,ju2018impact}. It is to be noted that the aforementioned studies and several others \citep{jung2011superhydrophobic,jung2012frost,huang2012effect,jung2012mechanism,boinovich2014effect,hao2014freezing,chaudhary2014freezing,zhang2016freezing} focused on the onset of freezing, freezing time and cusp formation at the tip of the droplet. They emphasised the influence of various factors (such as the wettability and roughness of the substrate and environmental conditions) on freezing dynamics. \citet{jung2012frost} investigated the formation of the frost halo and its subsequent freezing in the vicinity of the droplet by conducting experiments in controlled humidity conditions. \citet{jung2012frost} considered three different substrates, namely PMMA, polished titanium and copper substrates, which have heat conductivities ranging by three orders of magnitude but exhibit similar wettability. By carefully balancing the heat diffusion and vapour transfer, they observed frost halo formation on the substrate surrounding the droplet. A few studies also reported condensation frosting under saturation conditions around a freezing drop produced due to evaporation from the drop \citep{nath2017review,yancheshme2020mechanisms,castillo2021quantifying}. Further, the consequence of the frost halo formation during the freezing of multiple droplets was studied by \citet{graeber2018cascade}. They demonstrated that the vapour produced by the freezing process could alter the saturation conditions around neighbouring drops. This vapour can condense and freeze when the surrounding air is cooler than the saturation temperature. As a result, it forms tiny ice nuclei that significantly enhance the freezing dynamics. It has been recognised that drops do not freeze in isolation, and inter-droplet interactions, triggered by evaporation and halo formation, play a significant role in their frost spreading across the substrate \citep{nath2017review}. Moreover, the change in the nucleation mechanism due to the frost formation alters the icephobicity of a substrate.  

%Figure 1

A few researchers also studied the propagation of the freezing front, volume expansion and cusp formation at the tip of the droplet by conducting numerical simulations of the Navier-stokes equations in the framework of front-tracking \citep{vu2015numerical,vu2018numerical}, couple level-set and volume-of-fluid \citep{blake2015simulating} and lattice Boltzmann methods by incorporating the many-body dissipative particle dynamics with the energy conservation \citep{perez2021investigations,wang2022new}. \citet{tembely2018numerical} investigated the impact of a supercooled droplet on a cold substrate using a volume-of-fluid method. \citet{ajaev2004effect} conducted theoretical and experimental studies to examine the effect of tri-junction conditions corresponding to the fixed contact angle, fixed contact line, fixed growth angle and dynamic growth angle on the evolution of three interfaces. They showed that the dynamic growth angle condition could only predict the cusp formation observed in their experiments. \citet{zadravzil2006droplet} and \citet{Tembely2019} investigated the freezing dynamics of a two-dimensional sessile droplet by employing the lubrication approximation. They treated the shear stress singularity at the solid-liquid interface using a precursor layer technique and disjoining pressure formulation \citep{Schwartz1998}. \citet{zadravzil2006droplet} investigated the effect of a cooler porous substrate on the freezing of a sessile droplet. They analysed the spreading and imbibition that occurred along with the solidification inside the pores, resulting in the occlusion and contact line arrest of the droplet. Recently, \citet{Tembely2019} considered the curvature expression of the droplet-air and liquid-ice interfaces, disjoining pressure, the Gibbs-Thomson effect, natural convection, and substrate thermal and physicochemical properties, and investigated the impact of these parameters on the temporal evolution of the freezing front and cusp formation. It is to be noted that none of these numerical studies has considered evaporation and the recalescence phase during the freezing of sessile droplets. However, a few previous lubrication-based studies have examined only evaporation without considering freezing in sessile droplets on heated substrates (e.g., see Refs. \cite{Moosman1980,Ajaev2001,Craster2009c,Karapetsas2016}). \citet{stan2023rocket} observed that a frozen drop suspended in a vacuum migrates as a consequence of symmetry breaking driven by a localised increase in evaporation due to the heat release during the recalescence phase of freezing. In the experimental study conducted by \citet{sebilleau2021air}, the freezing of a water drop on a copper substrate was investigated under the conditions of a substrate temperature of $-15^{\circ}$C and an ambient temperature ranging from $15^{\circ}$C to $20^{\circ}$C. Their findings revealed a substantial impact of relative humidity on the freezing front propagation rates, while wettability did not significantly influence the process.

To summarize, the freezing behaviour, the propagation of the freezing front and the cusp formation at the apex of the frozen sessile droplet have been studied experimentally, numerically and theoretically. A sessile droplet undergoes frost halo formation around the droplet at the early recalescence phase. Although the frost halo was observed experimentally by \citet{jung2012frost}, it has not been theoretically investigated. To the best of our knowledge, the present study is the first attempt to model the formation of the frost halo using a lubrication approximation by incorporating evaporation and condensation of the vapor produced due to the release of the latent heat during freezing. We also estimate the temporal evolution of the width of the frost halo after the recalescence phase for different dimensionless parameters.      

The rest of this paper is organized as follows. In \S\ref{sec:model}, we present the problem formulation, governing equations, scalings and the numerical procedure adopted in the framework of lubrication approximation. After establishing the mechanism of the frost halo formation, a detailed parametric study is conducted in \S\ref{sec:Res}. Finally, concluding remarks are provided in \S\ref{sec:Conc}.

\section{Problem formulation} \label{sec:model}

\begin{figure}[h]
\centering
     \begin{subfigure}[b]{0.6\textwidth}
         \centering
         {\large (a)}
         \includegraphics[width=\textwidth]{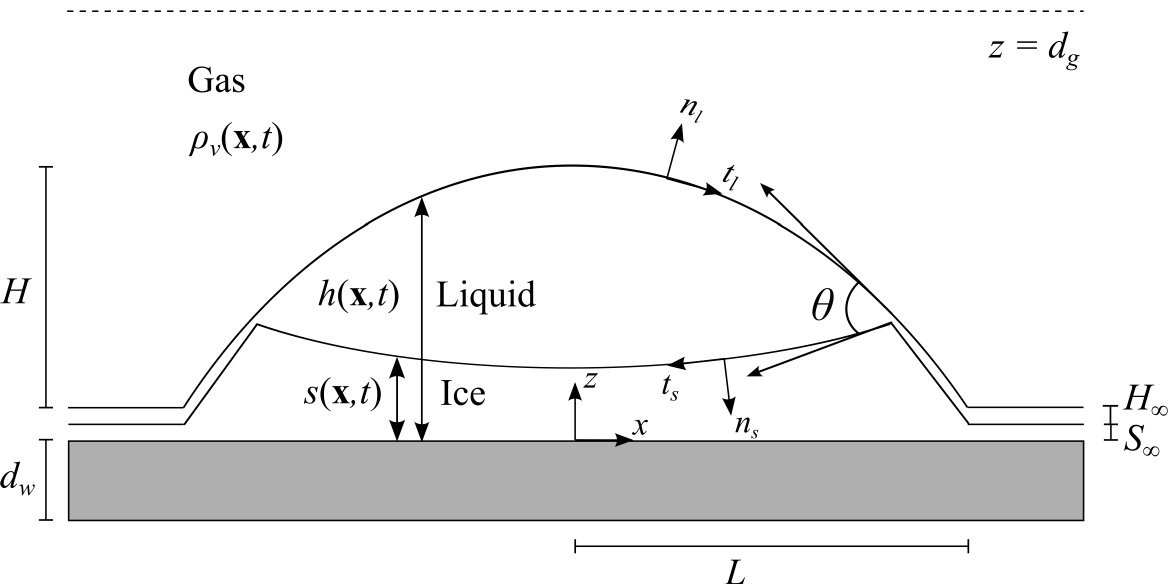}
     \end{subfigure}
     \vspace{3mm}
     \begin{subfigure}[b]{0.3\textwidth}
         \centering
         {\large (b)}
         \includegraphics[width=\textwidth]{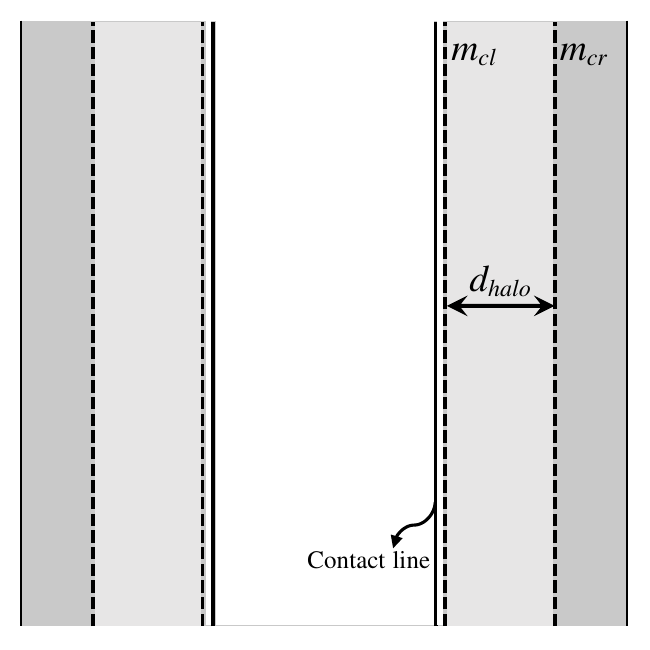}
     \end{subfigure}
\caption{(a) Schematic diagram of a sessile droplet undergoing freezing on a solid substrate. Here, $S_\infty$ and $H_\infty$ are the thickness of an initial microscopic ice layer and the thickness of the precursor layer, respectively. (b) Schematic representation of the left ($m_{cl}$) and right ($m_{cr}$) edges of the condensate mass and the thickness of the halo ($d_{halo}=m_{cr}-m_{cl}$) in the vicinity of the two-dimensional (2D) droplet.}
\label{geom}
\end{figure}

The freezing phenomenon of a sessile liquid droplet placed on a cold, horizontal solid substrate is investigated numerically using the lubrication approximation. Figure \ref{geom} depicts a schematic diagram of a two-dimensional (2D) droplet at a particular instant, $t$, during freezing, along with various parameters used in our modelling. The bottom of the substrate with thickness $(d_w)$, thermal conductivity $(\lambda_w)$ and specific heat $(C_{pw})$ is maintained at a constant temperature, $T_c$. We assume that, at $t=0$, the early recalescence phase has already occurred with the liquid reaching the melting temperature. At this point, the slower solidification step driven by the heat conduction ensues, and a very thin layer of ice of uniform thickness, $S_\infty$, has formed along the solid substrate. The height and half-width of the droplet are denoted by $H$ and $L$, respectively. The liquid is assumed to be incompressible and Newtonian, with constant density $(\rho_l)$, specific heat capacity $(C_{pl})$, thermal conductivity $(\lambda_l)$ and viscosity $(\mu_l)$. The surface tension of the liquid-gas interface $(\gamma_{lg})$, density of frozen solid phase $(\rho_s)$, specific heat capacity $(C_{ps})$ and thermal conductivity $(\lambda_s)$ are assumed to be constant. A Cartesian coordinate system $(x,z)$ with its origin at the centre of the droplet on the solid substrate is employed in our study as shown in figure \ref{geom}. Here, $z=s(x,t)$ and $z=h(x,t)$ represent the liquid-ice and liquid-gas interfaces, respectively. In the present work, we consider the drop to be very thin $(H \ll L)$. Thus, the aspect ratio of the droplet, $\epsilon=H/L$, is assumed to be very small. This assumption permits us to use the lubrication theory, employed below, to derive a set of evolution equations that govern the freezing dynamics of the sessile droplet considering the liquid, ice and gas phases. However, earlier studies demonstrated the validity of the lubrication model for contact angles up to $60^\circ$ \citep{charitatos2020thin,Tembely2019}.

\subsection{Dimensional governing equations}

\subsubsection{Liquid phase}

The dynamics in the liquid phase is governed by the mass, momentum and energy conservation equations, which are given by
\begin{eqnarray}
\rho_l \left( \frac{\partial \v}{\partial t} + \v \cdot \nabla \v \right) &=& - \nabla p + \mu_l \nabla^2 \v, \label{eq:mom} \\
\nabla \cdot \v &=& 0, \label{eq:cont} \\
\rho_l C_{pl} \left( \frac{\partial T_l}{\partial t} + \v \cdot \nabla T_l \right) &=& \lambda_l \nabla^2 T_l,\label{eq:energy_liquid}
\end{eqnarray}
where $\v$, $p$ and $T_l$ denote the velocity, pressure and temperature in the liquid phase, respectively; $\nabla$ represents the gradient operator. At the free surface ($z=h(x,t)$), the liquid velocity, $\v$, and the velocity of the liquid-gas interface, $\v_{lg}$, are related as
\begin{equation}
\v = \v_{lg} + (J_v/\rho_l) \; \n_l,
\end{equation}
where $J_v$ denotes the evaporative flux and $\n_l$ is the outward unit normal on the interface. However, the tangential components of both velocities, $\v_{\tau} = \v - (\v \cdot \n_l) \n_l = \v_{lg} - (\v_{lg} \cdot \n_l) \n_l$, are the same. Moreover, at $z=h(x,t)$, the velocity field satisfies the local mass, force and energy balance in the liquid and gas phases \citep{Karapetsas2016}. Thus,
\begin{equation}
\rho_l (\v -\v_{lg}) \cdot \n_l = \rho_g (\v_g -\v_{lg}) \cdot \n_l,
\end{equation}
where $\rho_g$ and $\v_g$ denote the density and velocity field of the gas phase, respectively. By ignoring the effect of vapour recoil and considering viscous effects from the gas phase to be negligible, the tangential and normal to the free surface components of the force balance at the liquid-gas interface are given by 
\begin{eqnarray}
- p + \n_l \cdot \tau \cdot \n_l &=& - p_g + \gamma_{lg} \kappa_{lg} + \Pi, \label{eq_normal-stress-balance} \\
\n_l \cdot \tau \cdot \t_l &=& 0 \label{eq_tangential-stress-balance}.
\end{eqnarray}
Here, $\tau=\mu_l \left( \nabla \v + \nabla \v ^T \right)$ denotes the extra stress tensor, $\kappa_{lg} = - \nabla_{s,l} \cdot \n_l$ denotes the mean curvature of the free surface, and $\nabla_{s,l} = (I-\n_l\n_l) \cdot \nabla$ represents the surface gradient operator. The disjoining pressure ($\Pi$) that accounts for the van der Waals interaction is defined as
\begin{equation} \label{eq:disj_press}
\Pi = A \left[ \left( \frac{B}{h-s} \right)^n - \left( \frac{B}{h-s} \right)^m \right],
\end{equation}
where $A= A_{Ham}/B^n \geq 0$ is a constant that describes the magnitude of the energy of the intermolecular interactions between the liquid-gas and liquid-ice interfaces, and $B$ denotes the precursor layer thickness. Here, $n > m > 1$ and $A_{Ham}$ denotes the dimensional Hamaker constant.

%Figure 2

The heat balance over the liquid-gas interface is given by
\begin{equation}
J_v L_v + \lambda_l \nabla T_l \cdot \n_l - \lambda_g \nabla T_g \cdot \n_l - h_{tc}(T_{g} - T_{lg}) = 0, \label{energy1}
\end{equation}
where $\lambda_g$ and $T_g$ denote the thermal conductivity and temperature of the gas phase, respectively. Here, $T_{lg}$ indicates the temperature at the liquid-gas interface, $h_{tc}$ represents the convective heat transfer coefficient, $L_v$ is the specific internal latent heat of vaporization.

At the liquid-ice interface $(z=s(x,t))$, the velocity is given by
\begin{equation} \label{eq:mass_bc_solid_ND}
\v =  \v_{ls} - (J_s/\rho_l) \; \n_s,
\end{equation}
where $J_s$ denotes the freezing mass flux and $\v_{ls}$ the velocity of the liquid-ice interface. As we impose the no-slip condition at the liquid-ice interface, the tangential component of the velocity is given by
\begin{equation} \label{eq:ice_no_slip}
\v \cdot \t_s = 0.
\end{equation}
Here, $\n_s$ and $\t_s$ are the outward unit normal and unit tangential vectors on the liquid-ice interface, respectively.

\subsubsection{Solid (ice) phase} \label{solid_ice_phase}

The energy conservation equation in the solid (ice) phase is given by
\begin{equation}\label{eq:energy_ice}
\rho_s C_{ps} \frac{\partial T_s}{\partial t} = \lambda_s \nabla^2 T_s,
\end{equation}
where $T_s$ denotes the temperature in the solid phase.

At the solid substrate ($z=0$), we impose continuity of temperature
\begin{equation}
T_s = T_w,
\end{equation}
where $T_w$ is the temperature of the substrate at $z=0$.

At the freezing front ($z=s(x,t)$), the boundary condition for the temperature is expressed as
\begin{equation} \label{eq:temp_bc_liquid_solid}
T_s = T_l = T_f. 
\end{equation}
We also assume that equilibrium temperature at the freezing front, $T_f$, is the same as the melting temperature, $T_m$.

At $z=s(x,t)$, the conservation of mass and energy between the liquid and solid phases leads to
\begin{equation} \label{eq:mass_bc_solid}
J_s = \rho_l (\v_{ls} - \v) \cdot \n_s = \rho_s (\v_{ls} - \v_s) \cdot \n_s,
\end{equation}
\begin{equation} \label{eq:energy_bc_solid}
\rho_s (\v_{ls} - \v_s) \cdot \n_s H_s - \lambda_s \nabla T_s \cdot \n_s = \rho_l (\v_{ls} - \v) \cdot \n_s H_l - \lambda_l \nabla T_l \cdot \n_s,
\end{equation}
where $\v_s$ denotes the velocity of the ice phase; $H_s$ and $H_l$ denote the enthalpy of the ice and liquid, respectively. By combining eqs. (\ref{eq:mass_bc_solid}) and (\ref{eq:energy_bc_solid}) and assuming that $\v_s=0$, we get
\begin{equation} \label{eq:energy_bc_solid_final}  
J_s \Delta H_{sl} - \lambda_s \nabla T_s \cdot \n_s + \lambda_l \nabla T_l \cdot \n_s = 0,
\end{equation}
where $\Delta H_{sl} = H_s - H_l$ denotes the enthalpy jump at the liquid-ice interface. Considering that $H_s = C_{ps} (T_f - T_m) + L_f(T_m)$ and $H_l = C_{pl} (T_f - T_m)$, we evaluate 
\begin{equation} \label{eq:DHsl}  
\Delta H_{sl} = (C_{ps} - C_{pl}) (T_f - T_m) + L_f,
\end{equation}
where $L_f$ denotes the latent heat of fusion considering the melting temperature, $T_m$ as the reference temperature. As will be shown below, eq. (\ref{eq:energy_bc_solid_final}) can be used to evaluate the position of the freezing front, $s(x,t)$.

\begin{table}
\centering
\caption{Typical values of the physical parameters considered in our simulations. These properties are for water-air system as considered by \citet{jung2012frost}.}
\begin{tabular}{ccc}
Property   &     Notation       & Value  \\  \hline
Density of the liquid phase   & $\rho_{l}$ &  1000 Kg m$^{-3}$ \\
Density of the frozen solid phase   & $\rho_{s}$ & 900 Kg m$^{-3}$  \\
Density of the gas phase   & $\rho_{g}$ &  $1.65 \times 10^{-3}$ Kg m$^{-3}$ \\
Viscosity of the liquid phase   & $\mu_{l}$ &  $10^{-3}$ Pa$\cdot$s\\
Viscosity of the gas phase   & $\mu_{g}$ &  $1.81 \times 10^{-5}$ Pa$\cdot$s \\
Melting temperature  & $T_m$ &  273 K \\
Ambient temperature	 & $T_g$ &  258.5 K \\
Temperature at the bottom of the substrate   & $T_c$ & 258.5 K  \\ 
Thickness of the substrate   & $d_w$ & $3 \times 10^{-3}$ m \\  
Thermal conductivity of the substrate   & $\lambda_w$ &  0.1 - 500 W m$^{-1}$ K$^{-1}$\\ 
Thermal conductivity of the liquid phase   & $\lambda_l$ &  0.57 W m$^{-1}$ K$^{-1}$ \\ 
Thermal conductivity of the frozen solid phase   & $\lambda_s$ &  2.21 W m$^{-1}$ K$^{-1}$ \\
Thermal conductivity of the gas phase   & $\lambda_g$ &  0.02 W m$^{-1}$ K$^{-1}$ \\ 
Specific heat capacity of the substrate   & $C_{pw}$ & 300 - 3000 J Kg$^{-1}$K$^{-1}$ \\ 
Specific heat capacity of the liquid phase   & $C_{pl}$ &  4220 J Kg$^{-1}$K$^{-1}$\\
Specific heat capacity of the frozen solid phase   & $C_{ps}$ &  2050 J Kg$^{-1}$K$^{-1}$ \\
Specific heat capacity of the gas phase   & $C_{pg}$ & 4220 J Kg$^{-1}$K$^{-1}$  \\
Surface tension of the liquid-gas interface at $T_m$  & $\gamma_{lgo}$ &  0.07 N m$^{-1}$ \\
Surface tension of the liquid-ice interface    & $\gamma_{ls}$ &  0.02 N m$^{-1}$ \\
Latent  heat  of  evaporation & $L_{v}$ &  $2.45\times10^6$ J Kg$^{-1}$ \\
Latent  heat  of  fusion   & $L_{f}$ &  $3.35\times10^5$ J Kg$^{-1}$\\
Diffusion coefficient of the vapour in the gas phase   & $D_m$ &  $1.89 \times 10^{-5}$ m$^2$/s\\
Universal  gas  constant   & $R_g$ & 8.314 J K$^{-1}$ mol$^{-1}$ \\
Accommodation  coefficient & $\alpha$ &  $\approx 1$ \\
Relative humidity & ${RH}$ &  $0 -1$ \\
Heat transfer coefficient & $h_{tc}$ &  10 - 7000 W m$^{-2}$ K$^{-1}$ \\
\hline
\end{tabular}
\label{T:wecr} 
\end{table}

\subsubsection{Gas phase}

In order to account for situations when a droplet freezes in an environment with varying humidity, we consider that the gas phase consists of both air and vapour. Based on the assumption that $\lambda_g \ll \lambda_l$ and thus we neglect thermal effects in the gas phase and thus consider that the gas has a constant bulk temperature, $T_g$. The gas phase velocity ($\v_{g}$) is assumed to be varied linearly between the liquid-gas interface and far away from the droplet, such that $\v_g \cdot \t_l=\v_{lg} \cdot \t_l$ at $z=h$ and $\v_{g}=0$ at $z=d_g$; the latter assumption is valid in the case where the gas phase can be considered to be quiescent far from the droplet. The concentration balance in the gas phase is given by
\begin{equation} \label{eq:Lap_vap_conc}
\frac{\partial \rho_v}{\partial t} + \v_g \cdot \nabla \rho_v = D_{m}\nabla^2\rho_v,
\end{equation}
where $\rho_v$ is the vapor concentration in the gas phase and $D_{m}$ represents the diffusion coefficient of the vapour in the gas phase. We consider the case of a well-mixed gas phase where proper ventilation maintains constant vapour concentration at some distance from the droplet $(\rho_{vi})$. Thus, at ${z=d_g}$, $\rho_v = \rho_{vi}$.

The liquid in the droplet and the vapour in the gas phase are coupled as a consequence of the evaporation and condensation at the liquid-gas interface. In the gas phase, the vapour mass flux is related to the departure from the uniform vapour density, which is given by
\begin{equation} \label{eq:ev_flux}
J_v = - D_m \left( \n_l \cdot \nabla \rho_v \right) \; \mbox{at} \; z = h(x,t).
\end{equation}

Additionally, the kinetic theory leads to a linear constitutive relation between the mass and the departure from equilibrium at the interface, which is known as the Hertz–Knudsen relationship \citep{Prosperetti1984,Sultan2005,Karapetsas2016}. This is given by
\begin{equation} \label{eq:Hertz_Knudsen}
J_v = \alpha \left( \frac{R_g T_{m}}{2 \pi M} \right)^{1/2} \left( \rho_{ve} (T_{lg}) - \rho_v \right),
\end{equation} 
where $R_g$ is the universal gas constant, $M$ is the molecular weight and $\alpha$ is the accommodation coefficient (close to unity). The equilibrium vapour concentration, $\rho_{ve} (T_{lg})$, can be obtained by employing a linear temperature-dependent equation of state:
\begin{equation}
\rho_{ve} (T_{lg}) = \rho_{ve} (T_m) \left[ 1 + \frac{M(p - p_{g})}{\rho_{l}R_gT_{m}} + \frac{ML_{v}}{R_{g}T_{m}}\left(\frac{T_{lg}}{T_{m}} - 1\right)\right].
\end{equation} 

Here, $L_{v}$ is the latent heat of evaporation. The combination of eqs. (\ref{eq:ev_flux}) and (\ref{eq:Hertz_Knudsen}) leads to
\begin{equation} \label{eq:int_vap_conc}
D_{m} \left( \n_l \cdot \nabla \rho_v \right) = - \alpha \left( \frac{R_g T_{m}}{2 \pi M} \right)^{1/2} \left(  \rho_{ve} (T_{lg}) - \rho_v  \right),
\end{equation} 
which can be used to evaluate the local interfacial vapour concentration, $\rho_v$. We use eq. (\ref{eq:int_vap_conc}) as the boundary condition at $z=h(x,t)$ to solve eq. (\ref{eq:Lap_vap_conc}).

\subsubsection{Solid substrate}

The energy conservation equation in the solid substrate is given by
\begin{equation}\label{eq:energy_solid}
\rho_w C_{pw} \frac{\partial T_w}{\partial t} = \lambda_w \nabla^2 T_w,
\end{equation}
where $\rho_{w}$ represents the density of the substrate. The above equation is subjected to the continuity of thermal flux at the ice-substrate interface ($z=0$): 
\begin{equation}
\lambda_s \frac{\partial T_s}{\partial z}  = \lambda_w \frac{\partial T_w}{\partial z},
\end{equation}
and at the bottom of the substrate, ${z=-d_w}$,
\begin{equation}
T_w = T_c.
\end{equation}

The properties of fluids and range of physical conditions considered in our study are listed in Table \ref{T:wecr}. 

\subsection{Scaling and evolution equations} \label{scaling}
The governing equations and boundary conditions are nondimensionalized using the following scalings (wherein tilde denotes the dimensionless variable):
\begin{equation} \label{eq:characteristic_scales}
\begin{gathered}
( x, z, h, d_{w}, d_{g} ) = L ( \tilde{x}, \epsilon \tilde{z}, \epsilon \tilde{h} , \epsilon \tilde{D_{w}}, \epsilon \tilde{D_{g}}), ~ t = \frac{L}{U} \tilde{t}, ~ ( u, w ) = U ( \tilde{u}, \epsilon \tilde{w} ),\\ 
(u_g, w_g ) = U ( \tilde{u_{g}}, \epsilon \tilde{w_{g}} ), ~ (p,\Pi) = \frac{\mu_l U L}{H^2} (\tilde{p},\tilde{\Pi}), ~ T_i = \Delta T \; \tilde{T}_i  + T_c \; (i=l,s,w), \\ 
\gamma_i =  \gamma_{lgo} \tilde{\gamma}_i \; (i=lg,ls), ~
J_s = \epsilon \rho_s U \; \tilde{J}_s, ~ J_v = \epsilon \rho_{ve}(T_m) U \tilde{J}_v, ~ \rho_v = \rho_{ve}(T_m) \; \tilde{\rho}_v , \\ 
\nabla = \frac{1}{L} \tilde{\nabla}, ~ \nabla_{s,i} = \frac{1}{L} \tilde{\nabla}_{s,i} \; (i=l,s),
\end{gathered}
\end{equation} 
where $\Delta T = T_m - T_c$, $\widetilde{\nabla} = \e_x\widetilde{\partial}_x +  \e_z \epsilon^{-1} \widetilde {\partial}_z$, $\widetilde{\nabla}_{s,i} = (I-\n\n) \cdot \widetilde{\nabla} \; (i=l,s)$ and $U = \epsilon^3 \gamma_{lgo} / \mu_l$ is the velocity scale. Henceforth, the tilde notation is suppressed, and ${\partial / \partial x}$, ${\partial / \partial z}$ and ${\partial / \partial t}$ are represented by the subscripts $x$, $z$ and $t$, respectively. By employing these scalings, the governing equations are rendered dimensionless, and the dimensionless groups that emerge are presented in Table \ref{T:dim_groups}.  

\begin{table}[h]
\renewcommand\arraystretch{1.5}
\caption{Dimensionless groups used in the present study.}
\centering
\begin{tabular}{cc}
Dimensionless group   &     Description  \\  \hline
$\epsilon = H / L$                                    & Droplet aspect ratio  \\
$D_w = d_w / H$                                       & Scaled wall thickness \\
$D_g = d_g / H$                                       & Scaled height of the gas phase domain \\
$T_v = (T_g - T_c)/(T_m - T_c)$                       & Scaled gas temperature \\
$D_v = \rho_{ve}(T_m)/ \rho_l$                        & Density ratio (vapour-liquid) \\
$D_s =\rho_s/\rho_l$                 & Density ratio (solid-liquid) \\
$\Lambda_s = \lambda_s / \lambda_l$                              & \quad Thermal conductivity ratio (ice-liquid) \\
$\Lambda_w = \lambda_w / \lambda_l$                              & \quad Thermal conductivity ratio (wall-liquid) \\
$\chi = \epsilon \rho_{ve}(T_m) U L_v H / (\lambda_l \Delta T)$   & \quad Scaled latent heat of vaporization \\
$Bi = h_{tc}H/\lambda_l$                                         & Biot number \\
$A_{n} = H A / \gamma_{lgo}$                                     & Scaled Hamaker constant \\
$Ste = \lambda_l \Delta T / ( \epsilon \rho_s U L_f H )$         & Stefan number \\
$Pe_v = {\epsilon U H/D_m}$                                & P\'eclet number \\
$V_{\rho,r} = \rho_{vi} / \rho_{ve}(T_m)$          & Vapour density ratio \\
$K = \epsilon U / \alpha \sqrt{ 2 \pi M / \left( R_g T_{m} \right) }$ & Dimensionless number associated with equilibrium condition at the interface\\
$\Delta = \mu_{l}ULM / \left( H^{2}\rho_{l}R_gT_m \right)$ &  Dimensionless number for pressure effect on evaporation \\
$\Psi = L_{v}M\Delta T / \left( R_{g}T_{m}^2 \right)$ &  Dimensionless number associated with evaporative cooling \\
\hline
\end{tabular}
\label{T:dim_groups} 
\end{table}
Incorporating the lubrication approximation ($\epsilon \ll 1$) and following the typical procedure (given in detail in the Appendix \S\ref{sec:lub}), we obtain the following evolution equations for the height of the liquid-gas and the liquid-ice interface:
\begin{equation}
\partial_t h - \partial_t s = - \partial_x q_l - D_v J_v - D_s J_s,
\end{equation}
\begin{equation}
\partial_t s = Ste \left(  \frac{\Lambda_w T_f}{D_w+s \Lambda_w/\Lambda_s} + \chi J_v - Bi(T_{v}-T_{l}\big|_{h}) \right),
\end{equation}
where
\begin{equation}
q_l = \frac{\partial_x p}{2} \left (\frac{h^3}{3} - s^2 h + \frac{2s^3}{3} \right) - h \partial_x p  \left (\frac{h^2}{2}-s h  + \frac{s^2}{2} \right).
\end{equation}
Moreover, the temperature profiles in the liquid, ice and solid substrate are given by 
\begin{eqnarray}
T_l &=& \left (-\chi J_v+Bi(T_{v}-T_{l}\big|_{h}) \right) (z-s) + T_f,
\\
T_s &=& \frac{T_f}{D_w+s \Lambda_w/\Lambda_s} \left( D_w + z  \frac{\Lambda_w}{\Lambda_s} \right),
\\
T_w &=& \frac{T_f}{D_w + s \; \Lambda_w/\Lambda_s}(z+D_w).
\end{eqnarray}

\subsubsection{Gas phase - K\'{a}rm\'{a}n-Pohlhausen approximation}

Special care is needed in order to evaluate the vapour concentration profile in the gas phase in order to retain the advection terms. To this end, we employ the K\'{a}rm\'{a}n-Pohlhausen integral approximation and define the integrated form of $\rho_v$, which is given by
\begin{equation}
\int_{h}^{D_g} \rho_v dz = f.
\label{eq:f_int}
\end{equation}
In order to be able to evaluate this integral, we need to prescribe the form of $\rho_v$ as a function of the vertical coordinate. To this end, we assume that $\rho_v$
can be approximated by a polynomial of the form 
\begin{equation}
    \rho_v = c_1 z^2 + c_2 z + c_3.
\end{equation}
By substituting the corresponding polynomial in eq. (\ref{eq:f_int}) and applying the appropriate boundary conditions at the liquid-gas interface and far from the droplet (see eqs. (\ref{bc:1_mass_bal}) and (\ref{eq:vap_conc_hum_scaled}) in the Appendix \S\ref{sec:lub}), it is possible to evaluate the polynomial constants and eventually derive the following expressions for the constants $c_1$, $c_2$ and $c_3$.
\begin{eqnarray}
    c_1 & = & \frac{f-\frac{Pe_v J_v}{2}(D_g-h)^2-V_{\rho,r}(D_g-h)}{\frac{2}{3}(h-D_g)^3},\\[3pt]
    c_2 & = & -Pe_ vJ_v - 2h\left[\frac{f-\frac{Pe_v J_v}{2}(D_g-h)^2-V_{\rho,r}(D_g-h)}{\frac{2}{3}(h-D_g)^3}\right],\\[3pt]
    c_3 & = & V_{\rho,r}-{D_g}^2\left[\frac{f-\frac{Pe_vJ_v}{2}(D_g-h)^2-V_{\rho,r}(D_g-h)}{\frac{2}{3}(h-D_g)^3}\right] + Pe_v J_vD_g \nonumber\\
    && + 2h D_g\left[\frac{f-\frac{Pe_v J_v}{2}(D_g-h)^2-V_{\rho,r}(D_g-h)}{\frac{2}{3}(h-D_g)^3}\right],
\end{eqnarray}
where $V_{\rho,r} = \rho_{vi} / \rho_{ve}(T_m)$ denotes that vapour density ratio, which is related with the relative humidity ($RH$) as
\begin{equation}\label{eq:RH_Vrhor}
RH = \frac{V_{\rho,r}}{1+\Psi(T_{v}-1)}.
\end{equation}
By integrating the vapour concentration balance (eq. \ref{eq:rho_v_scaled}) and using the boundary conditions, we get the following integrated form of the concentration equation which can eventually be solved with respect to $f$:
\begin{eqnarray}
     Pe_v\left[\frac{\partial f}{\partial t} + \frac{\partial g_v}{\partial x}  - \rho_{v}\big|_{h}(D_v J_v)\right]  &=& \nonumber \\
     \frac{\partial \rho_v}{\partial z}\big|_{D_g} - \frac{\partial \rho_v}{\partial z}\big|_{h} &+& \epsilon^2 \left[\frac{\partial }{\partial x}\left(\frac{\partial f}{\partial x} + \rho_v\big|_{h}h_x\right) + \frac{\partial \rho_v}{\partial x}\big|_{h}h_x\right].
\end{eqnarray}
In the above equation $g_v$ is defined as
\begin{equation} \label{eq:g_int}
\int_{h}^{D_g} u_g\rho_v dz = g_v.
\end{equation}
To evaluate eq. (\ref{eq:g_int}), we need to prescribe the $x$-component of the velocity profile in the gas phase, which can be approximated by considering the following linear profile
\begin{equation}\label{eq:u_g_scaled}
u_g = a z + b.
\end{equation}
The constants $a$ and $b$ can be evaluated by simply considering that, at the liquid-gas interface ($z=h(x,t)$), the velocity of the gas is equal to the velocity of liquid
\begin{equation}\label{eq:vap_velocity_bc1}
u_{g} = u,~ {\rm and} ~ w_{g} = w,
\end{equation}
and at the far-field ($z = D_g$) the gas phase is quiescent, 
\begin{equation}\label{eq:vap_velocity_bc2}
u_{g} = 0,~ {\rm and} ~ w_{g} = 0.
\end{equation}
Thus,
\begin{eqnarray}
    a & = & \frac{\frac{\partial_x p}{2}  (h^2 - s^2) - h \partial_x p (h-s)}{(h-{D_g})},\\[3pt]
    b & = & -{D_g}\left[\frac{\frac{\partial_x p}{2}   (h^2 - s^2) - h \partial_x p (h-s)}{(h-{D_g})}\right].
\end{eqnarray}

\subsection{Initial and boundary conditions - Numerical procedure and validation}

In our modelling, the droplet is deposited on a thin precursor layer that resides on top of a thin layer of ice. The precursor layer in our study can be related to the quasi-liquid layer, typically characterized by a thickness ranging from 1-100 nm, as discussed by \citet{nagata2019surface, slater2019surface}. While the quasi-liquid layer is a recognized phenomenon in supersaturated conditions, its existence in undersaturated conditions remains unestablished. Nevertheless, we note that a similar approach has successfully been used in earlier theoretical studies of droplet freezing (see \cite{zadravzil2006droplet, Tembely2019}) to alleviate the contact line singularity. The following initial conditions are imposed on the domain. 
\begin{eqnarray}
& h(x,t=0) = \max \left(h_{\infty}+s_{\infty}+1-x^{2},h_{\infty}+s_{\infty} \right), \\
& f(x,t=0) = V_{\rho,r}\left({D_{g}}-h(x,t=0)\right).
\end{eqnarray}
The dimensionless equilibrium precursor layer thickness ($h_{\infty}=(h-s_{\infty})= H_{\infty}/{H}$) far from the droplet can be estimated by considering that in this region, the fluid is flat with zero mean curvature and sufficiently thin such that the attractive van der Waals forces suppress evaporation. Therefore, by taking the constitutive equation for the evaporation flux and setting it to zero, the dimensionless equilibrium precursor thickness $h_{\infty}$ can be evaluated by solving the following nonlinear equation
\begin{equation}
1-{V_{\rho,r}} + \Delta\left(- \epsilon^{-2} A_{n} \left[ \left( \frac{\beta}{h_\infty} \right)^n - \left( \frac{\beta}{h_\infty} \right)^m \right]\right) + \Psi\left(\frac{1+Bi({T_{v}}h_\infty)}{1+Bi(h_\infty)}-1\right)= 0.
\end{equation}
In all our simulations, we have taken $\beta=0.01$. The initial thickness of the thin ice layer is taken to be $s_{\infty}=10^{-3}$, but we have verified that our findings remain unchanged when considering, e.g. one or two orders of magnitude smaller values of $s_{\infty}$. To prevent the precursor layer from freezing, we adopt a similar approach to that of \citet{zadravzil2006droplet} and introduce the thickness-dependent Stefan number $(Ste (x))$, which is given by the following expression.
\begin{equation}
Ste(x) = \frac{1}{2}(1+\tanh[4\times10^{3}((h-s)-1.4\beta)])Ste. 
\label{eq_penalty}
\end{equation}

Finally, we impose the following set of boundary conditions
\begin{eqnarray}
h_{x}(0,t) = h_{xxx}(0,t) &=& h_{x}(x_{\infty},t) = h_{xxx}(x_{\infty},t)= 0,
\\
s_{x}(0,t) = s_{xxx}(0,t) &=& s_{x}(x_{\infty},t) = s_{xxx}(x_{\infty},t)= 0,
\\
h(x_{\infty},t)-s(x_{\infty},t) &=& h_{\infty}, ~ s(x_{\infty},t) = s_{\infty}, 
\\
f_x(0,t) = 0, ~
f(x_{\infty},t) &=& {V_{\rho,r}} \left({D_{g}}-s(x_{\infty},t)-h_{\infty} \right).
\end{eqnarray}
where $x_{\infty}$ represents the end of the domain.

The Galerkin Finite Element Method is utilized to discretize all dimensionless governing equations. To mitigate boundary effects, our simulations incorporate a computational domain that is sufficiently large, extending six times the dimensionless initial radius. The optimum size of this computational domain is determined by investigating the effect of its size. A grid convergence test has been conducted for a representative parameter set considered in our study. We found that the results obtained with 2001, 4801 and 8001 grid points are indistinguishable (figure \ref{fig:Grid_Convergence} in Appendix \S\ref{sec:grid}). Consequently, we selected 4801 grid points for the remaining simulations. The solutions are obtained using the Newton-Raphson scheme, which is applied after the simulation has progressed in time using an implicit Euler method and an adaptive time step.

\subsubsection{Validation}
\begin{figure}[h]
\centering
\includegraphics[width=0.75\textwidth]{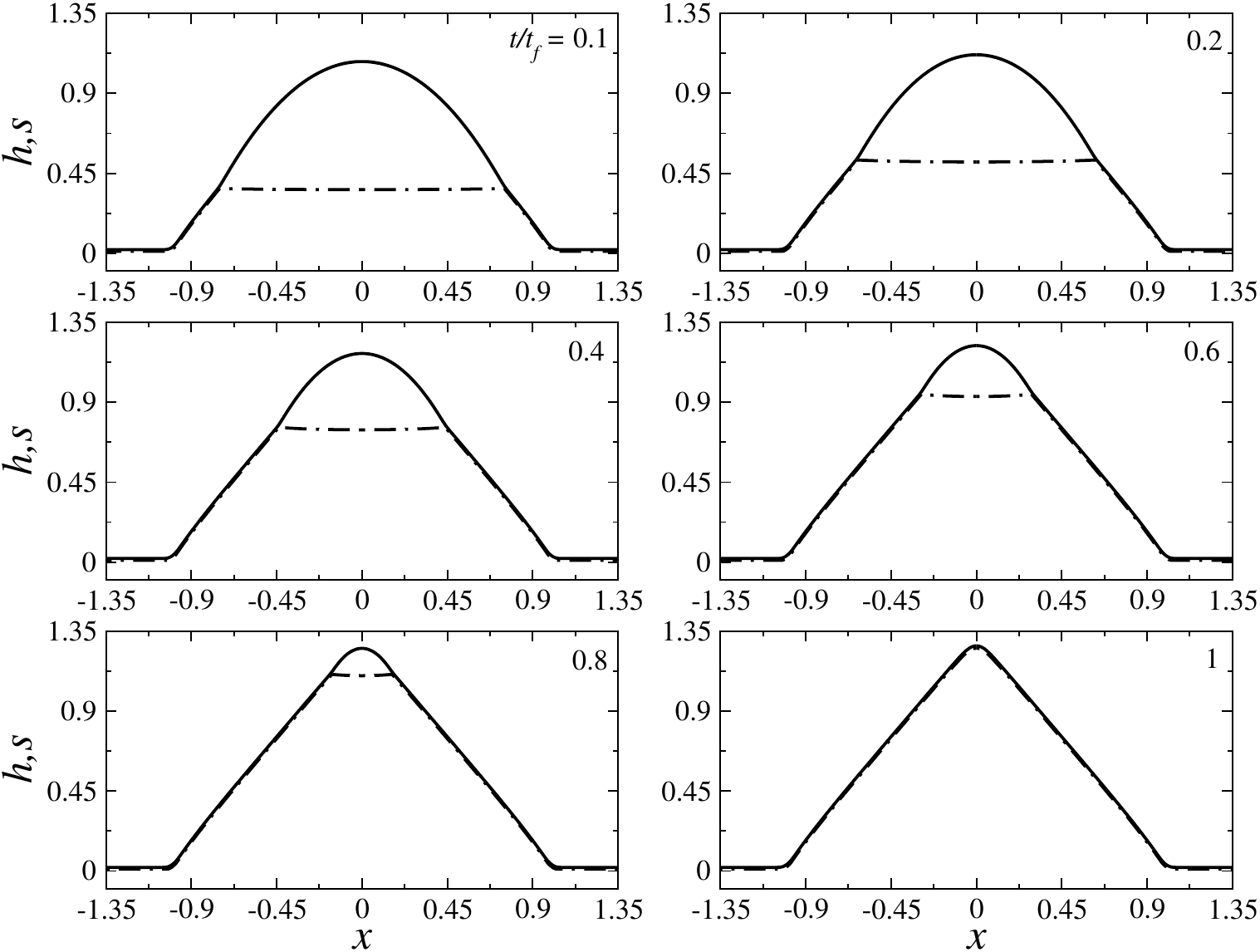}
\caption{Evolution of the freezing front, $s$ (dot-dashed lines) and shape of the droplet, $h$ (solid line) placed on a cold substrate. The rest of the dimensionless parameters are $\epsilon=0.2$, $Ste = 0.04$, $T_{v} = 0.5$, $A_{n} = 6.25$, $D_{s} = \Lambda_{S} = \Lambda_{W} = V_{\rho,r} = K = Bi = 1$, and $D_{w} = D_{v} = \Delta = \Psi = \chi = Pe_{v} = 0$. This set of parameters is similar to that of \citet{zadravzil2006droplet}. The value of the dimensionless total freezing time $(t_{f})$ of the droplet is 15. Here, the evaporation is neglected. }
\label{fig2_hvsx}
\end{figure}

In order to validate our solver, we simulate a case that was considered by \citet{zadravzil2006droplet} in figure \ref{fig2_hvsx}. This result mimics figure 15 of \citet{zadravzil2006droplet}. It shows the evolution of the freezing front, $s$ (dot-dashed line) and the shape of the droplet, $h$ (solid line), placed on a cold substrate for $Ste = 0.04$, $T_{v} = 0.5$, $A_{n} = 6.25$, $D_{s} = \Lambda_{S} = \Lambda_{W} = V_{\rho,r} = K = Bi = 1$ and $ D_{w} = D_{v} = \Delta = \Psi = \chi = Pe_{v} = 0$. Under this limit, we recover the formulation of \citet{zadravzil2006droplet} for a non-porous substrate without gravity effect. As depicted in figure \ref{fig2_hvsx}, the ice-front and the morphology of the liquid-gas interface in our study exhibit qualitative similarities to those observed in \citet{zadravzil2006droplet}.

%Figure 3

An important feature of the freezing process, typically observed in experiments, is the formation of a cusp at the drop apex when freezing ends. \citet{marin2014universality} experimentally observed a tip angle of $\sim 139^{\circ}$. They also reported that the tip angle remains independent of substrate temperature, wettability, and solidification rate. In figure \ref{fig:Tip_angle} of Appendix \S\ref{sec:tip}, we consider a case for a typical set of parameters considered in the present study ($Ste = 1.22 \times 10^{-3}$, $T_{v} = 1$, $A_{n} = 6.25$, $D_{g} = 2$, $D_{s} = 0.9$, $\Lambda_{S} = 3.89$, $\Lambda_{W} = 698$, $\chi = 0.01$, $K = 8\times10^{-4}$, $Bi = 0.16$, $D_{w} = 15$, $V_{\rho,r} = 0.70$, $\epsilon=0.2$, $D_{v} = 1.65 \times 10^{-6}$, $\Delta = 10^{-4}$, $\Psi = 0.94$ and $Pe_{v} = 1$). At this point, it is important to comment on the limitations of our numerical model. As mentioned above, to maintain the thickness of the precursor layer far from the drop and ensure accurate predictions for condensation and evaporation, we employ a penalty function, see eq. (\ref{eq_penalty}). However, this approach introduces a constraint on the freezing rate as the liquid layer thickness approaches that of the precursor layer. Consequently, our model is expected to fail at the end of the freezing process and thus cannot entirely capture the formation of a cusp at the conclusion of freezing. With this limitation in mind, and also considering that the cusp is expected to form only at the very end of the freezing process, we can infer the tip angle by the shape of the ice-gas interface in the vicinity of the tip. Thus, as shown in figure \ref{fig:Tip_angle} of Appendix \S\ref{sec:tip}, we provide an estimation for the tip angle by drawing the tangents at the interface near the cusp and evaluate the tip angle to be $\sim 145^{\circ}$, a prediction very close to the experimentally reported value.

\section{Results and Discussion} \label{sec:Res}

\subsection{Comparison with earlier studies - effect of evaporation} \label{sec:validation}

%Figure 4
\begin{figure}
\centering
\hspace{0.8cm}{\large (a)}   \hspace{6.5cm}  {\large (b)} \\
\includegraphics[width=0.4\textwidth]{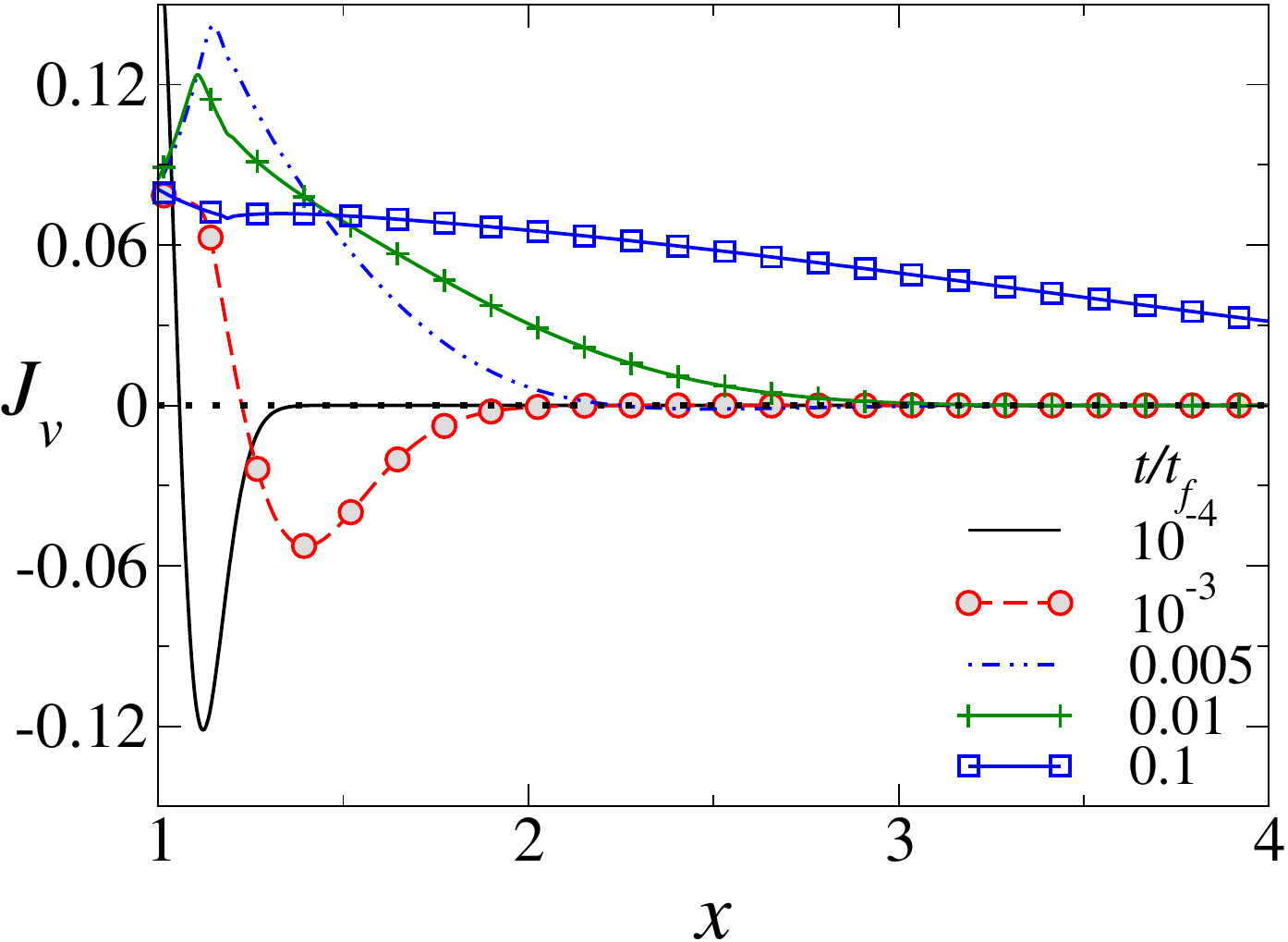}\hspace{0.0mm}
\includegraphics[width=0.41\textwidth]{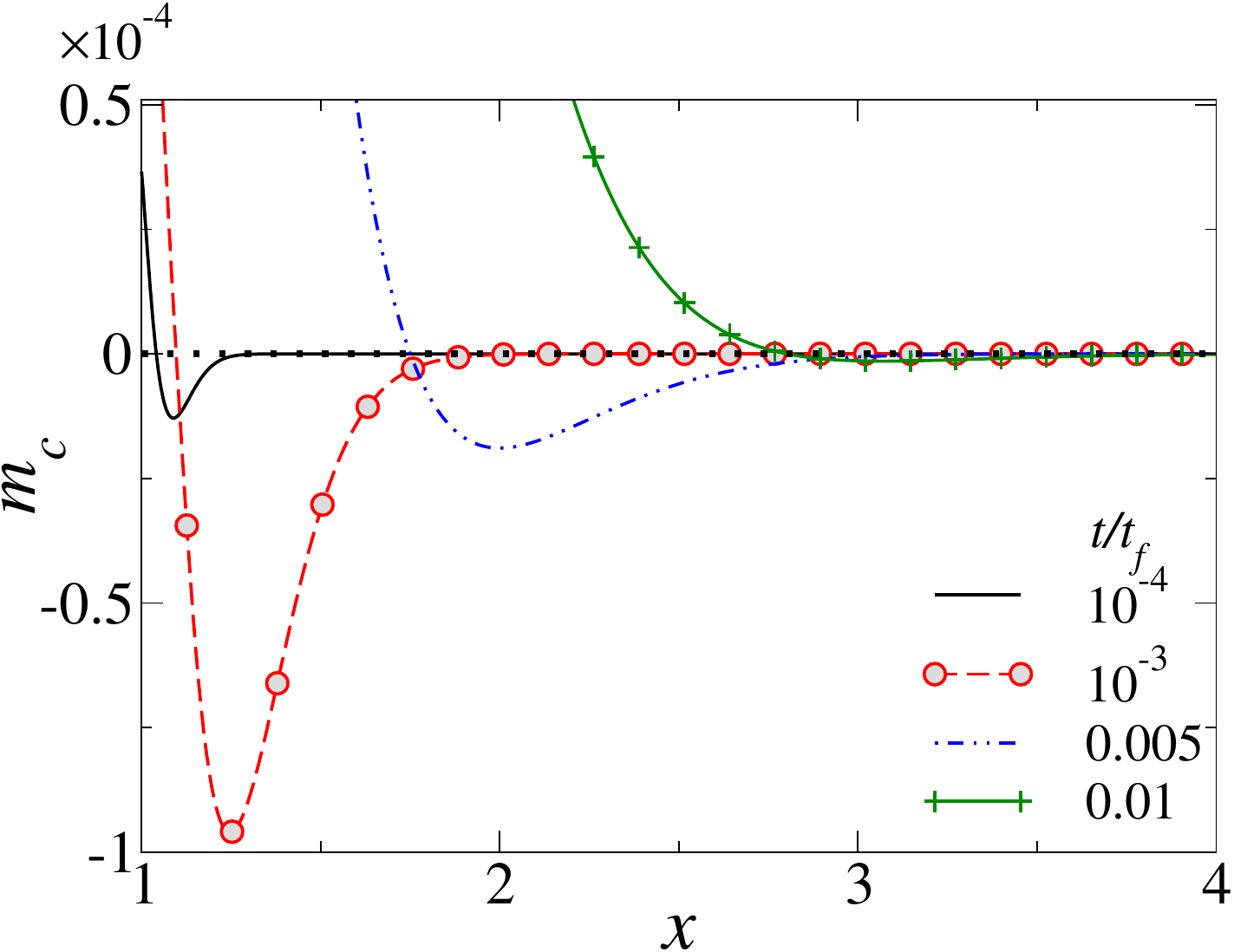} \\  
\hspace{0.8cm}{\large (c)}   \hspace{6.5cm}  {\large (d)}\\
\includegraphics[width=0.41\textwidth]{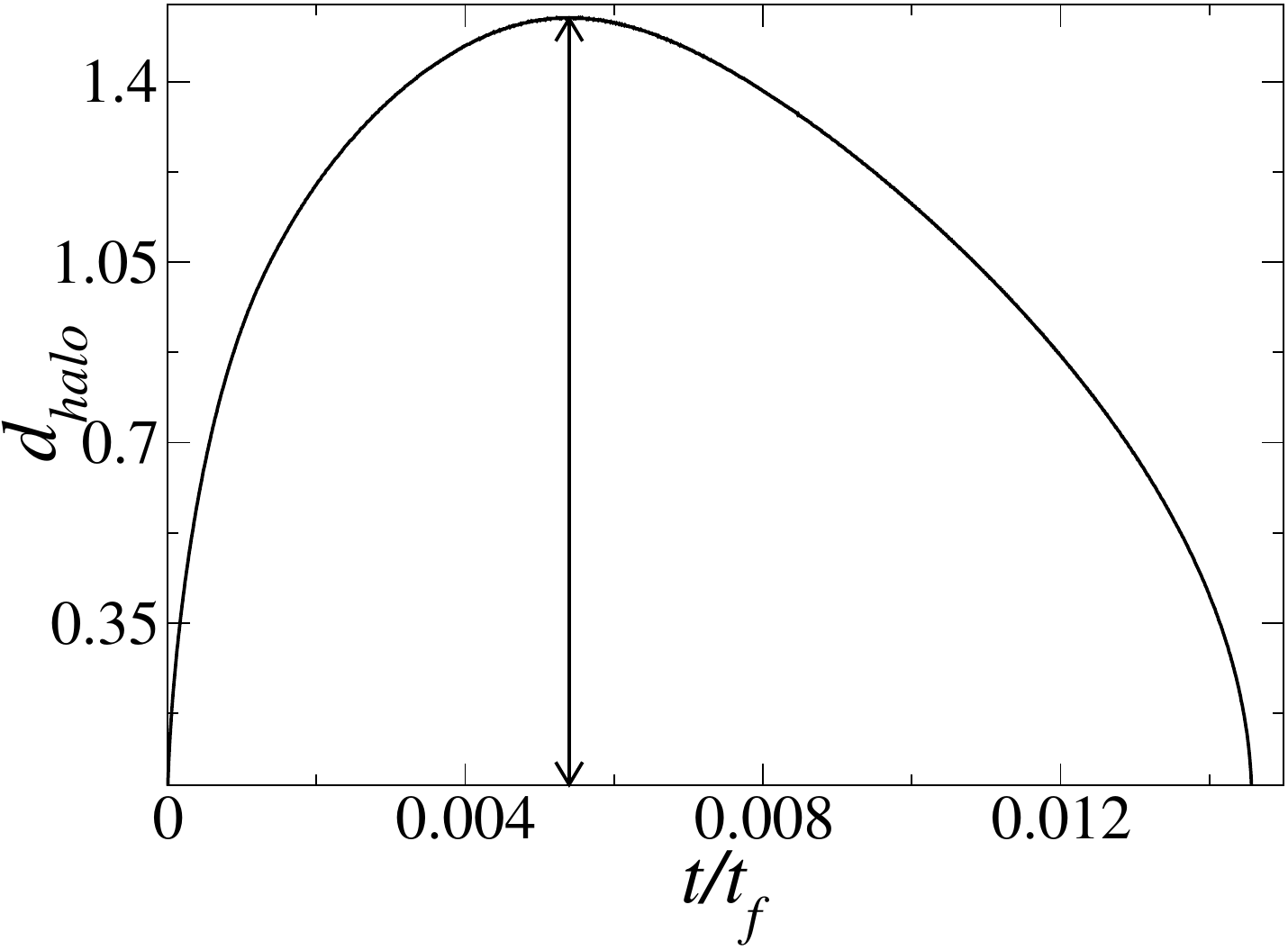} \hspace{0mm}
\includegraphics[width=0.4\textwidth]{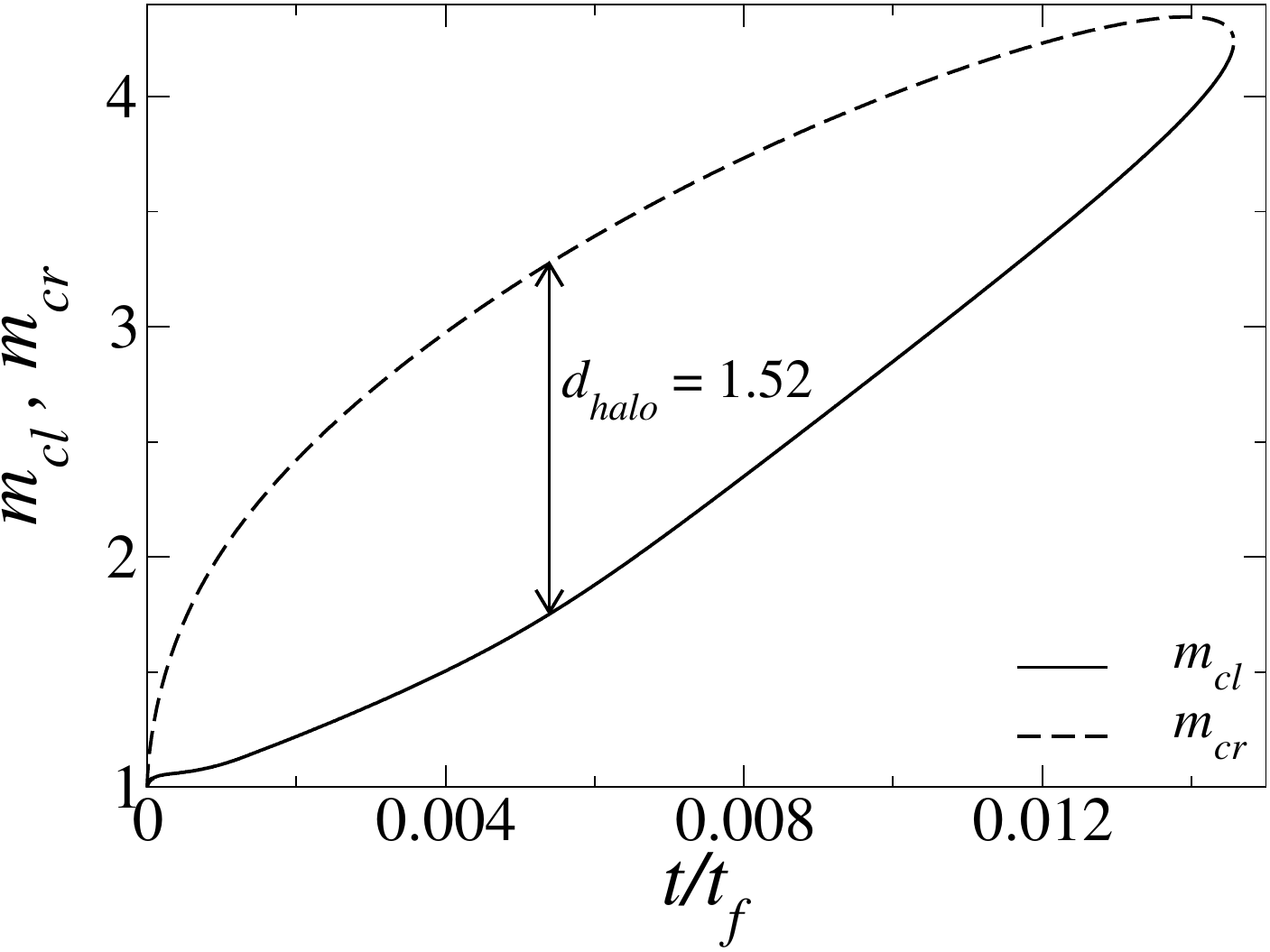}
\caption{(a) Variation of (a) the evaporation flux $(J_v)$ and (b) the total mass of the condensate $(m_c)$ along the substrate for different values of $t/t_f$.  Temporal variation of (c) the width of the frost halo $(d_{halo})$ and (d) the left $(m_{cl})$ and right $(m_{cr})$ ends of the condensation halo over the substrate. The maximum width of the halo is also marked in panels (c) and (d). The rest of the dimensionless parameters are $D_{w} = 7.5$, $\epsilon=0.2$, $Ste = 1.7 \times 10^{-4}$, $T_{v} = 0.2$, $A_{n} = 6.25$, $D_{g} = 2$, $D_{s} = 0.9$, $\Lambda_{S} = 3.82$, $\Lambda_{W} = 191$, $K = 8\times10^{-4}$, $Bi = 2.29$ and $D_{v} = 4.85 \times 10^{-6}$. The value of the dimensionless total freezing time $(t_{f})$ of the droplet is 1220, respectively.}
\label{fig:Tembely_Evap_Halo}
\end{figure}

We begin our investigation by examining the consequences of including evaporation in our model, which was neglected by the earlier theoretical investigations \citep{Tembely2019,zadravzil2006droplet}. To this end, we consider a case with the following set of parameters: $D_{w} = 7.5$, $\epsilon=0.2$, $Ste = 1.7 \times 10^{-4}$, $T_{v} = 0.2$, $A_{n} = 6.25$, $D_{g} = 2$, $D_{s} = 0.9$, $\Lambda_{S} = 3.82$, $\Lambda_{W} = 191$, $K = 8\times10^{-4}$, $Bi = 2.29$ and  $D_{v} = 4.85 \times 10^{-6}$. This set of parameters corresponds to a system with similar physical properties to that examined by \citet{Tembely2019}. While we observe (not shown) that the freezing front and droplet shape with and without evaporation obtained by setting $V_{\rho,r}= 1.0$, $\chi = 0$, $\Delta = 0$, $\Psi = 0$ and $Pe_{v} = 0$ (as considered by \citet{Tembely2019}) are qualitatively the same, the droplet freezes faster when evaporation is accounted for in the model. We also found that in the evaporation situation, as opposed to when it is neglected, the freezing front propagates towards the apex of the sessile droplet more quickly. For the set of parameters considered in figure \ref{fig:Tembely_Evap_Halo}, the values of the dimensionless total freezing time $(t_{f})$ of the droplet without and with evaporation are 1349 and 1220, respectively. This behaviour can be attributed to evaporative cooling and the effect of latent heat of evaporation mitigating the effect of latent heat of fusion in the droplet. It is also to be noted that the model incorporating evaporation exhibits a reduction in the expansion volume, albeit small, due to the difference in densities between the ice and water phases, as compared to a model without evaporation \citep{Tembely2019}. Note that our results cannot be directly compared to \citet{Tembely2019} since our simulations correspond to a thin ($\epsilon=0.2$) two-dimensional droplet, taking into account the limitations of lubrication approximation, in contrast to a thick ($\epsilon=0.7$) axisymmetric droplet considered by \citet{Tembely2019}. Another important difference is the fact that \citet{Tembely2019} introduced a modified latent heat into their formulation. As reported therein, it differs by almost a factor of three, with the normal latent heat, significantly affecting the results of their simulations. It was reported that this modification is necessary to achieve a good agreement with experiments. Nevertheless, it is important to note that the introduction of the modified latent heat is, in fact, incorrect and stems from an inconsistent use by these authors of the reference temperature in the evaluation of the enthalpy jump at the interface. As explained in \S\ref{solid_ice_phase}, the enthalpy jump is given by eq. (\ref{eq:DHsl}). Considering that the local temperature at the liquid-solid interface is very close to the melting temperature of ice, i.e. $T_f \approx T_m$ (even when considering the Gibbs-Thomson effect as was done by \citet{Tembely2019}), the effect of change in heat capacity between the two phases in eq. (\ref{eq:DHsl}) should actually be negligible. Additionally, \citet{Tembely2019} account differently for the heat loss due to convection in the gas phase, considering the dependence of the local gas temperature on the droplet height. The latter effect has been neglected in our formulation. Both these effects significantly impact the local temperature profile inside the droplet, generating a super-cooling of the droplet, which affects the predicted shape of the liquid-gas and liquid-ice interfaces, as will be discussed in detail in the parametric study that follows. To account for the super-cooling of the droplet, in figure \ref{fig:Tembely_Evap_Halo}, we consider a lower ambient temperature $(T_v=0.2)$ than that of \citet{Tembely2019}. The effect of $T_v$ is also investigated in \S\ref{effectTg}.

The most distinct difference between the situations with and without evaporation occurs during the early stages of freezing. It can be seen in figure \ref{fig:Tembely_Evap_Halo}(a) that the droplet exhibits negative values of $J_v$ at the early times in the region around the droplet. This indicates that the vapour generated during the freezing process condenses on the substrate in the vicinity of the contact line, forming a frost halo surrounding the freezing drop. To determine the size and temporal evolution of the condensation halo, we introduce the parameter $d_{halo}$, which represents the width of the frost halo formed around the droplet. This width corresponds to the region where the net condensate is accumulated, as determined numerically by integrating the evaporation flux over time. Figure \ref{fig:Tembely_Evap_Halo}(b) displays the mass of the net condensate near the droplet on the substrate at different instants. In this figure, the area where ($m_c$) becomes negative indicates the presence of the net condensate at that particular time. Moreover, figure \ref{fig:Tembely_Evap_Halo}(b) reveals that initially, the condensate accumulates near the contact line ($x \approx 1$), and as time progresses, this region moves away from the contact line. Considering the diffusive nature of water vapor, we expect minimal condensation far away from the droplet, which is also supported by our simulations shown in figures \ref{fig:Tembely_Evap_Halo}(a) and \ref{fig:Tembely_Evap_Halo}(b), where the magnitude of the negative evaporation flux and net condensate decreases as we move away from the contact line. The interplay between condensation and evaporation processes influences the location and width of the frost halo. The condensation halo expands due to net condensation and contracts when net evaporation occurs. As the mass of the condensate increases in a particular region of the substrate, the condensate starts to re-evaporate, causing a shift in the area where the evaporation flux is negative. Consequently, the region where the net condensate is present changes as time progresses. In order to better comprehend this phenomenon, we have plotted the temporal variation of ($d_{halo}$) in figure \ref{fig:Tembely_Evap_Halo}(c). The temporal variation of the leftmost ($m_{cl}$) and rightmost ($m_{cr}$) ends of the net condensate are shown in figure \ref{fig:Tembely_Evap_Halo}(d). It is evident from figure \ref{fig:Tembely_Evap_Halo}(c) that $d_{halo}$ initially increases, reaches a maximum and then decreases before reaching the end of its lifetime. Moreover, as shown by figure \ref{fig:Tembely_Evap_Halo}(d), the location of the frost halo also gradually shifts away from the contact line over time, as shown by the evolution of the leftmost ($m_{cl}$) and rightmost ($m_{cr}$) ends of frost halo. ($m_{cl}$) and ($m_{cr}$) are initially close to each other, then move apart as $d_{halo}$ grows, and finally come closer to each other towards the end of the lifetime of the halo. It can be seen in figure \ref{fig:Tembely_Evap_Halo}(d) that, for the set of parameters considered, the leftmost ($m_{cl}$) and rightmost ($m_{cr}$) ends of the halo approach each other at ($x \approx 4.3$) just before eventually disappearing. These results confirm that the droplet exhibits a frost halo during the early stages of the freezing process $(0 < t/t_f < 0.014)$. 

In the discussion that follows, we conduct a parametric investigation to comprehend the underlying physics and effects of various parameters on the formation of frost halo. Since the formation of the frost halo has been studied in detail in the experimental work of \citet{jung2012frost}, we will henceforth focus our attention on the range of parameters considered by these authors and consider a `base' case with the typical values of $Ste = 1.22 \times 10^{-3}$, $T_{v} = 0$, $A_{n} = 6.25$, $D_{g} = 2$, $D_{s} = 0.9$, $\Lambda_{S} = 3.89$, $\Lambda_{W} = 0.33$, $\chi = 0.01$, $K = 8\times10^{-4}$, $Bi = 0.16$, $D_{w} = 15$,  $D_{v} = 1.65 \times 10^{-6}$, $\Delta = 10^{-4}$, $\Psi = 0.94$ and $Pe_{v} = 1$. These parameters correspond to the freezing of a slender supercooled water droplet at $-14.5^{\circ}\textrm{C}$ on a PMMA substrate and are kept unchanged in the subsequent figures unless noted otherwise. Under the limit of lubrication approximation, we consider a thin droplet with an equilibrium contact angle of $14.3^{\circ}$. A similar system was examined in figure S5(A) of \citet{jung2012frost}, albeit for a more hydrophobic system.

\subsection{Effect of $RH$}

\begin{figure}
\centering
\hspace{0.8cm}{\large (a)}   \hspace{6.5cm}  {\large (b)} \\
\includegraphics[width=0.4\textwidth]{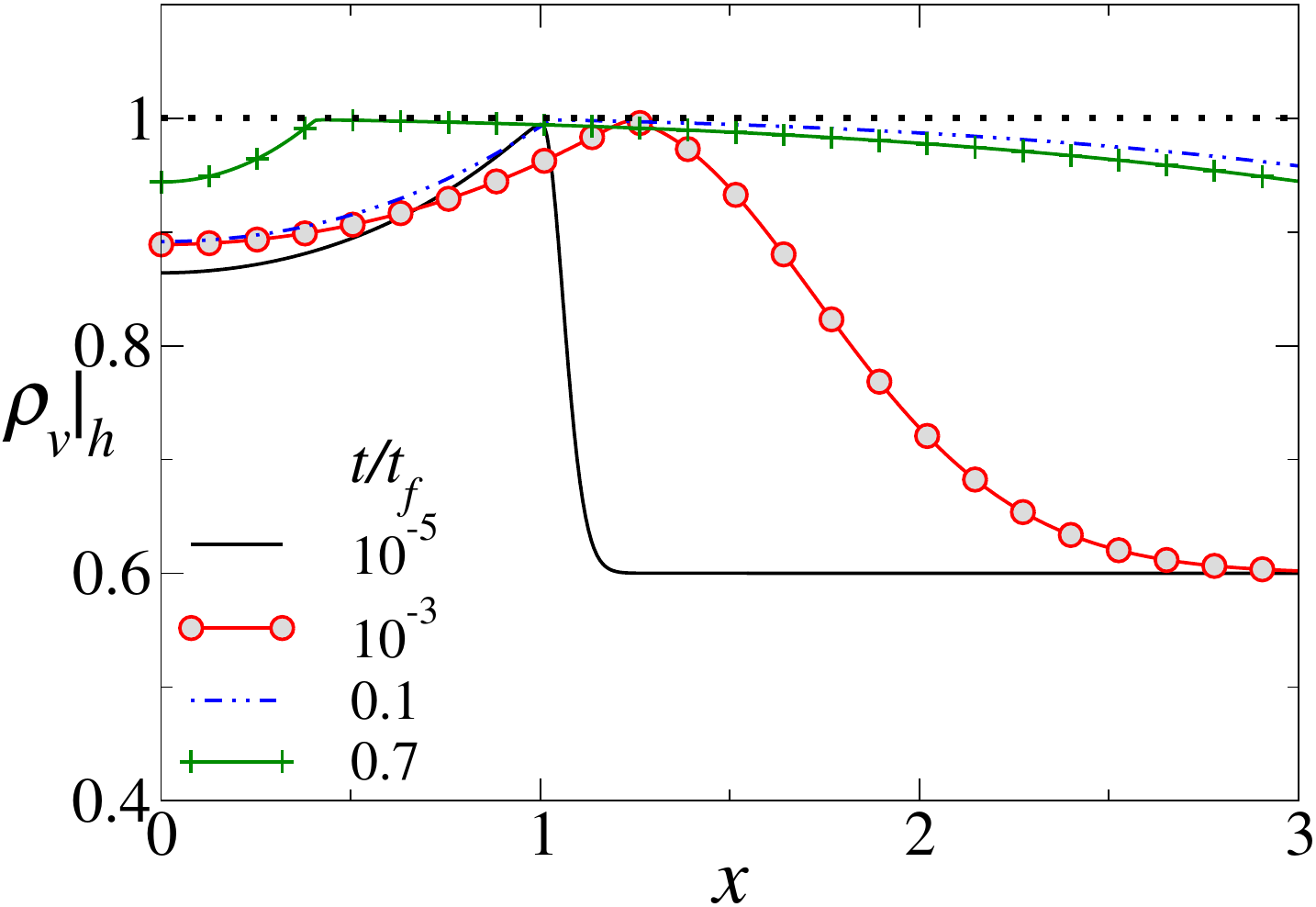} 
\includegraphics[width=0.4\textwidth]{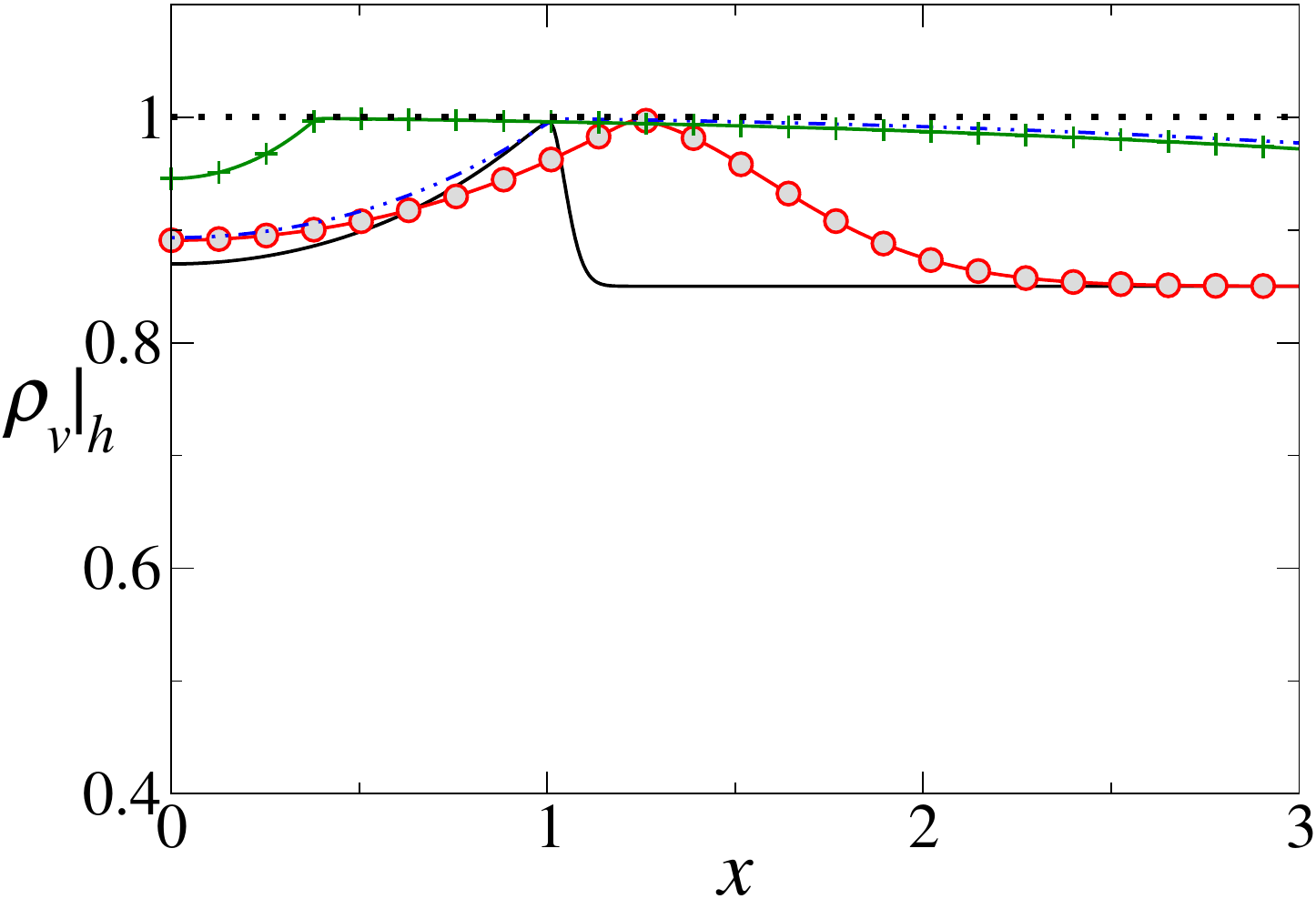} 
\caption{Variation of the vapour concentration $(\rho_{v}\big|_{h})$ along the substrate at different values of $t/t_f$ for (a) $V_{\rho,r} = 0.60$ and (b) $V_{\rho,r} = 0.85$. The values of the rest of the dimensionless parameters are $Ste = 1.22 \times 10^{-3}$, $T_{v} = 0$, $A_{n} = 6.25$, $D_{g} = 2$, $D_{s} = 0.9$, $\Lambda_{S} = 3.89$, $\Lambda_{W} = 0.33$, $\chi = 0.01$, $K = 8\times10^{-4}$, $Bi = 0.16$, $D_{w} = 15$, $\epsilon=0.2$, $D_{v} = 1.65 \times 10^{-6}$, $\Delta = 10^{-4}$, $\Psi = 0.94$ and $Pe_{v} = 1$. The dotted line represents the maximum vapour concentration attainable at the liquid-gas interface.}
\label{fig:RH_vapour}
\end{figure}

In this section, we investigate the effect of relative humidity on the freezing behaviour of the droplet by varying the value of the vapour density ratio $V_{\rho,r}$. The relationship between the vapour density ratio $V_{\rho,r}$ and relative humidity is given in eq. (\ref{eq:RH_Vrhor}). It can be noted that evaporation is only anticipated when the gas phase contains unsaturated humid air. Figure \ref{fig:RH_vapour} shows the profile of the vapour concentration along the liquid-gas interface at different values of $t/t_f$ for $V_{\rho,r}= 0.6$ and $0.85$ to examine the effect of vapour density ratio. The predicted total freezing times $(t_{f})$ for $V_{\rho,r}= 0.6$ and $0.85$ are 4700 and 5121, respectively. We note that at very early times  ($t/t_f=10^{-5}$), since the droplet environment is not saturated, water evaporates with vapour concentrating above the droplet; far from the droplet vapour concentration remains equal to the vapour density ratio of the ambient gas phase. A local maximum of the vapour concentration arises at the contact line since volatile droplets evaporate preferentially. Later, vapour diffuses ahead of the droplet, coming in contact with the supercooled precursor film ahead of the droplet and condenses there, as shown below.

%Figure 5

%Figure 6
\begin{figure}
\centering
\hspace{0.8cm}{\large (a)}   \hspace{6.5cm}  {\large (b)} \\
\includegraphics[width=0.4\textwidth]{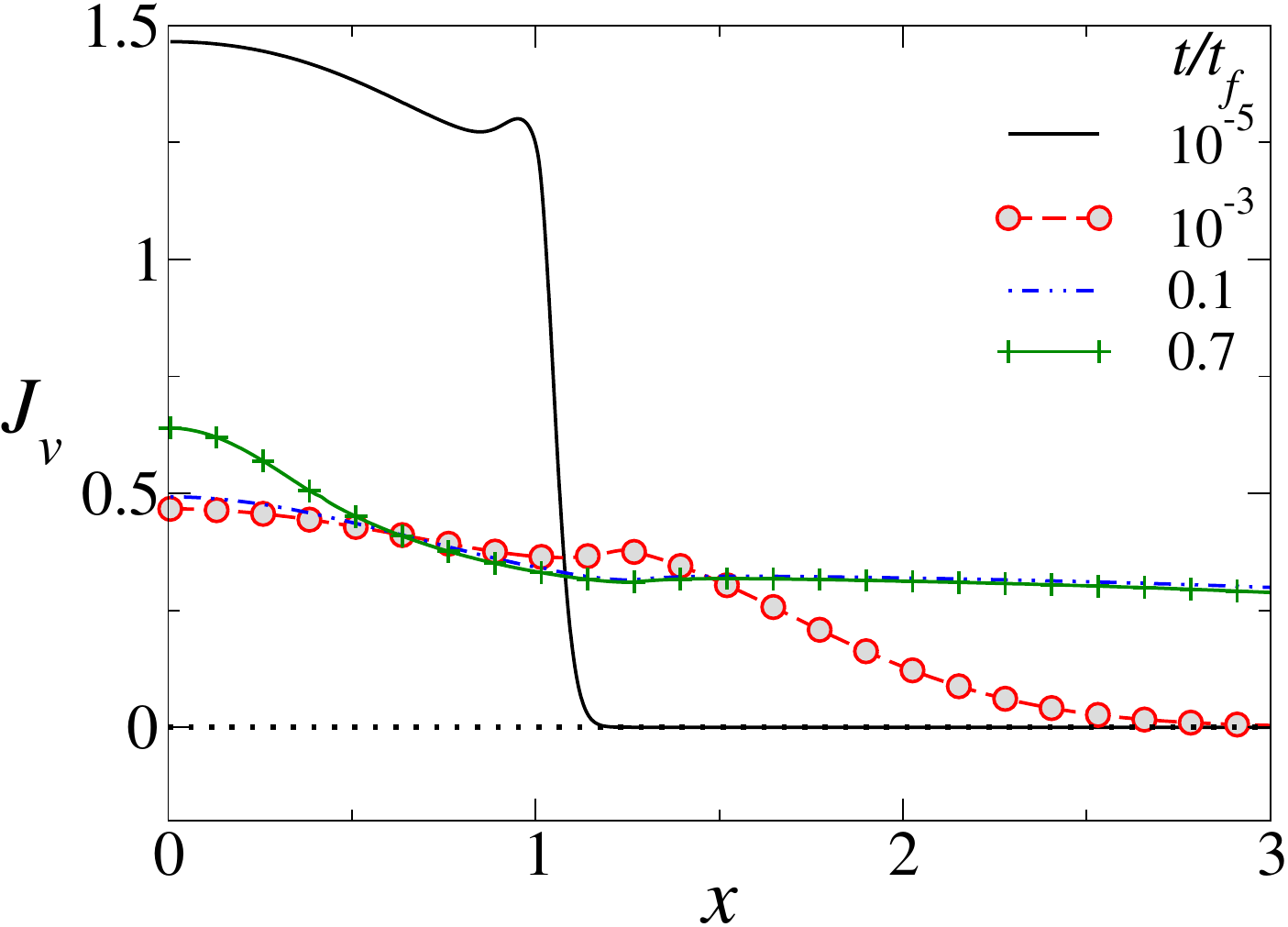} \hspace{0mm}
\includegraphics[width=0.4\textwidth]{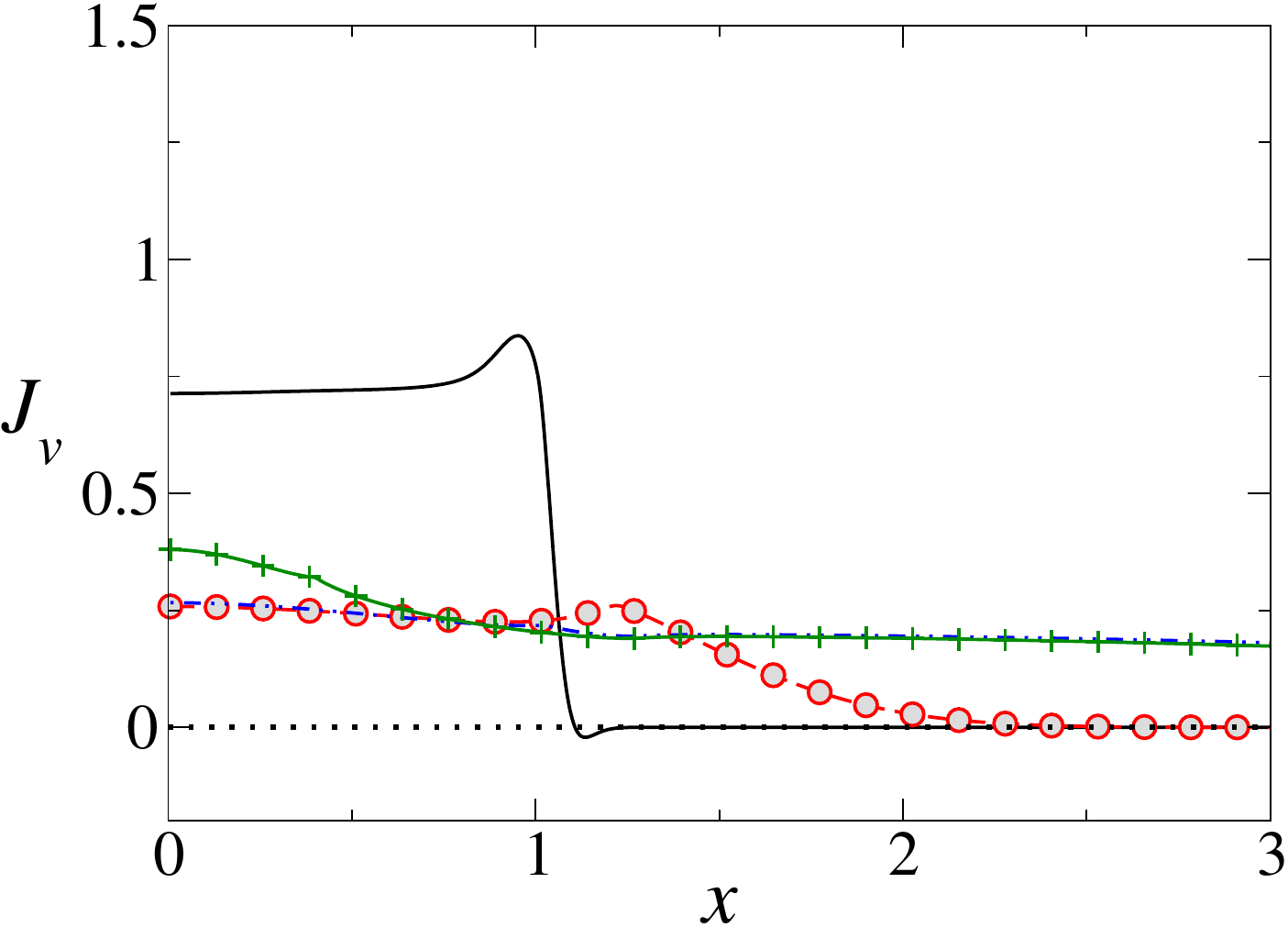} \\ 
\hspace{0.8cm}{\large (c)}   \hspace{6.5cm}  {\large (d)} \\
\includegraphics[width=0.4\textwidth]{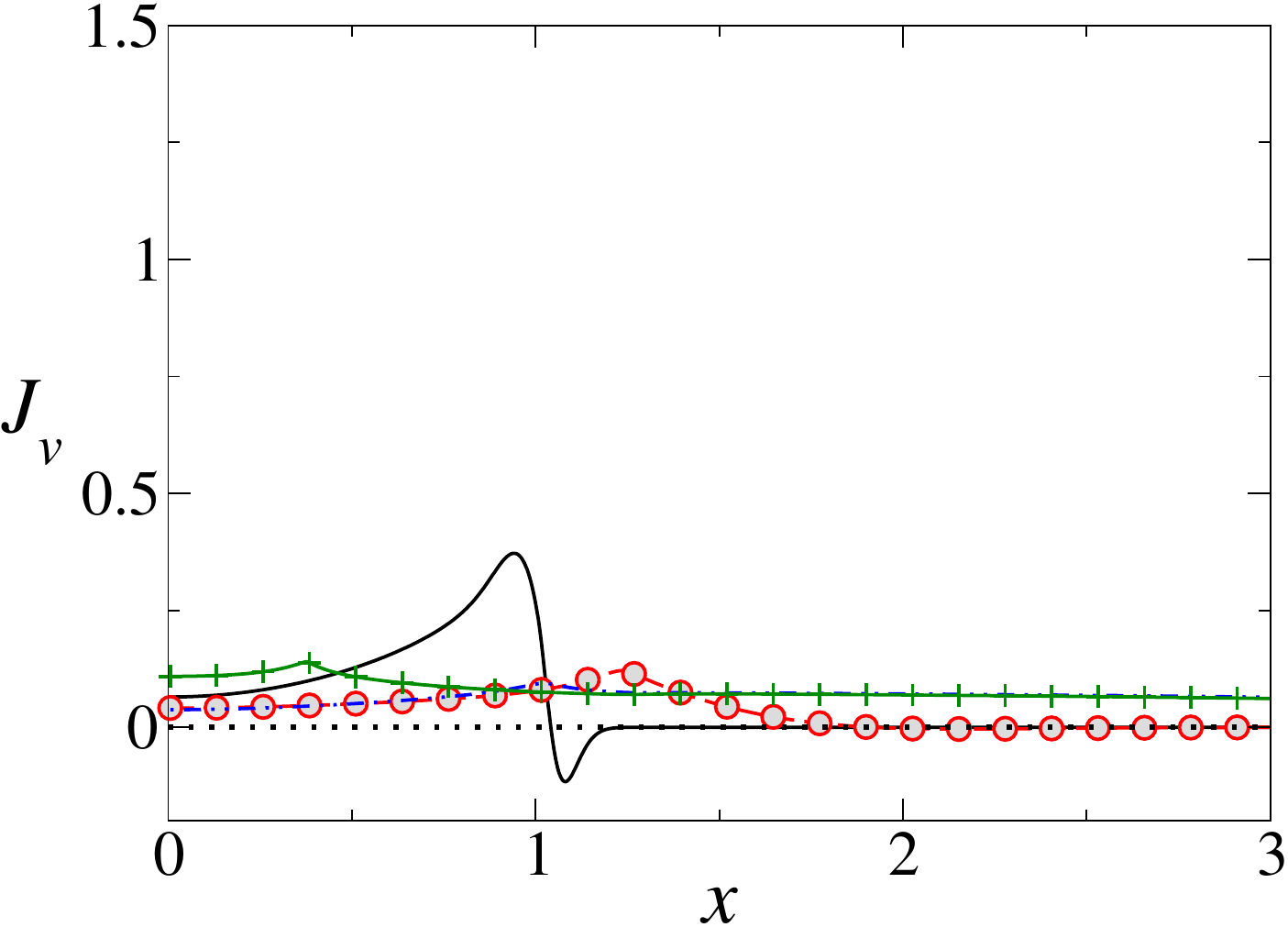} \hspace{0mm}
\includegraphics[width=0.4\textwidth]{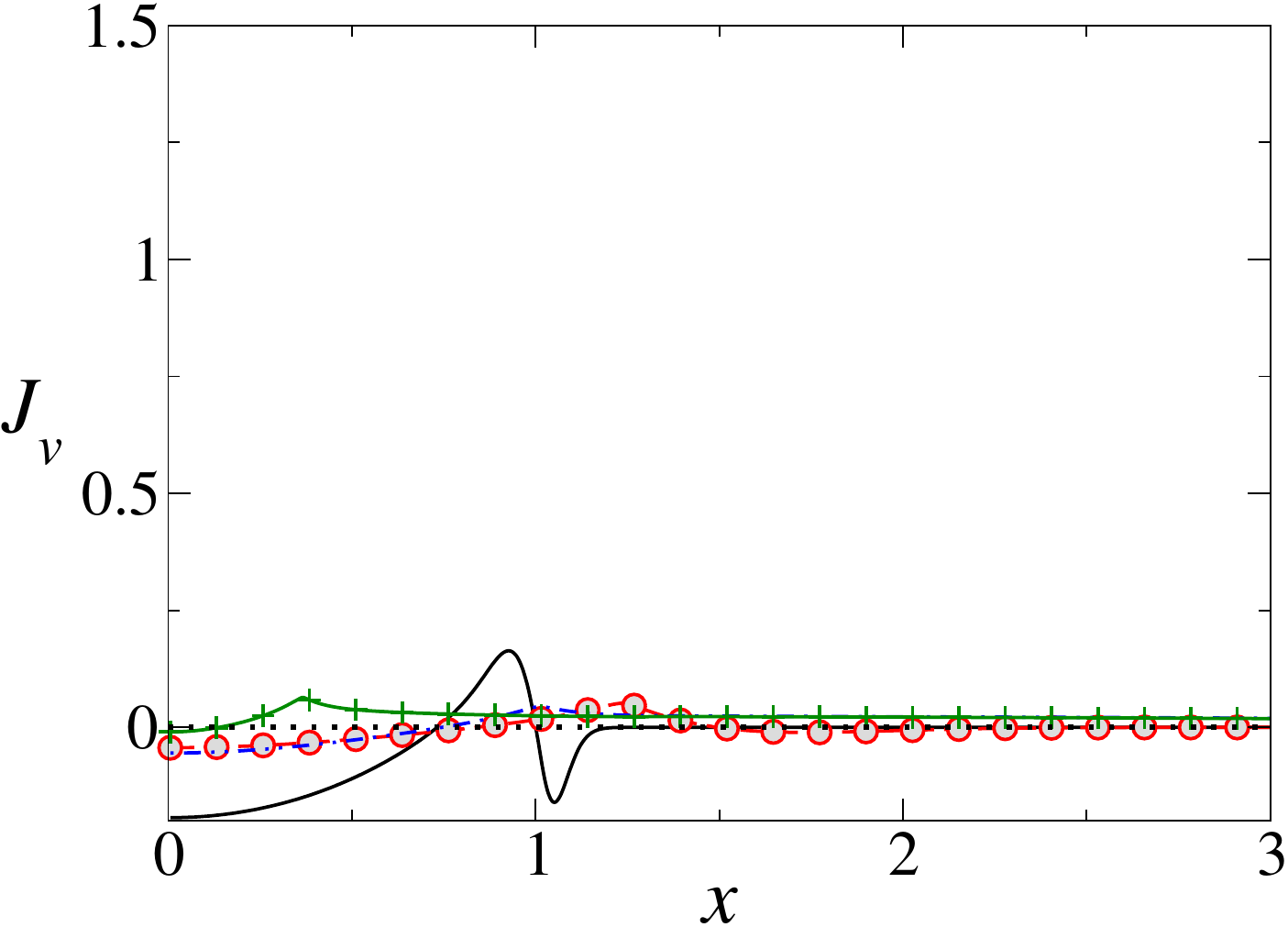}
\caption{Variation of the evaporative flux $(J_v)$ along the substrate for (a) $V_{\rho,r} = 0.35$, (b) $V_{\rho,r} = 0.60$, (c) $V_{\rho,r} = 0.85$ and (d) $V_{\rho,r} = 0.95$. The rest of the dimensionless parameters are the same as figure \ref{fig:RH_vapour}. The dotted line is plotted to distinguish the positive and negative values of the evaporation flux. The values of $t_{f}$ for $V_{\rho,r} = 0.35$, 0.60, 0.85 and 0.95 are $t_f=4473$, 4700, 5121 and 5550, respectively.}
\label{fig:RH_Evaporation_flux}
\end{figure}

%Figure 7
\begin{figure}
\centering
\hspace{0.8cm}{\large (a)}   \hspace{6.5cm}  {\large (b)} \\
\includegraphics[width=0.4\textwidth]{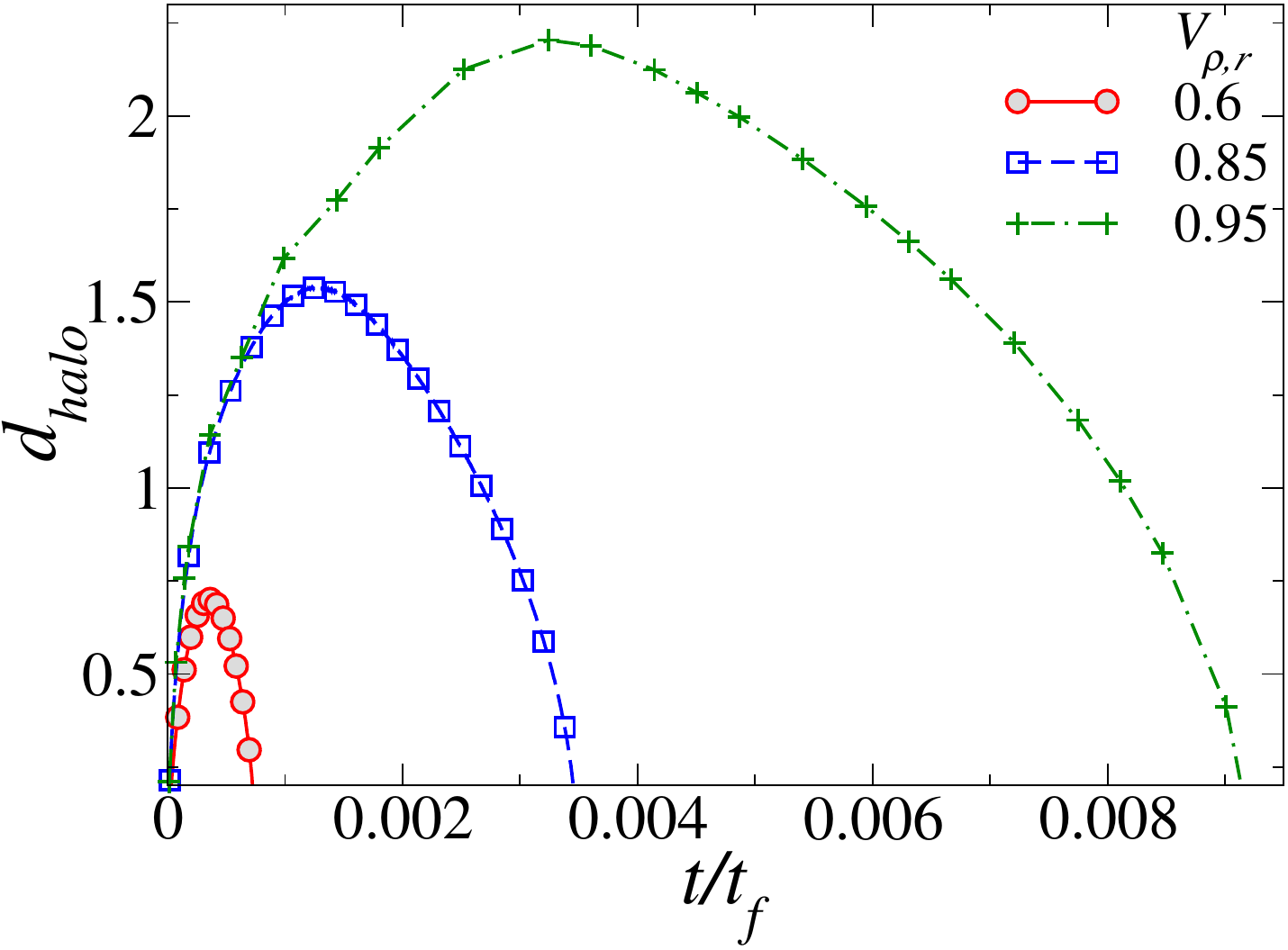}
\includegraphics[width=0.4\textwidth]{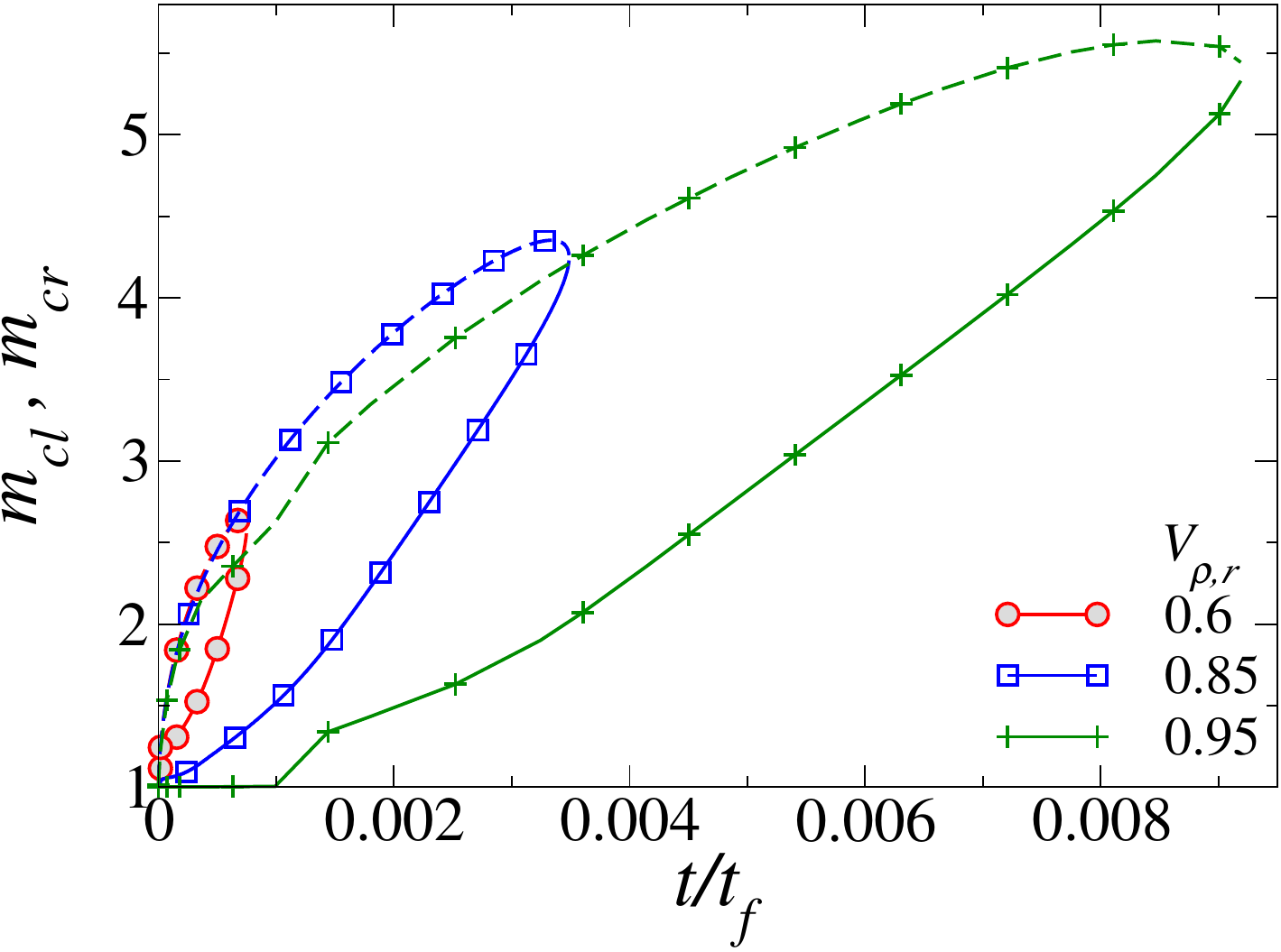}
\caption{Temporal variation of (a) the width of the frost halo $(d_{halo})$ and (b) the left $(m_{cl})$ and right $(m_{cr})$ ends of the condensation halo over the substrate for different values of the vapour density ratio $(V_{\rho,r})$. In panel (b), the solid and dashed lines represent the variation of $m_{cl}$ and $m_{cr}$, respectively. The rest of the dimensionless parameters are the same as figure \ref{fig:RH_vapour}.}
\label{fig:RH_Halo}
\end{figure}

To examine the effect of vapour density ratio on the evaporation/condensation rate, in figure \ref{fig:RH_Evaporation_flux}, we depict the profile of the evaporation flux at different values of $t/t_f$ for $V_{\rho,r} = 0.35-0.95$. It can be seen that increasing the value of $V_{\rho,r}$ considerably decreases the evaporation flux $(J_v)$ at the early times $(t/t_f<10^{-3})$. It is clear from eq. (\ref{eq:Hertz_Knudsen}) that if the vapour concentration at any point along the liquid-gas interface exceeds saturation, the evaporation flux becomes negative, indicating the condensation of vapour on the substrate. The opposite happens (evaporation occurs) when the vapour concentration falls below the saturation value. For $V_{\rho,r}=0.35$, no condensation takes place, and water continues to evaporate while the droplet undergoes freezing (figure \ref{fig:RH_Evaporation_flux}a). The evaporation flux is highest at the apex of the droplet, with a local maximum arising in the contact line region of the droplet. This is evident at $t/t_f=10^{-5}$ in figure \ref{fig:RH_Evaporation_flux}a. As expected, the evaporation flux at the apex of the droplet decreases as time progresses. For $V_{\rho,r} = 0.6$, we observe condensation near the contact line of the droplet at $t/t_f=10^{-5}$ (figure \ref{fig:RH_Evaporation_flux}a). However, subsequently (for $t/t_f \ge 10^{-3}$), the evaporation flux turns positive along the entire substrate, indicating that the condensation is eventually suppressed. Inspection of figure \ref{fig:RH_Evaporation_flux}(a-d) reveals that the condensation in the contact line region is enhanced with increasing humidity of the droplet atmosphere. Nevertheless, the increased humidity also leads to significantly lower evaporation rates. In fact, for very humid environments (see figure \ref{fig:RH_Evaporation_flux}d for $V_{\rho,r}=0.95$), evaporation takes place only in a small part of the droplet (i.e. near the contact line), and vapour mostly condenses along the liquid-gas interface. We also note that in the early stages of the freezing process, the relative importance between evaporation and condensation changes drastically with increasing humidity.

As discussed in \S \ref{sec:validation}, the vapour condensation in the contact line region forms the frost halo surrounding the frozen droplet. As shown in figure \ref{fig:RH_Evaporation_flux}, the droplet exhibits a frost halo for $V_{\rho,r} \ge 0.6$ for the base parameters. Thus, to evaluate the effect of the vapour density ratio on the estimated size of halo $(d_{halo})$, the temporal variation of the thickness of the region beyond the contact line with net condensate is plotted in figure \ref{fig:RH_Halo}(a) for different values of vapour density ratio ranging from $V_{\rho,r} = 0.60$ to $V_{\rho,r} = 0.95$. The corresponding temporal variation of the left $(m_{cl})$ and right $(m_{cr})$ ends of the condensation halo over the substrate are shown in figure \ref{fig:RH_Halo}(b). It can be seen that the width of the frost halo, $d_{halo}$, initially increases with time for all the values of $V_{\rho,r}$, reaches a maximum before disappearing at later stages of the freezing process. Inspection of these results reveals that an increase in the vapour density ratio $V_{\rho,r}$ has a significant impact on the total lifetime ($t/t_f$) of the frost halo. This effect is primarily attributed to a substantial reduction in evaporation, which prolongs the presence of the net condensate on the substrate. Figure \ref{fig:RH_Halo}(b) further illustrates that as the value of $V_{\rho,r}$ increases, the extent of the halo increases significantly.

\begin{figure}[h]
\centering
\hspace{0.8cm}{\large (a)}   \hspace{6.5cm}  {\large (b)} \\
\includegraphics[width=0.4\textwidth]{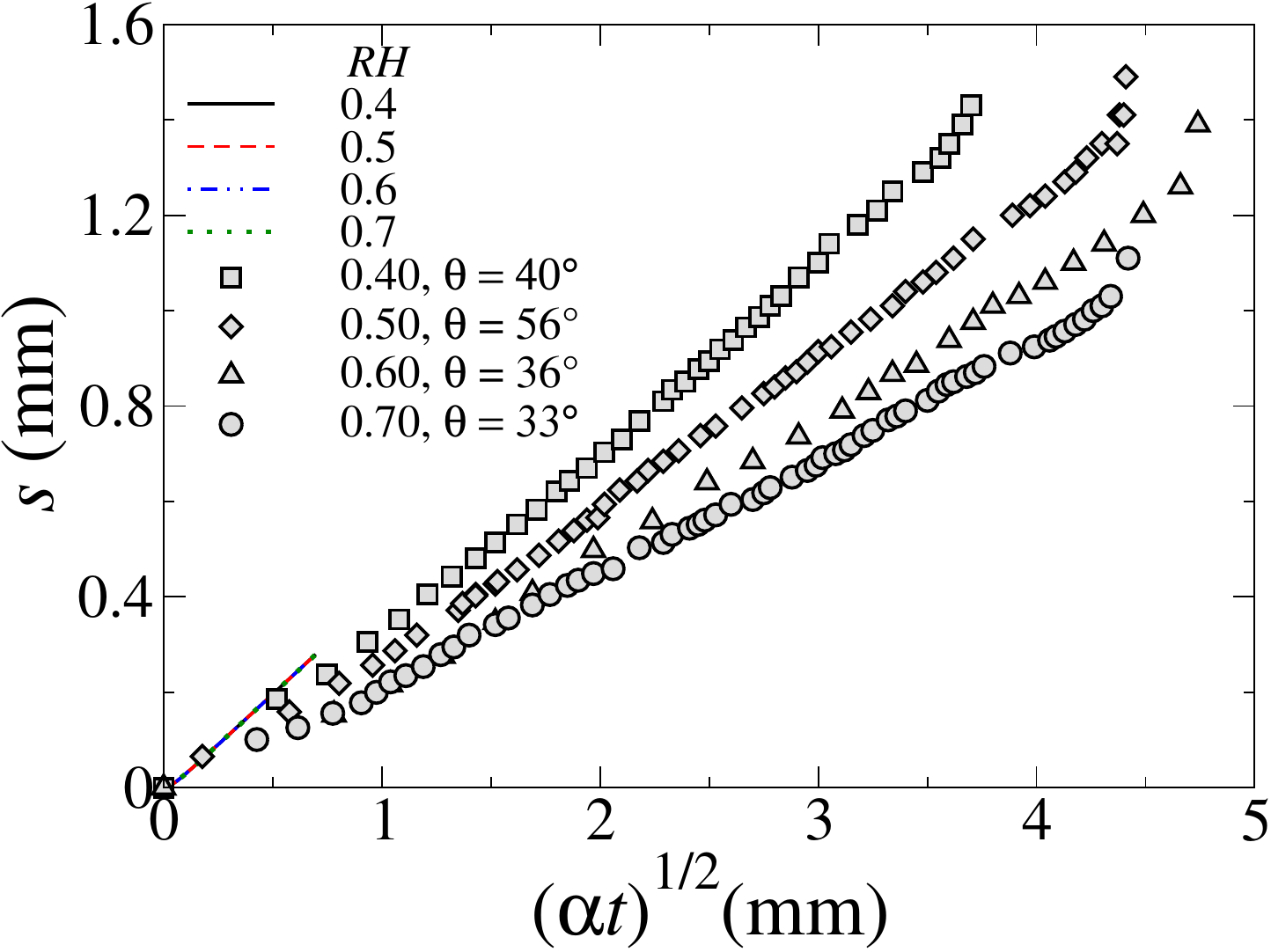} \hspace{0mm}
\includegraphics[width=0.4\textwidth]{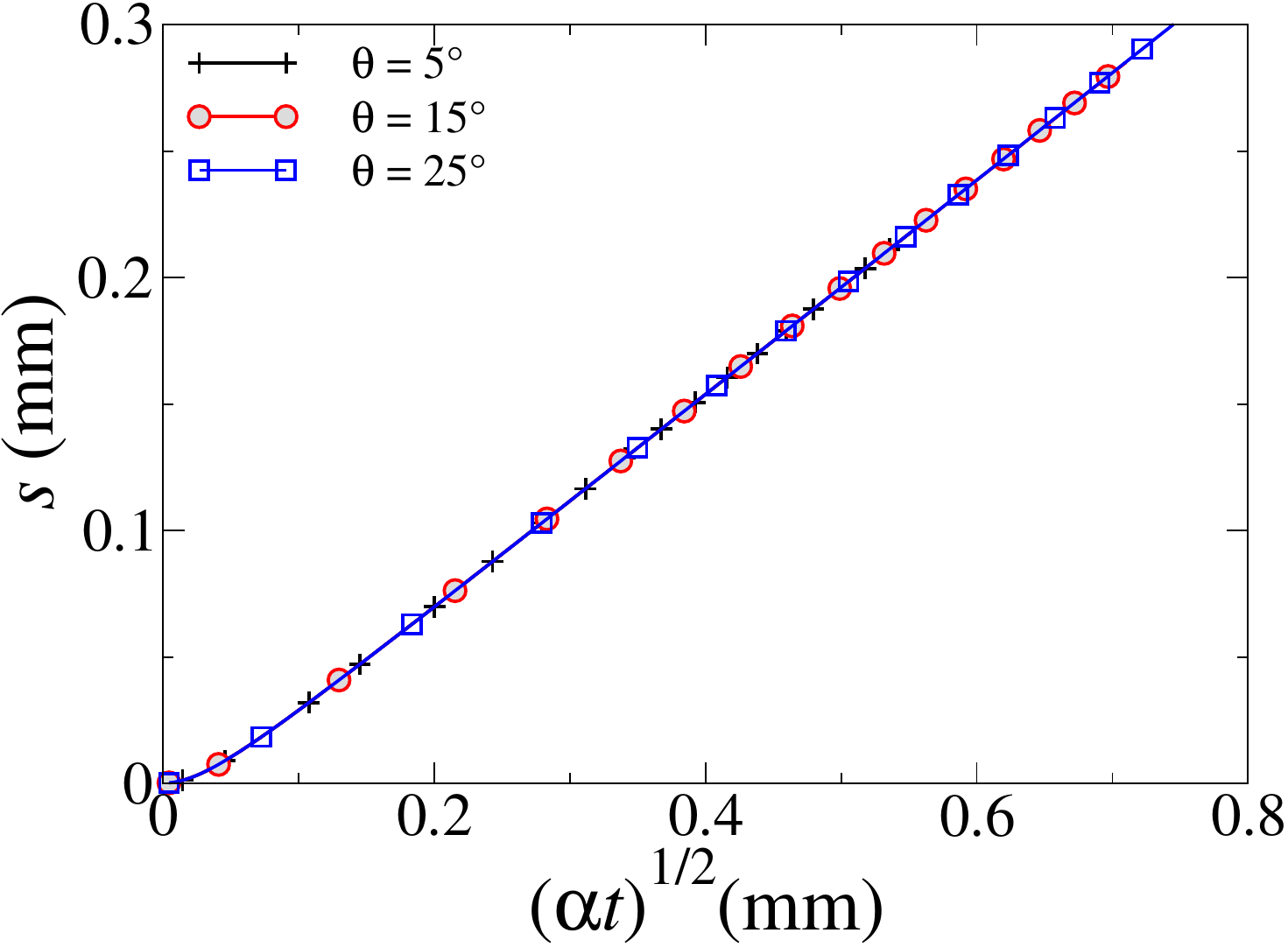} \\
\caption{\ks{(a) Comparison of the evolution of the freeing front obtained from our simulation with the experimental result of  \citet{sebilleau2021air}. Here, the results shown by lines are from our simulations with different values of the relative humidity for a droplet with a contact angle of $14.3^{\circ}$. The symbols are associated with the experimental results of \citet{sebilleau2021air}. (b) Variation of the freezing front with $(\alpha t)^{1/2}$ for different values of the contact angle for $RH=0.7$. The values of the dimensionless parameters are $Ste = 1.22 \times 10^{-3}$, $T_{v} = 1$, $A_{n} = 6.25$, $D_{g} = 2$, $D_{s} = 0.9$, $\Lambda_{S} = 3.89$, $\Lambda_{W} = 698$, $\chi = 0.01$, $K = 8\times10^{-4}$, $Bi = 0.16$, $D_{w} = 15$, $V_{\rho,r} = 0.70$, $\epsilon=0.2$, $D_{v} = 1.65 \times 10^{-6}$, $\Delta = 10^{-4}$, $\Psi = 0.94$ and $Pe_{v} = 1$.}}
\label{fig:Sellibleau_comp}
\end{figure}

\ks{At this point, it would be useful to compare the evolution of the freezing front obtained from our model with the experimental observations of \citet{sebilleau2021air}. In the latter study, the freezing of a water drop on a copper substrate was investigated under the conditions of a substrate temperature of $-15^{\circ}$C and an ambient temperature ranging from $15^{\circ}$C to $20^{\circ}$C. Their findings revealed a substantial impact of relative humidity on the freezing front propagation rates, while wettability did not significantly influence the process. On the other hand, our study explores a considerably colder ambient temperature range ($-15^{\circ}$C to $0^{\circ}$C). It is important to note that the freezing front dynamics depend on the heat transfer from the water-ice interface to the ice, substrate, and the remaining water and atmosphere, as suggested by \citet{jung2012frost,sebilleau2021air}. Thus, in contrast to our investigation, \citet{sebilleau2021air} emphasizes the substantial impact of evaporative cooling, leading to a significant variation in the freezing front propagation rate under different relative humidity conditions. By performing a simulation for a scenario where the ambient temperature aligns with the melting temperature ($T_v = 1$) and a smaller droplet with a contact angle of $14.3^{\circ}$, while keeping the remaining parameters the same as those of \citet{sebilleau2021air}, we observe a close match agreement between our simulated freezing front propagation and experimental result of \citet{sebilleau2021air} at $40\%$ relative humidity (see figure \ref{fig:Sellibleau_comp}a). However, given that the heat flux primarily traverses into the ice and subsequently to the substrate, we do not observe significant variations in the freezing front propagation rates for different relative humidity values. Furthermore, we explore the impact of the contact angle of the droplet in figure \ref{fig:Sellibleau_comp}(b) for a typical set of parameters. Our findings indicate that changes in wettability do not significantly affect the freezing front propagation. A similar behaviour was observed by \citet{sebilleau2021air}.}

%Figure 8

%\gk{I WOULD SUGGEST TO MOVE THE FIGURE TO THE MAIN TEXT AND ALSO INCLUDE THE SECOND PANEL WHICH SHOWS THE INDEPENDENCE WITH CONTACT ANGLE - SEE RESPONSE TO 2ND REVIEWER}

\subsection{Effect of liquid volatility} \label{effect_Dv}

%Figure 8
\begin{figure}[h]
     \centering
\includegraphics[width=0.9\textwidth]{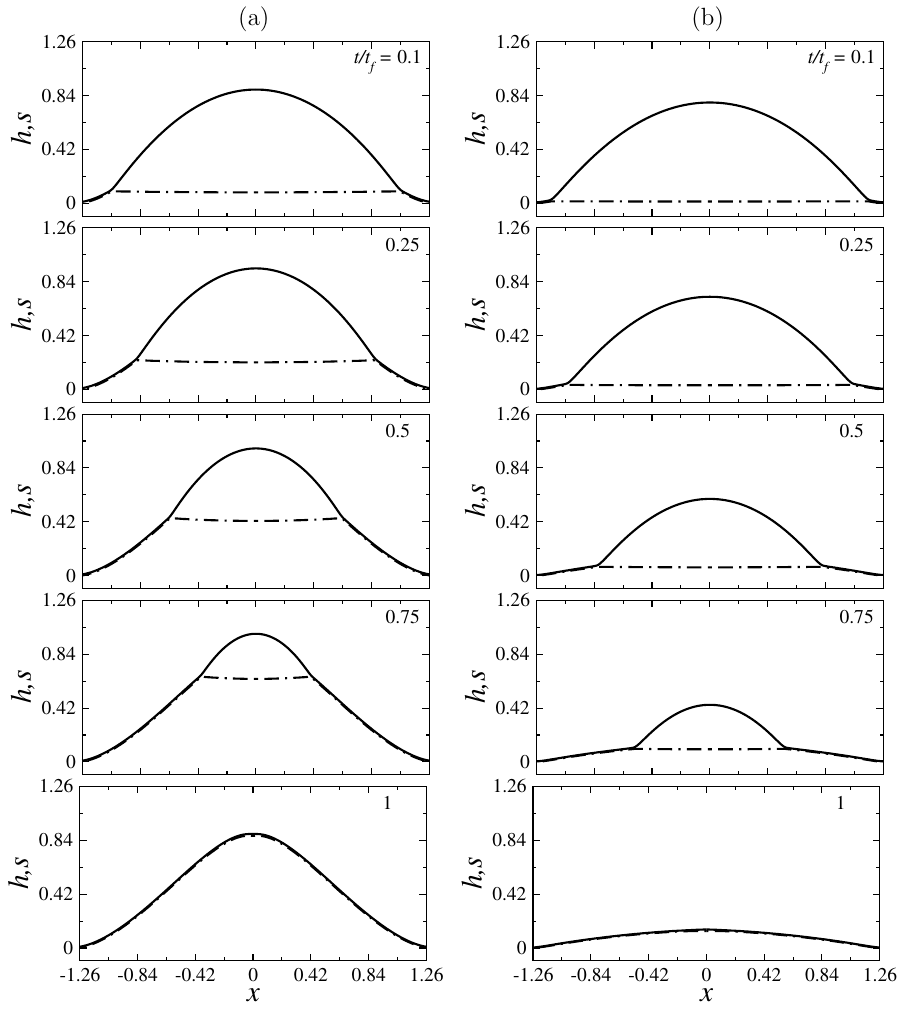}
\caption{Evolution of the freezing front, $s$ (dot-dashed lines) and shape of the droplet, $h$ (solid line) for (a) $D_{v} = 10^{-4}$ and (b) $D_{v} = 10^{-2}$. The values of the rest of the dimensionless parameters are $Ste = 1.22 \times 10^{-3}$, $T_{v} = 0$, $A_{n} = 6.25$, $D_{g} = 2$, $D_{s} = 0.9$, $\Lambda_{S} = 3.89$, $\Lambda_{W} = 0.33$, $\chi = 0.01$, $V_{\rho,r} = 0.85$, $K = 8\times10^{-4}$, $Bi = 0.16$, $D_{w} = 15$, $\epsilon=0.2$, $\Delta = 10^{-4}$, $\Psi = 0.94$ and $Pe_{v} = 1$. The values of $t_{f}$ for $D_{v} = 10^{-4}$ and $10^{-2}$ are $t_f = 4210$ and 634, respectively.}
\label{fig:Dv_profile}
\end{figure}

%Figure 10
\begin{figure}[h]
\centering
\hspace{0.8cm}{\large (a)}   \hspace{6.5cm}  {\large (b)} \\
\includegraphics[width=0.4\textwidth]{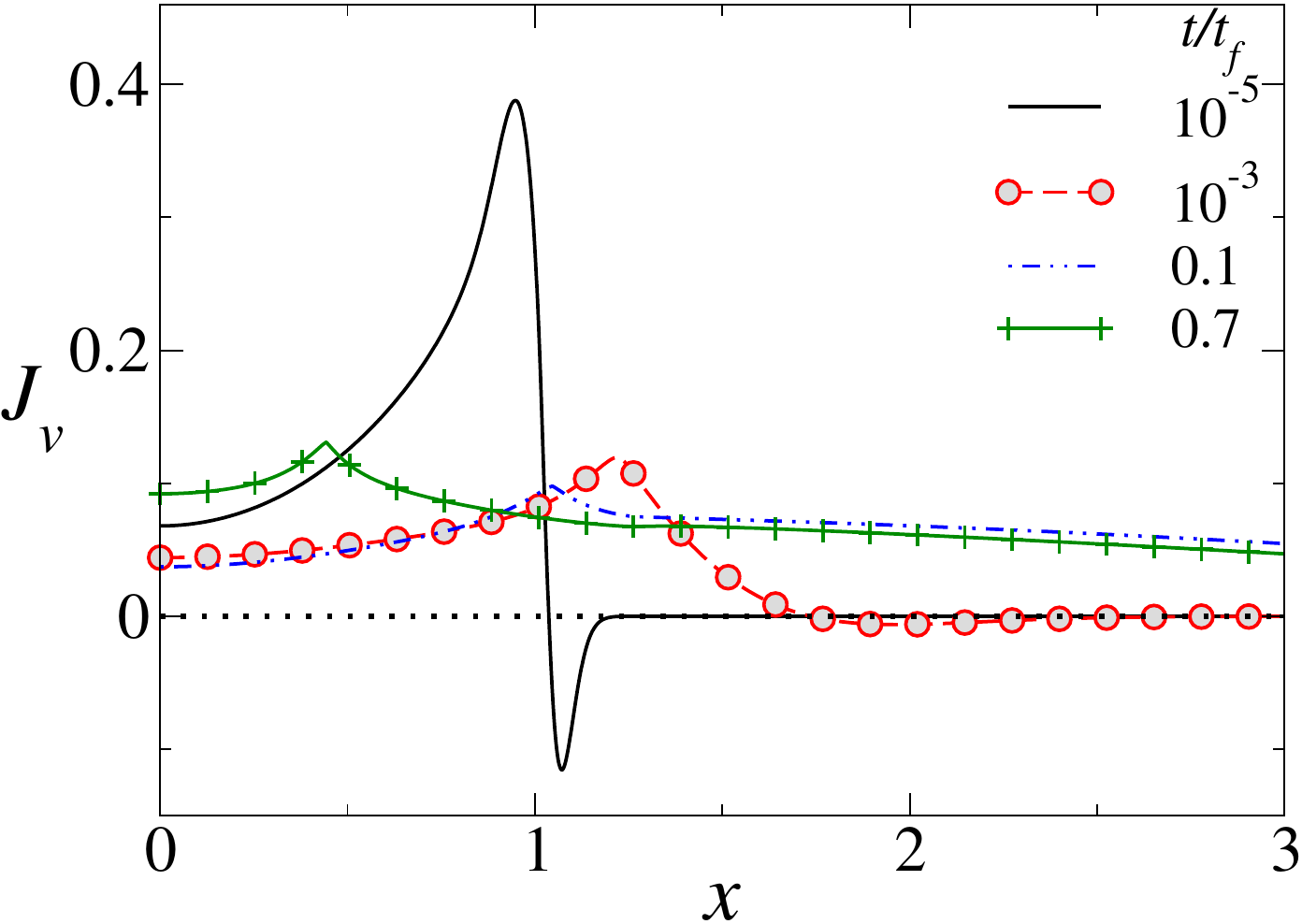}
\includegraphics[width=0.4\textwidth]{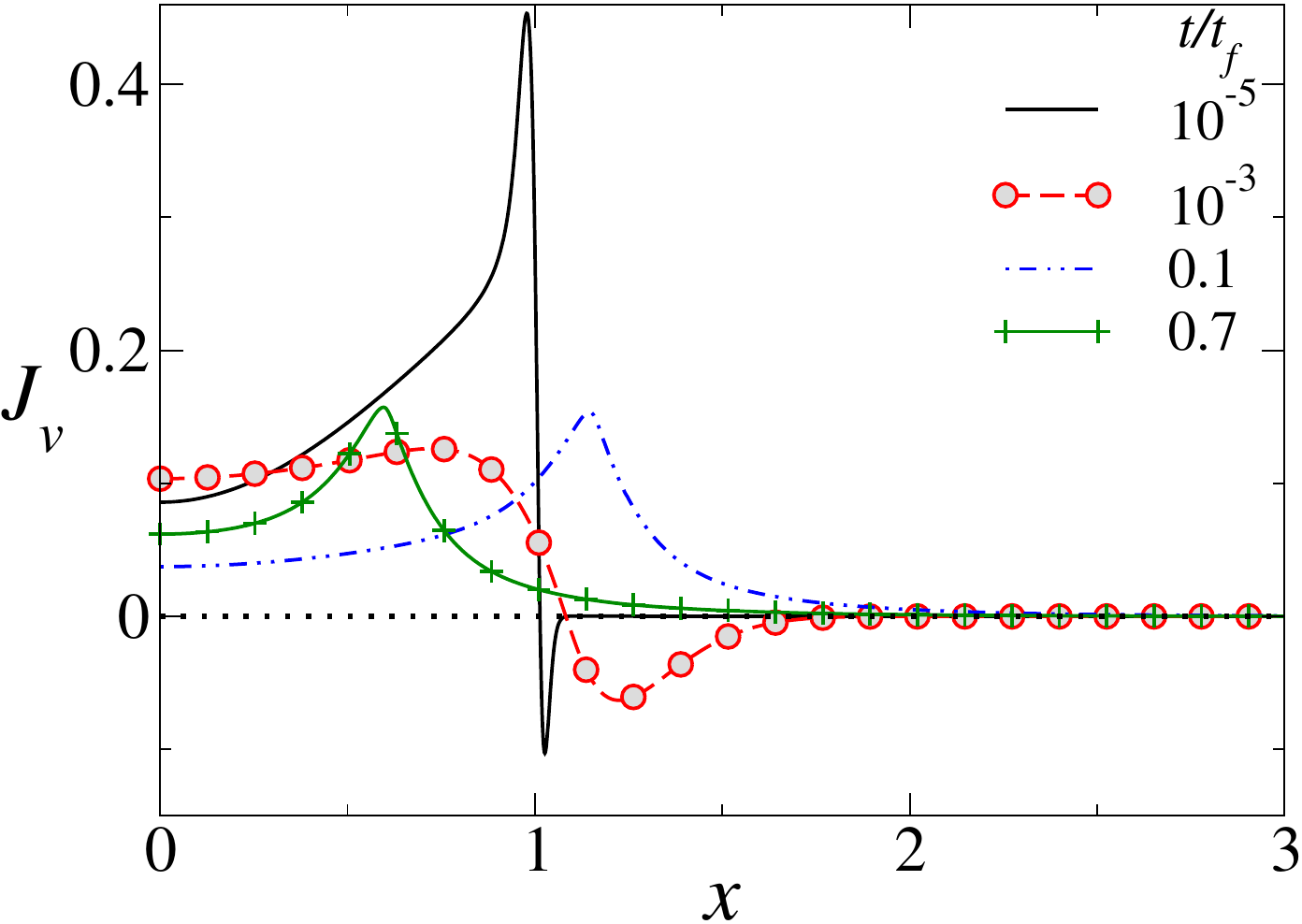} \\
\hspace{0.8cm}{\large (c)}   \hspace{6.5cm}  {\large (d)} \\
\includegraphics[width=0.4\textwidth]{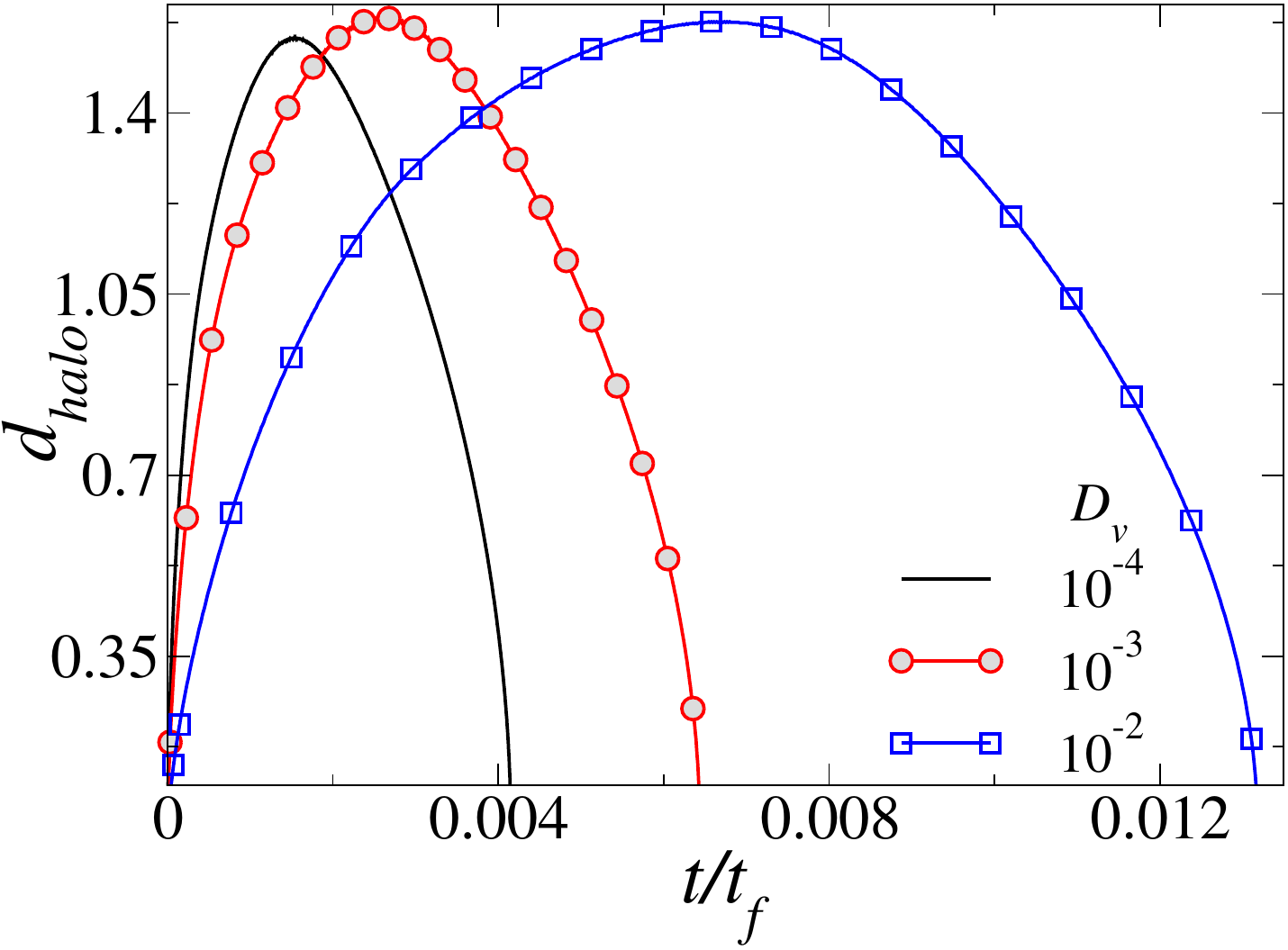}
\includegraphics[width=0.4\textwidth]{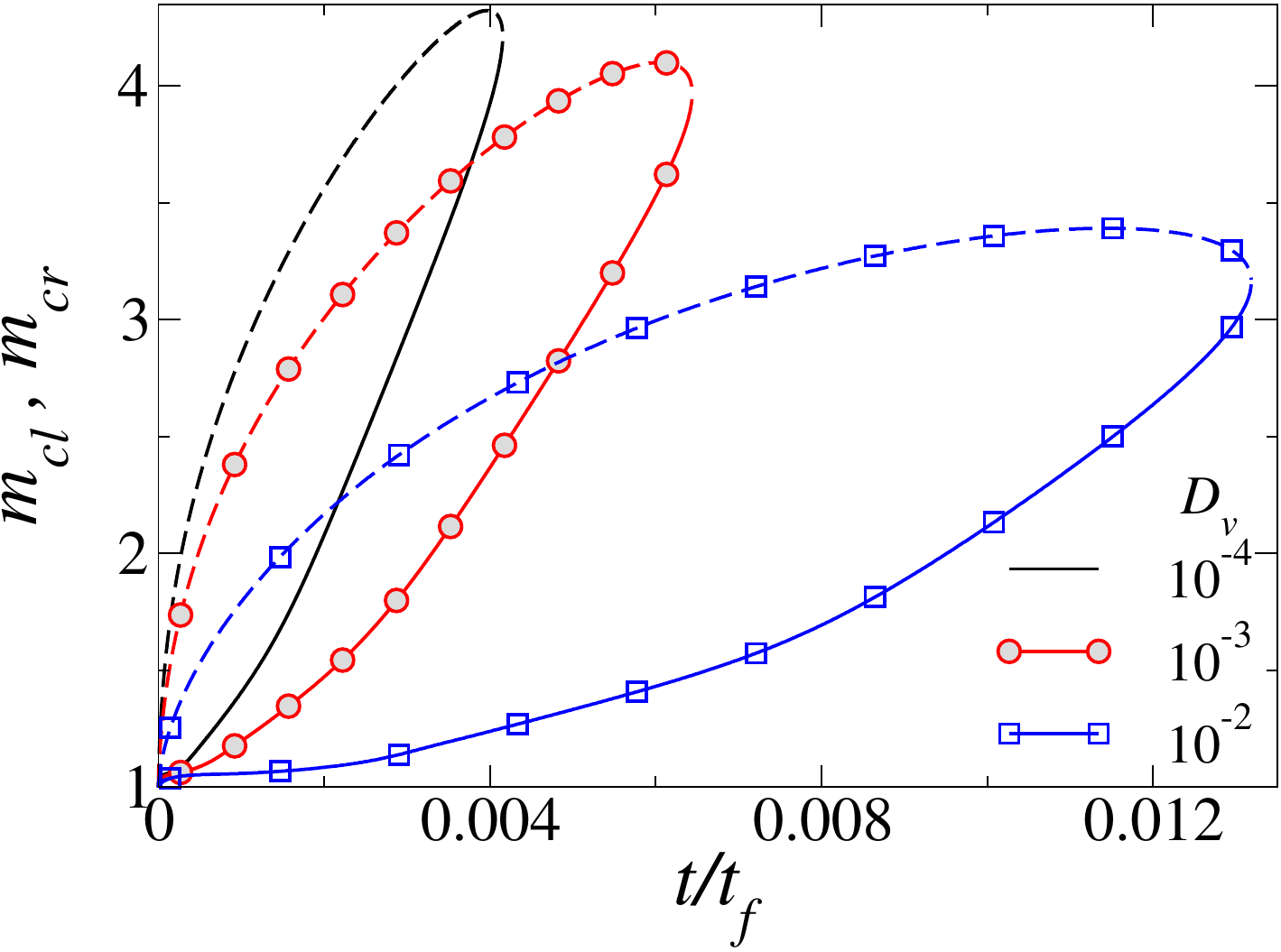}\\
\hspace{0.8cm} {\large (e)}\\
\includegraphics[width=0.4\textwidth]{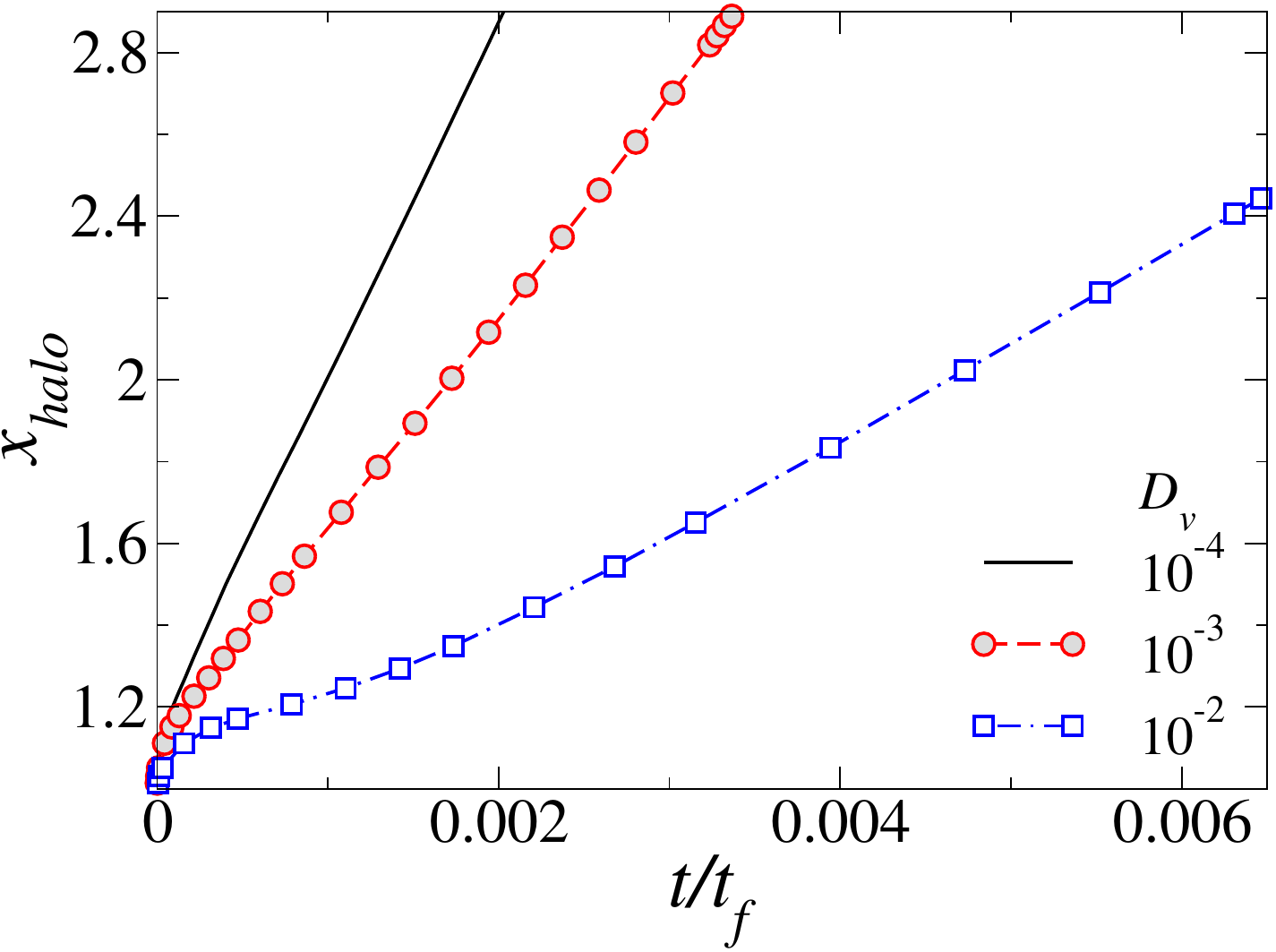}
\caption{Variation of the evaporative flux $(J_v)$ along the substrate for (a) $D_{v} = 10^{-4}$ and (b) $D_{v} = 10^{-2}$. Temporal variation of (c) the width of the frost halo $(d_{halo})$, (d) the left $(m_{cl})$ and right $(m_{cr})$ ends of the condensation halo over the substrate, and (e) the location of maximum condensation flux $(x_{halo})$ for different values of $D_{v}$. In panel (d), the solid and dashed lines represent the variation of $m_{cl}$ and $m_{cr}$, respectively. The rest of the dimensionless parameters are the same as figure \ref{fig:Dv_profile}. The values of $t_{f}$ for $D_{v} = 10^{-4}$, $10^{-3}$ and $10^{-2}$ are $t_f = 4210$, 2318 and 634, respectively.}
\label{fig:Dv_Halo}
\end{figure}

To further examine the effect of liquid volatility, we vary the parameter, $D_v$, which is the ratio between the equilibrium vapour concentration and liquid density, in figure \ref{fig:Dv_profile}. We consider a wide range for this parameter (not applicable to a typical water droplet) in order to examine cases covering the entire range between non-volatile and highly volatile liquids. A few examples of highly volatile liquids are butane, toluene and octane. The total freezing time for highly volatile liquids decreases drastically ($t_f=634$ for $D_v=10^{-2}$ and $t_f=4210$ for $D_v=10^{-4}$) due to both the effect of evaporative cooling as well as the faster depletion of the liquid layer which vaporizes in the presence of an unsaturated atmosphere. In addition, the difference in liquid volatility impacts the final shape of the frozen droplet. Highly volatile droplets assume a much thinner profile with a lower total mass of ice than less volatile ones.

To examine the effect of $D_v$ on the frost halo formation, we plot, in figure \ref{fig:Dv_Halo}(a) and (b), the profiles of the evaporation flux ($J_v$) for $D_v=10^{-4}$ and $10^{-2}$ at various time instants. As expected, the evaporation flux is higher for highly volatile liquids, but we also notice more substantial condensation in the periphery of the droplet, driven by a higher amount of vapour in that area.  In order to quantify this effect, in figure \ref{fig:Dv_Halo}(c) and (d), we plot the temporal variation of the width of the frost halo ($d_{halo}$), the leftmost point $(m_{cl})$ and the rightmost point of the condensation halo $(m_{cr})$ for different values of $D_v$. It can be seen that both the normalised lifespan ($t/t_{f}$) and the width of the halo increase considerably with the increasing volatility of the liquid. Moreover, we notice that the condensation occurs closer to the droplet with increasing volatility (see figure \ref{fig:Dv_Halo}(a) and (b)). This is shown more clearly in figure \ref{fig:Dv_Halo}(e), where we plot the position of the maximum condensation flux $(x_{halo})$ for different values of $D_v$. Evaporation is accompanied by cooling due to the effect of latent heat. We observe (not shown) that the concavity of the freezing front increases with increasing $\chi$ due to the increased influence of evaporative cooling near the tip of the droplet. This leads to a local decrease in the temperature, promoting freezing in that region. The total freezing times for $\chi$ to $0.01$, $0.1$, and $1.0$ are $t_f = 4210$, 4110, and 3503, respectively. %The increase in condensation for higher values of $\chi$ can be attributed to the fact that the vapor in the gas phase comes into contact with a cooler substrate, undergoing more intense cooling due to the higher latent heat of evaporation. However, we did not observe any significant alteration in the halo size for different values of $\chi$.

\subsection{Effect of substrate thermal resistance} \label{effect_cond}

%Figure 11
\begin{figure}
     \centering
\includegraphics[width=0.9\textwidth]{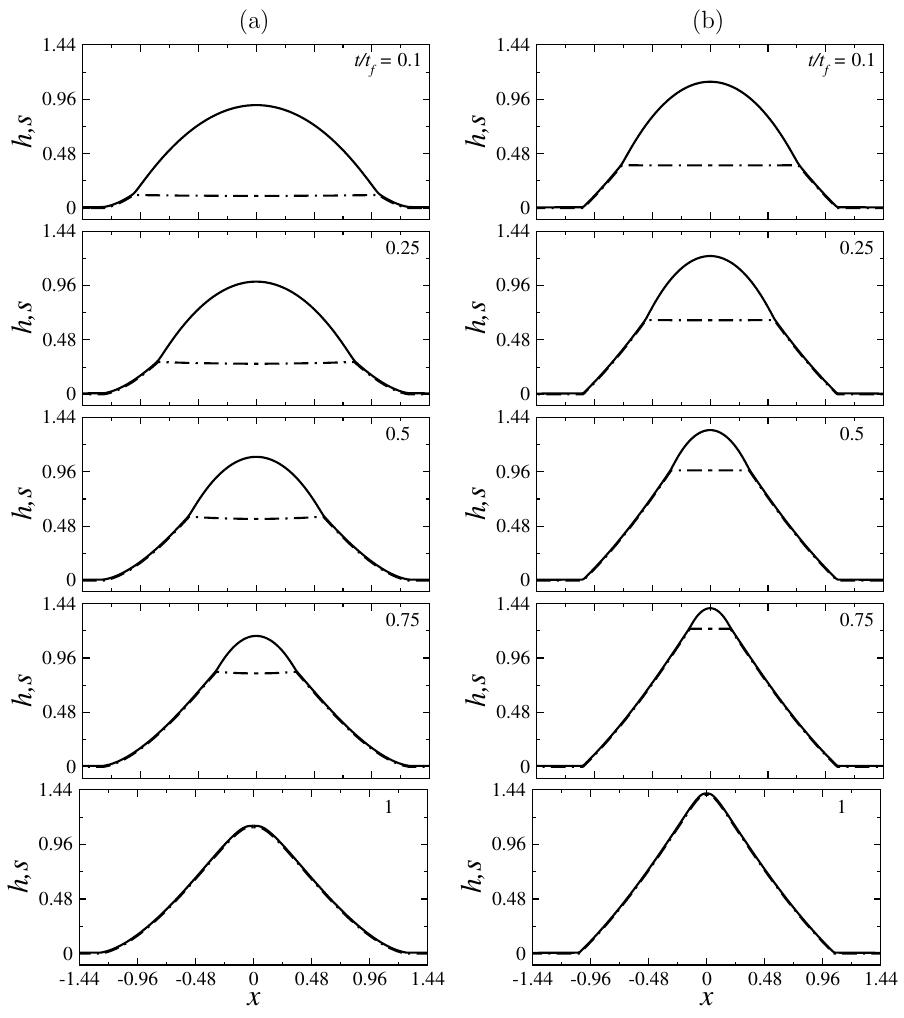}
\caption{Evolution of the freezing front, $s$ (dot-dashed lines) and shape of the droplet, $h$ (solid line) for (a) $\Lambda_{W} = 1$ and (b) $\Lambda_{W} = 500$. The values of the rest of the dimensionless parameters are $Ste = 1.22 \times 10^{-3}$, $T_{v} = 0$, $A_{n} = 6.25$, $D_{g} = 2$, $D_{s} = 0.9$, $\chi = 0.01$, $\Lambda_{S} = 3.89$, $V_{\rho,r} = 0.85$, $K = 8\times10^{-4}$, $Bi = 0.16$, $D_{w} = 15$, $\epsilon=0.2$,  $D_{v} = 1.65 \times 10^{-6}$, $\Delta = 10^{-4}$, $\Psi = 0.94$ and $Pe_{v} = 1$.}
\label{fig:LW_Profile}
\end{figure}

%Figure 12
\begin{figure}
\centering
\hspace{0.8cm}{\large (a)}   \hspace{6.5cm}  {\large (b)} \\
\includegraphics[width=0.4\textwidth]{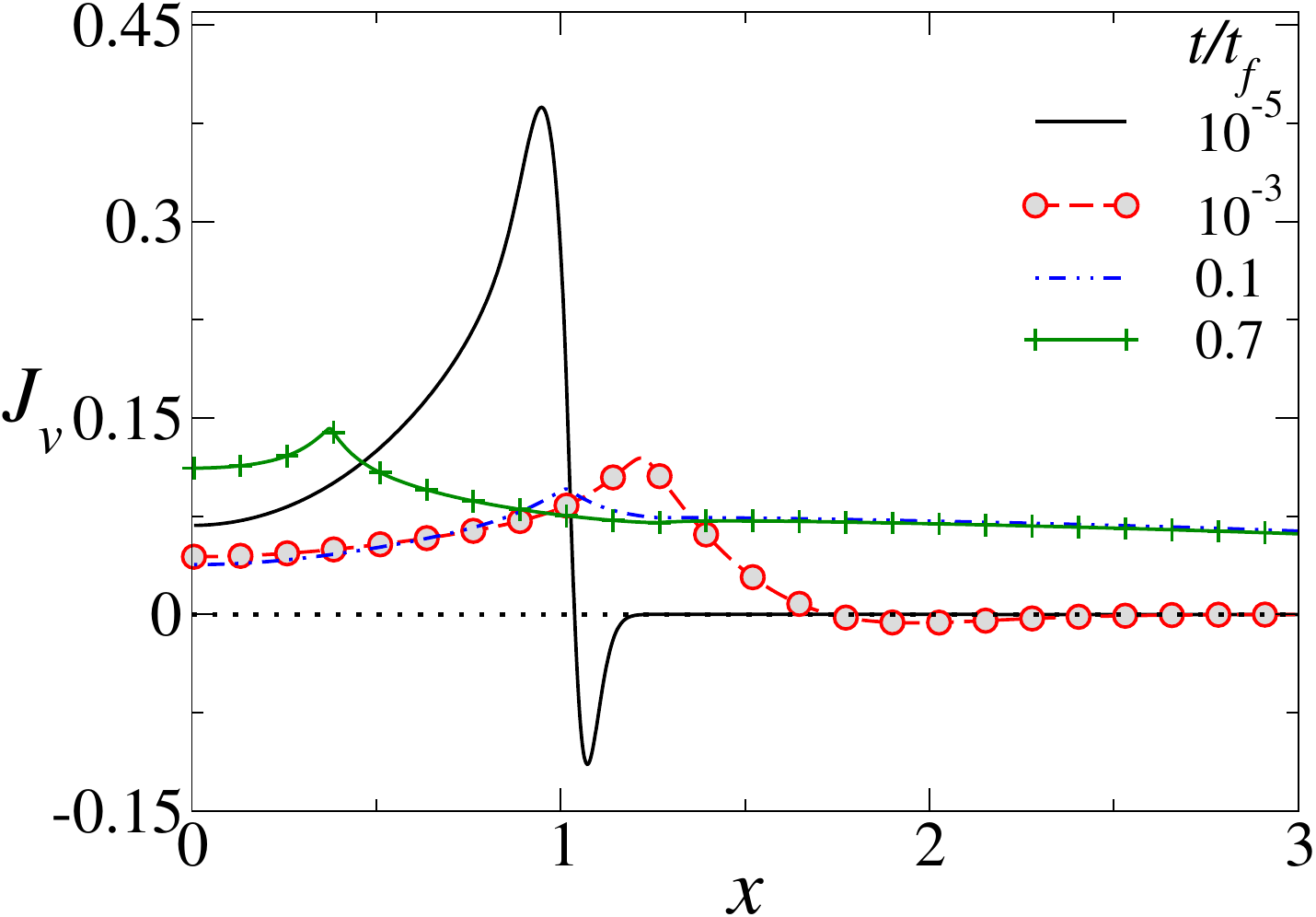} \hspace{0mm}
\includegraphics[width=0.4\textwidth]{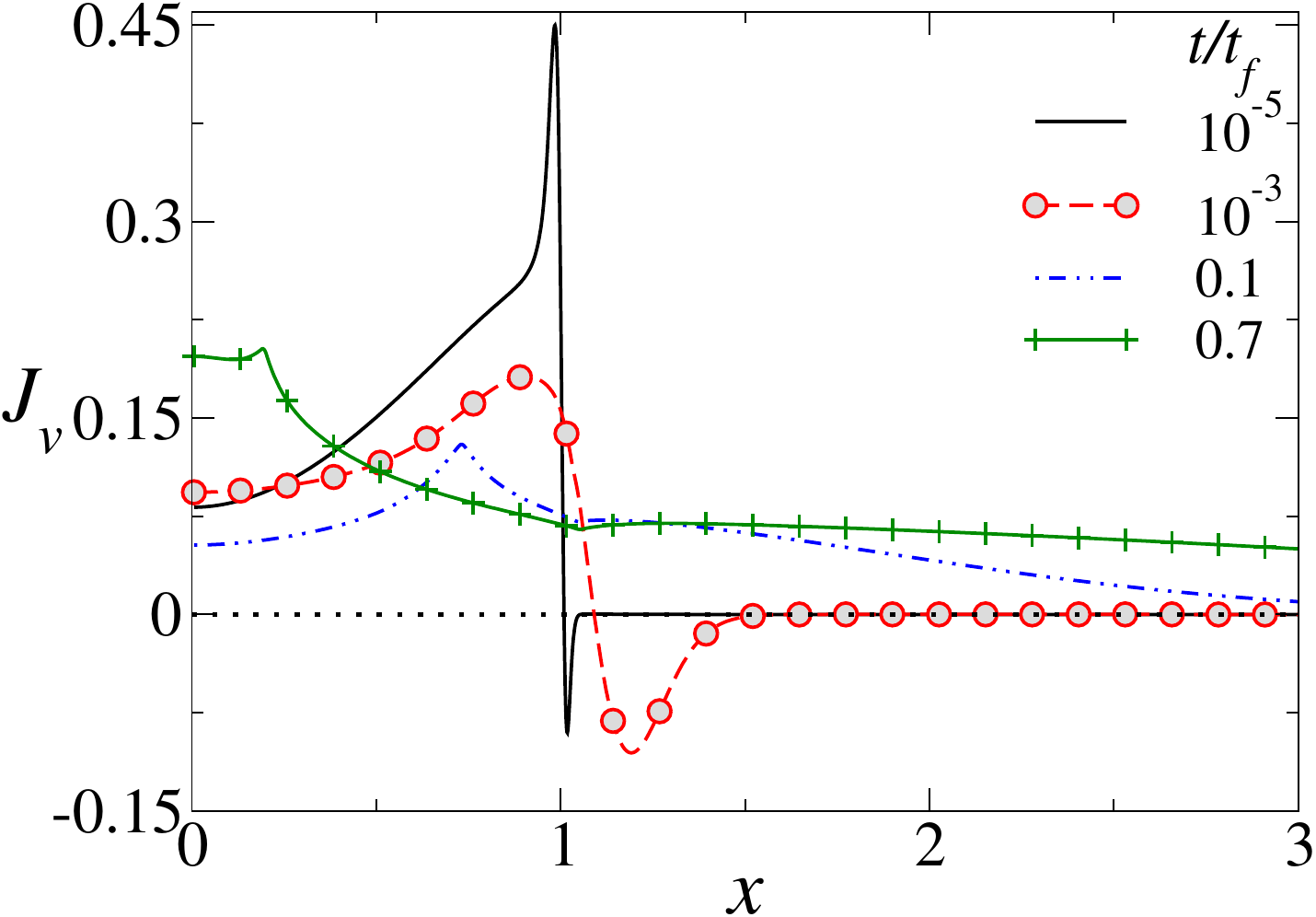} \\ 
\hspace{0.8cm}{\large (c)}   \hspace{6.5cm}  {\large (d)} \\
\includegraphics[width=0.4\textwidth]{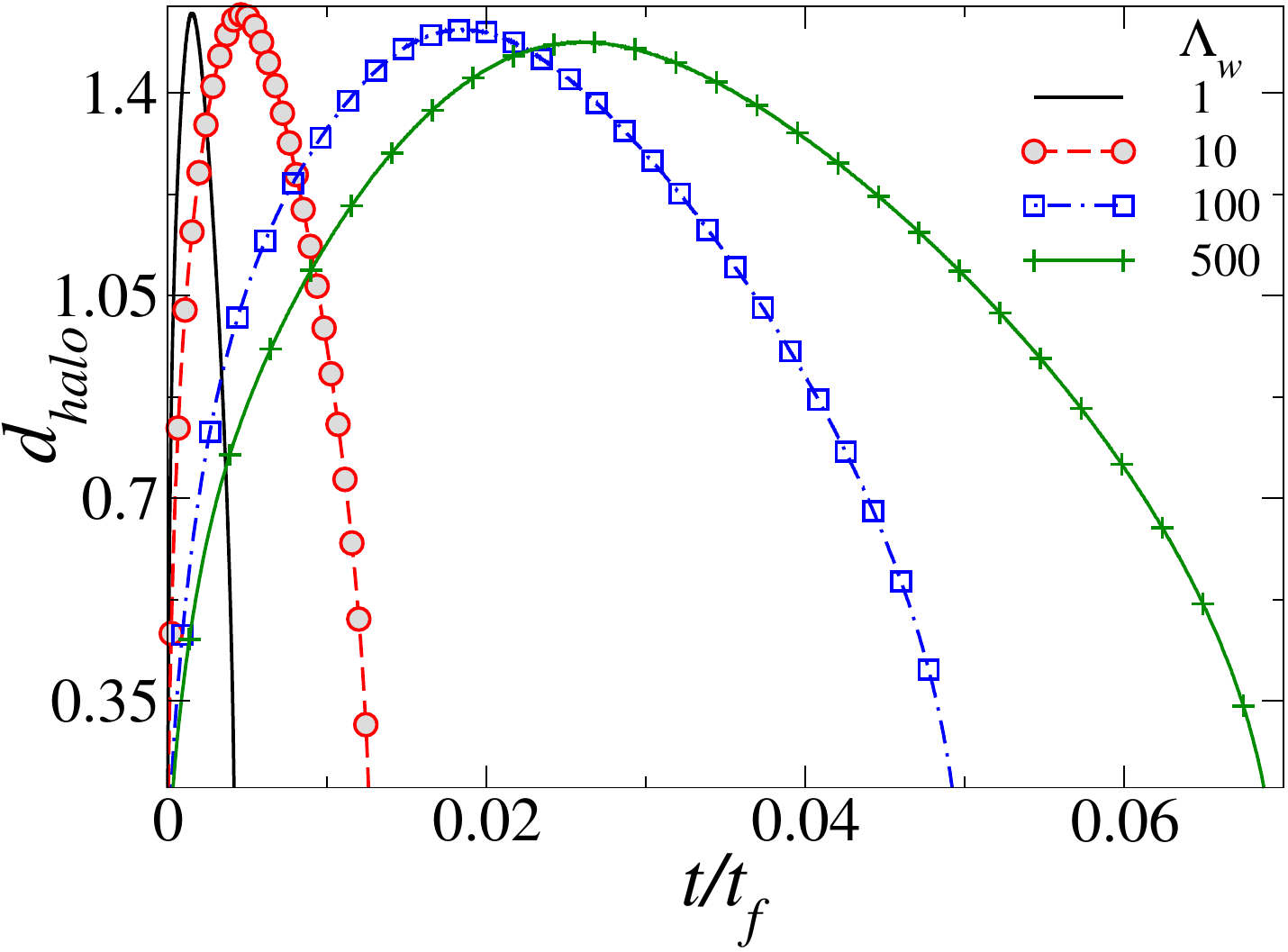}
\hspace{0mm}
\includegraphics[width=0.4\textwidth]{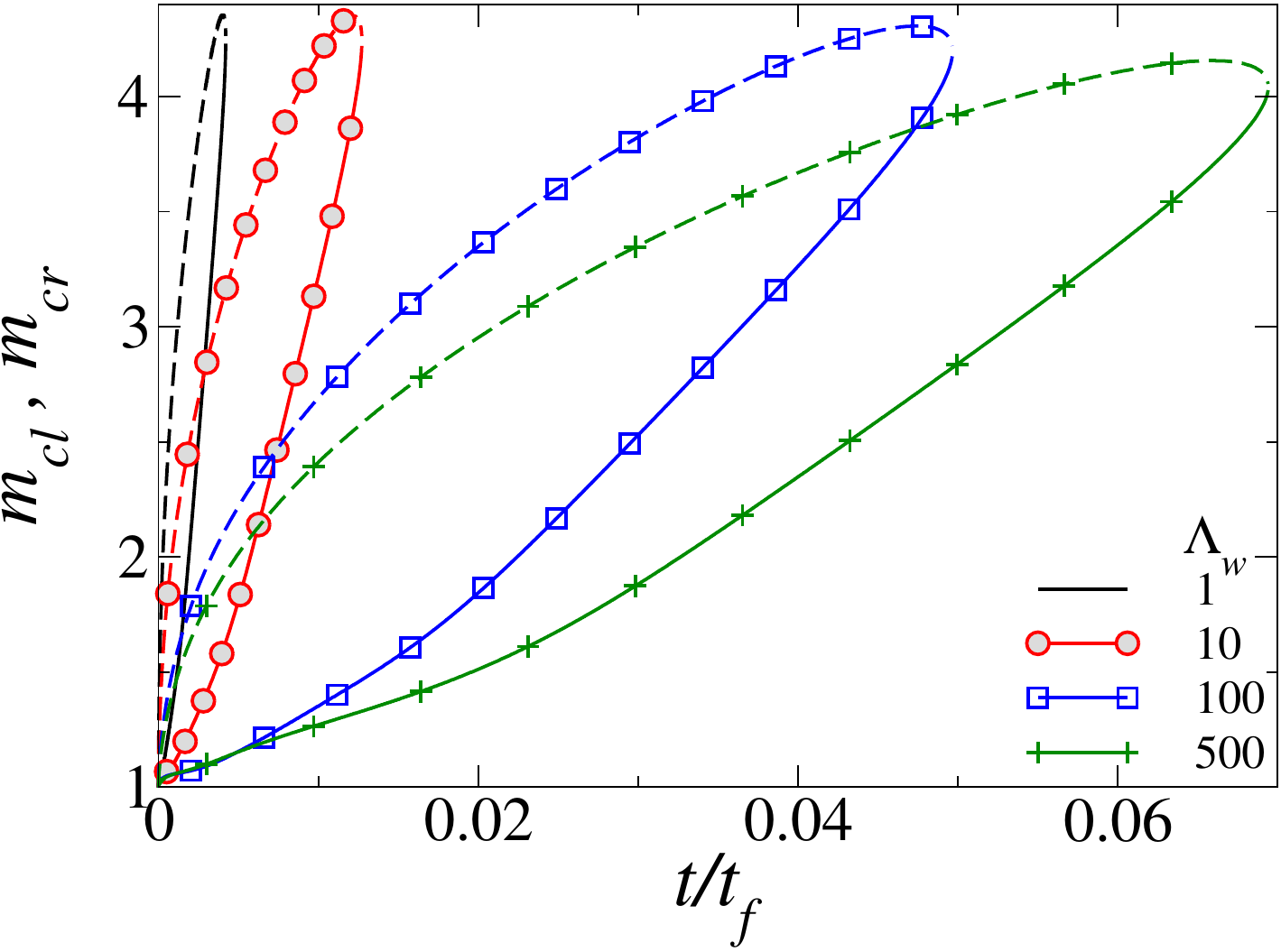}\\
\hspace{0.8cm} {\large (e)}\\
\includegraphics[width=0.4\textwidth]{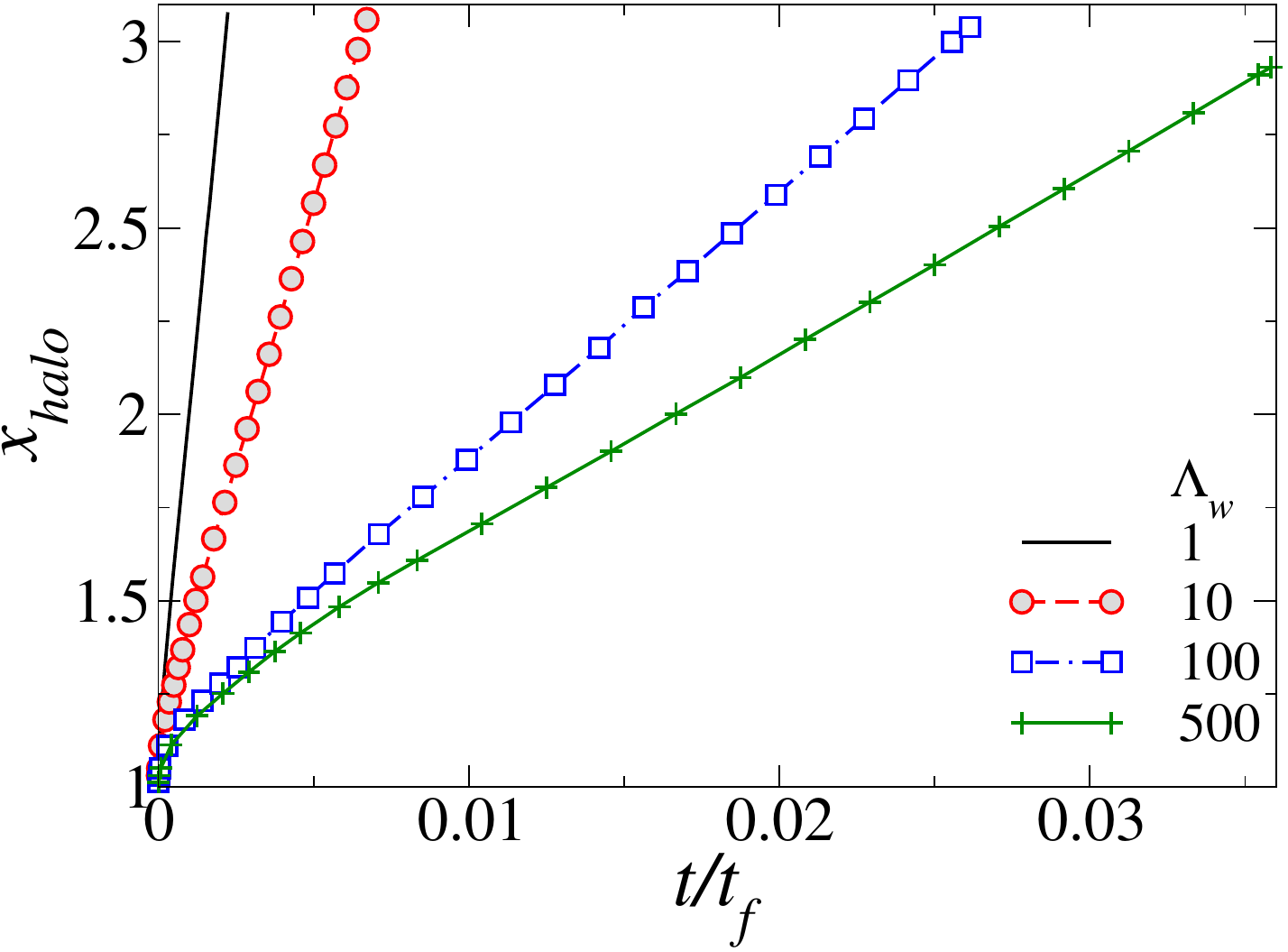}
\caption{Variation of the evaporative flux along the substrate for (a) $\Lambda_{W} = 1$ and (b) $\Lambda_{W} = 500$. Temporal variation of (c) the width of the frost halo $(d_{halo})$, (d) the left $(m_{cl})$ and right $(m_{cr})$ ends of the condensation halo over the substrate, and (e) the location of maximum condensation flux $(x_{halo})$ for different values of $\Lambda_{W}$. In panel (d), the solid and dashed lines represent the variation of $m_{cl}$ and $m_{cr}$, respectively. The rest of the dimensionless parameters are the same as figure \ref{fig:LW_Profile}. The values of $t_{f}$ for $\Lambda_{W} = 1$, $10$, $100$ and $500$ are $t_f=4240$, 1400, 352 and 240, respectively.}
\label{fig:LW_Evaporation_Halo}
\end{figure}

%Figure 13
\begin{figure}
\centering
\includegraphics[width=0.4\textwidth]{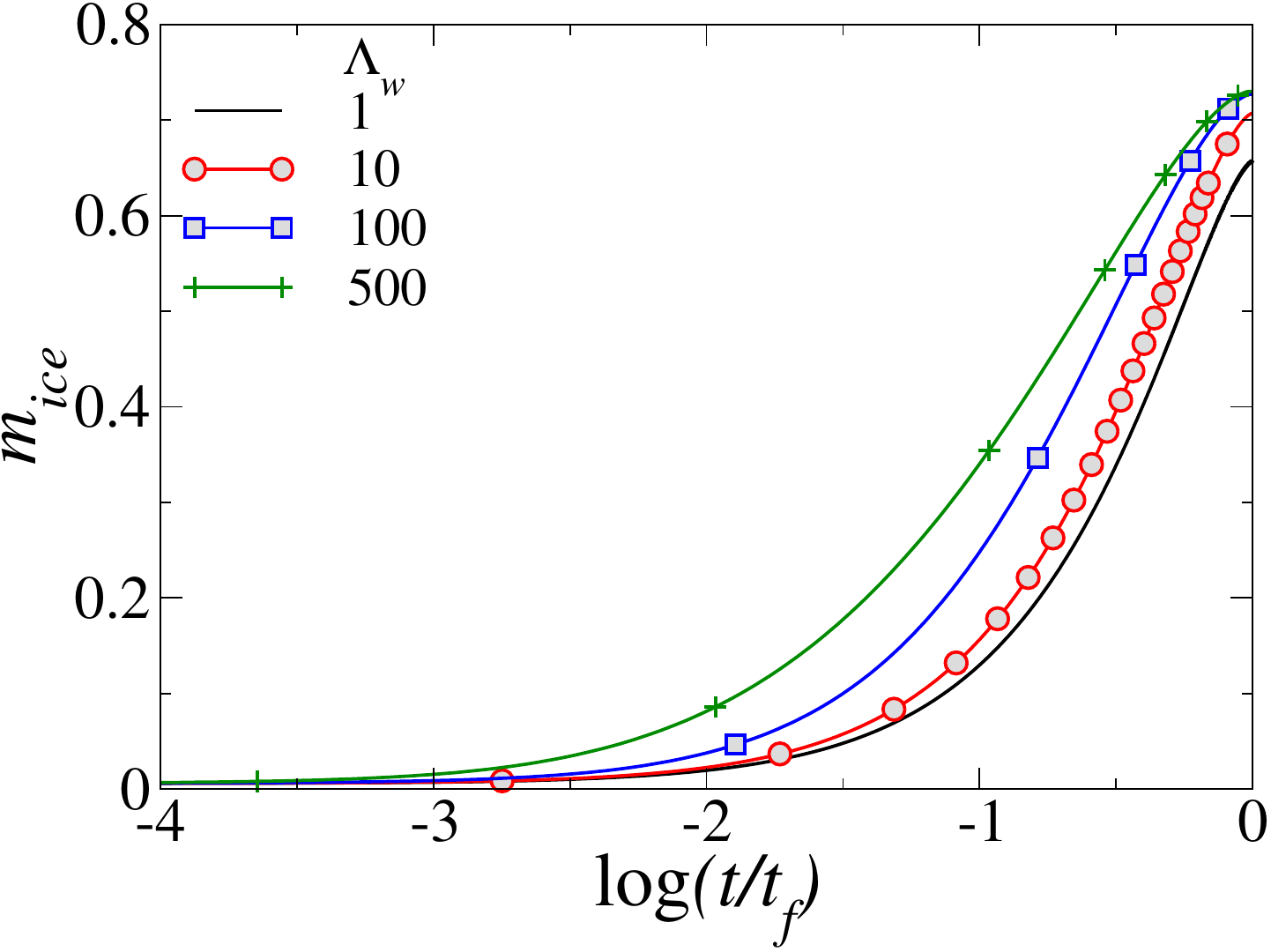}
\caption{Variation of the mass of ice layer, $m_{ice}$ with $\log(t/t_f)$ for different values of $\Lambda_{W}$. The rest of the dimensionless parameters are the same as figure \ref{fig:LW_Profile}.}
\label{fig:LW_Contact_line}
\end{figure}

Next, we investigate the effect of substrate thermal resistance on the freezing behavior of drop by changing the thermal conductivity of the substrate, for a constant thickness of the substrate and keeping $\lambda_{l}=0.57$ Wm$^{-1}$K$^{-1}$ fixed. Here, $V_{\rho,r}=0.85$ and the rest of the parameters are the same as those in figure \ref{fig:RH_vapour}. The temporal evolutions of the freezing front and the droplet shape (figures \ref{fig:LW_Profile}) reveal that increasing the thermal conductivity of the substrate promotes freezing and influences the formation of the freezing front by making the final droplet shape more cusp-like morphology. A close inspection of figure \ref{fig:LW_Profile}(a) and (b) reveals that the droplet exhibits a concave freezing front for a low thermal conductivity substrate. This is because, for low conductive substrates, the liquid-gas interface dissipates more heat through convection and evaporative cooling than the substrate itself, resulting in a lower temperature near the triple contact line than at the centre of the droplet. Quantitatively, we observe that the total freezing time of the droplet for $\Lambda_W=1$ is almost 18 times that for $\Lambda_W=500$. A similar result was also observed in the experiments of \citet{jung2012frost}.

To examine the effect of the thermal conductivity ratio on the frost halo formation, the variation of the evaporation flux $(J_{v})$ along the substrate for different values of $t/t_f$ are plotted in figure \ref{fig:LW_Evaporation_Halo}(a) and (b) for $\Lambda_W=1$ and 500, respectively. In order to quantify this effect, in figure \ref{fig:LW_Evaporation_Halo}(c), we plot the temporal variation of the width of the frost halo ($d_{halo}$), the leftmost ($m_{cl}$) and rightmost point ($m_{cr}$) of the net condensate for different values of $\Lambda_W$. Figure \ref{fig:LW_Evaporation_Halo}(c) shows that, as the thermal conductivity of the substrate ($\Lambda_W$) increases, the width of the halo decreases. It is also evident from figure \ref{fig:LW_Evaporation_Halo}(d) that the halo region is closer to the contact line for the substrate with higher conductivity. As the freezing front evolves, the temperature at the liquid-gas interface increases, leading to a slight rise in evaporation for the substrate with higher thermal conductivity. It is essential to note that the total freezing time ($t_f$) increases as the thermal conductivity of the substrate decreases. This leads to a significantly higher amount of vaporized liquid during the whole freezing process for the droplet on a substrate with a low thermal conductivity than the higher thermal conductivity (see figure ~\ref{fig:LW_Contact_line}). Due to this evaporation, the resulting vapour condenses in the droplet's vicinity, with its concentration reaching supersaturation. In figure \ref{fig:LW_Evaporation_Halo}(a), it can be observed that the condensation occurs far away from the droplet (near $x=2$) at $t/t_{f} = 10^{-3}$. In contrast, it occurs near the contact line of the droplet in figure \ref{fig:LW_Evaporation_Halo}(b). To better quantify this effect, in figure \ref{fig:LW_Evaporation_Halo}(e), we plot the position of the maximum condensation flux $(x_{halo})$ for different values of $\Lambda_w$. In our configuration, as the condensation happens at an earlier time of the freezing, the scaled time $(t/t_f)$ for the substrate with lower conductivity represents a higher value in terms of the absolute dimensionless time $(t)$ when compared to the substrate with higher conductivity leading to condensation far away from the substrate. Regarding the effect of the substrate thickness, it can be noted that numerically $D_w = 0$ resembles a substrate with infinite thermal conductivity. Increasing the thickness of the substrate increases the thermal resistance of the substrate, and thereby, it has the same effect as decreasing the thermal conductivity of the substrate. 

\subsection{Effect of $T_v$} \label{effectTg}

%Figure 14
\begin{figure}[h]
\centering
\hspace{0.8cm}{\large (a)}   \hspace{6.5cm}  {\large (b)} \\
\includegraphics[width=0.4\textwidth]{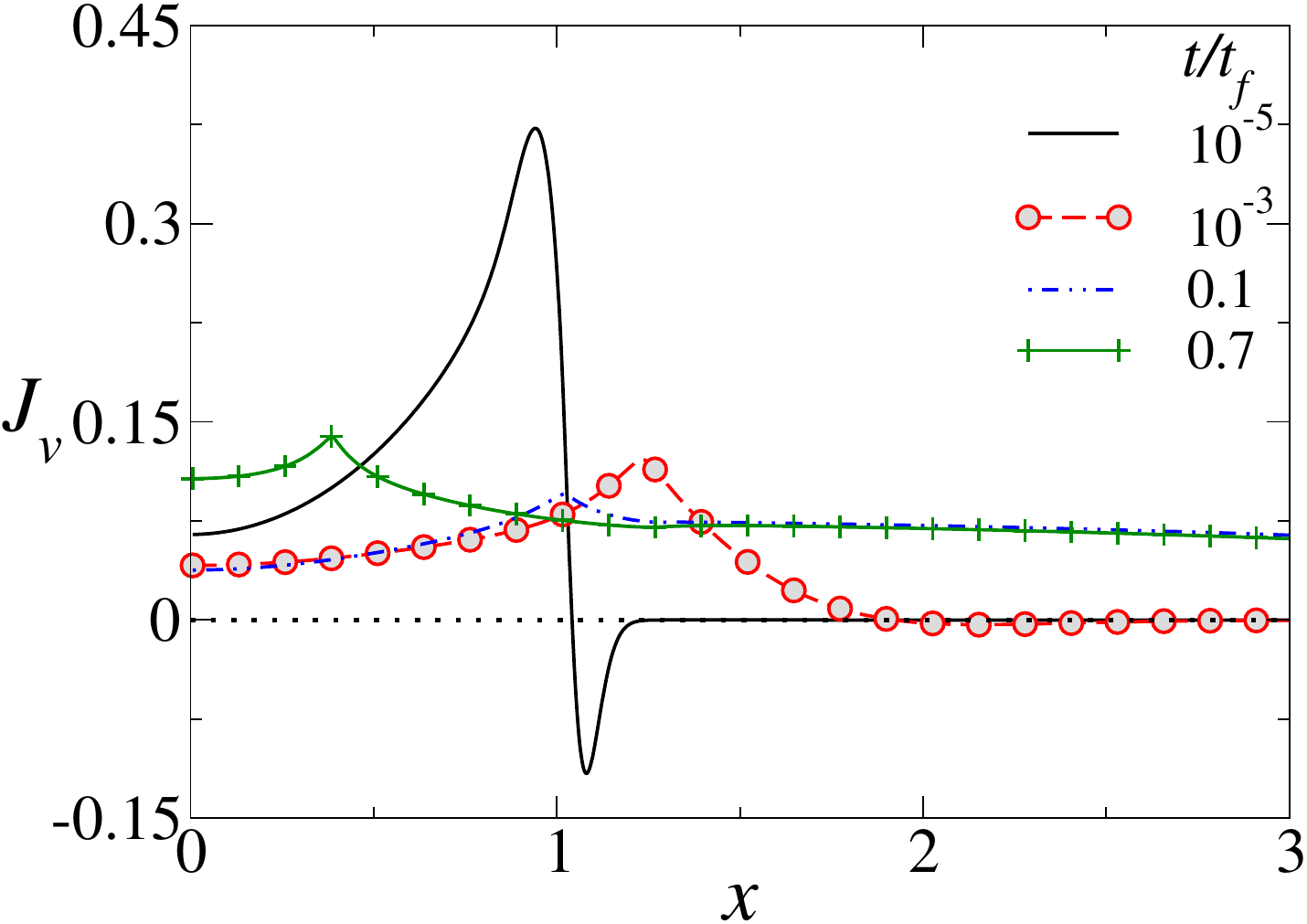} \hspace{0mm}
\includegraphics[width=0.4\textwidth]{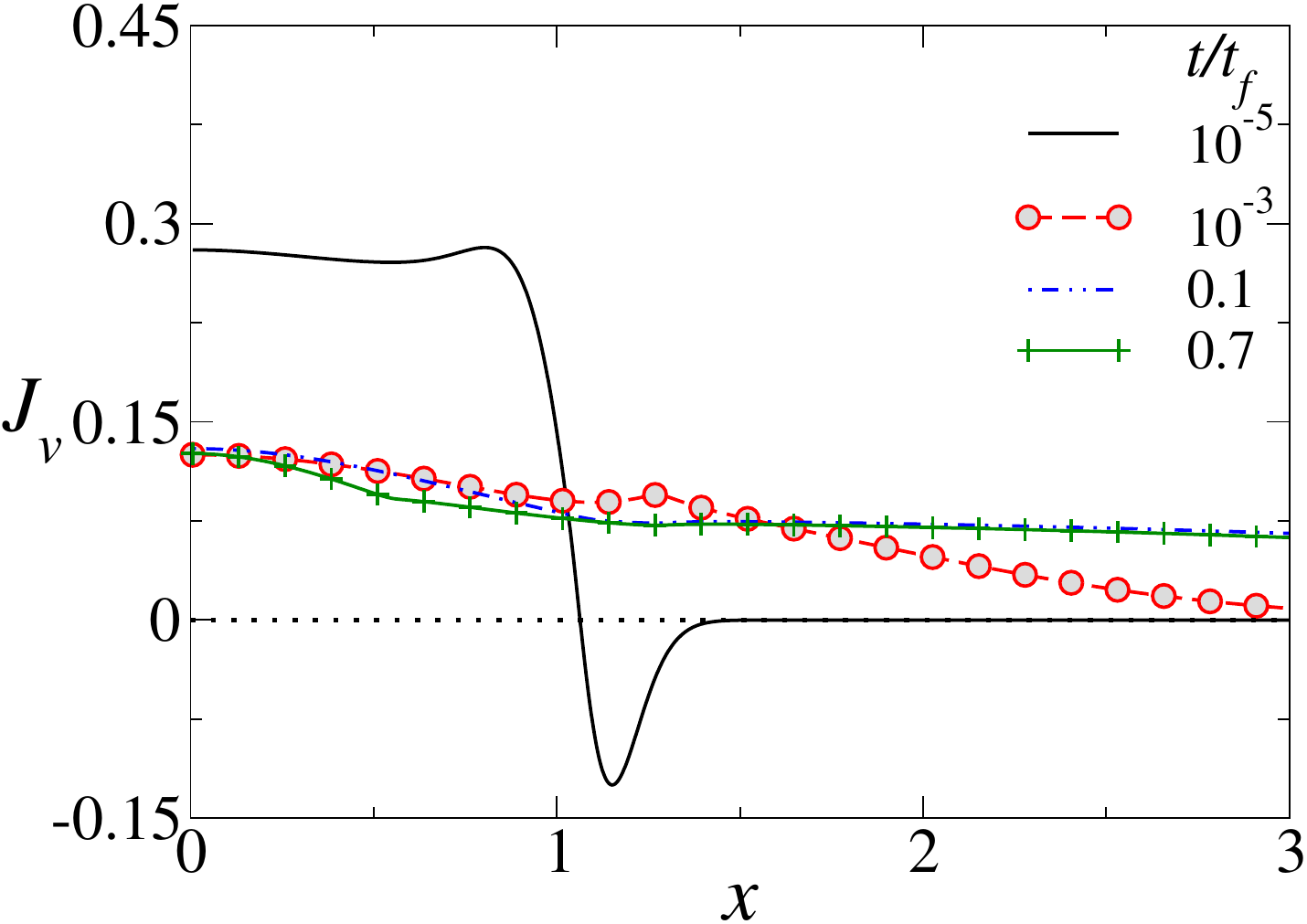} \\ 
\hspace{0.8cm}{\large (c)}   \hspace{6.5cm}  {\large (d)} \\
\includegraphics[width=0.4\textwidth]{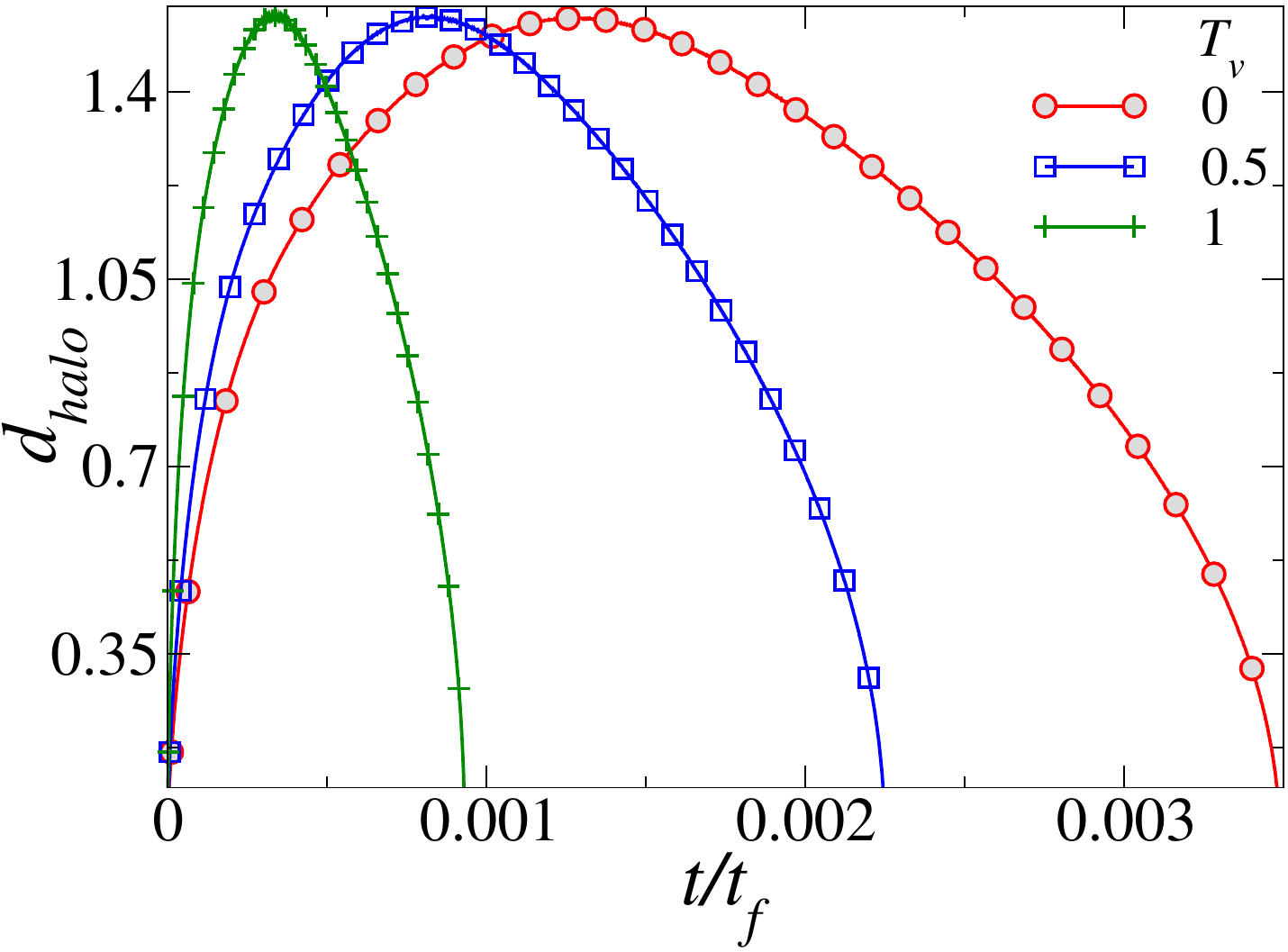} \hspace{0mm}
\includegraphics[width=0.4\textwidth]{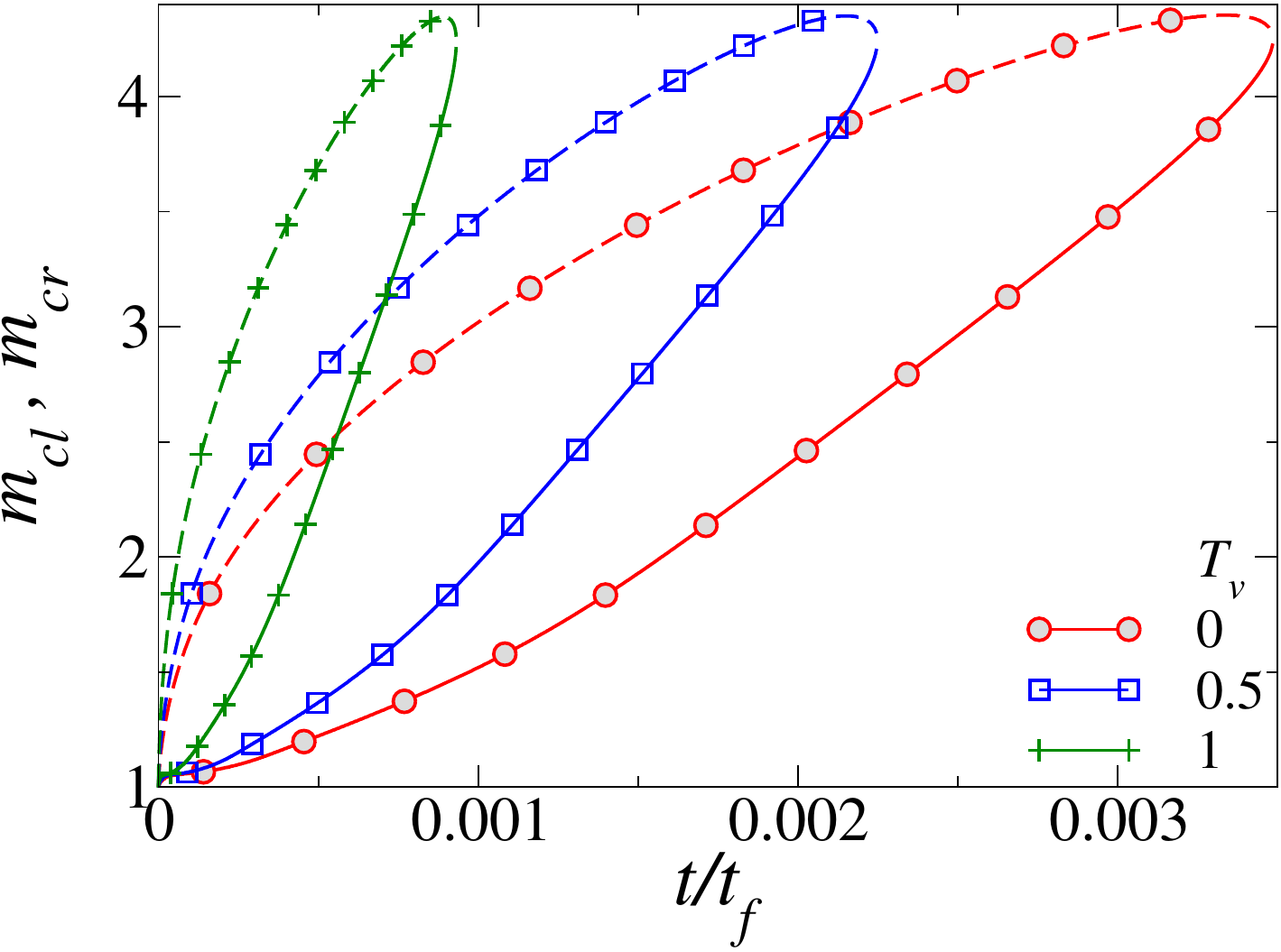}\\
\hspace{0.8cm} {\large (e)}\\
\includegraphics[width=0.4\textwidth]{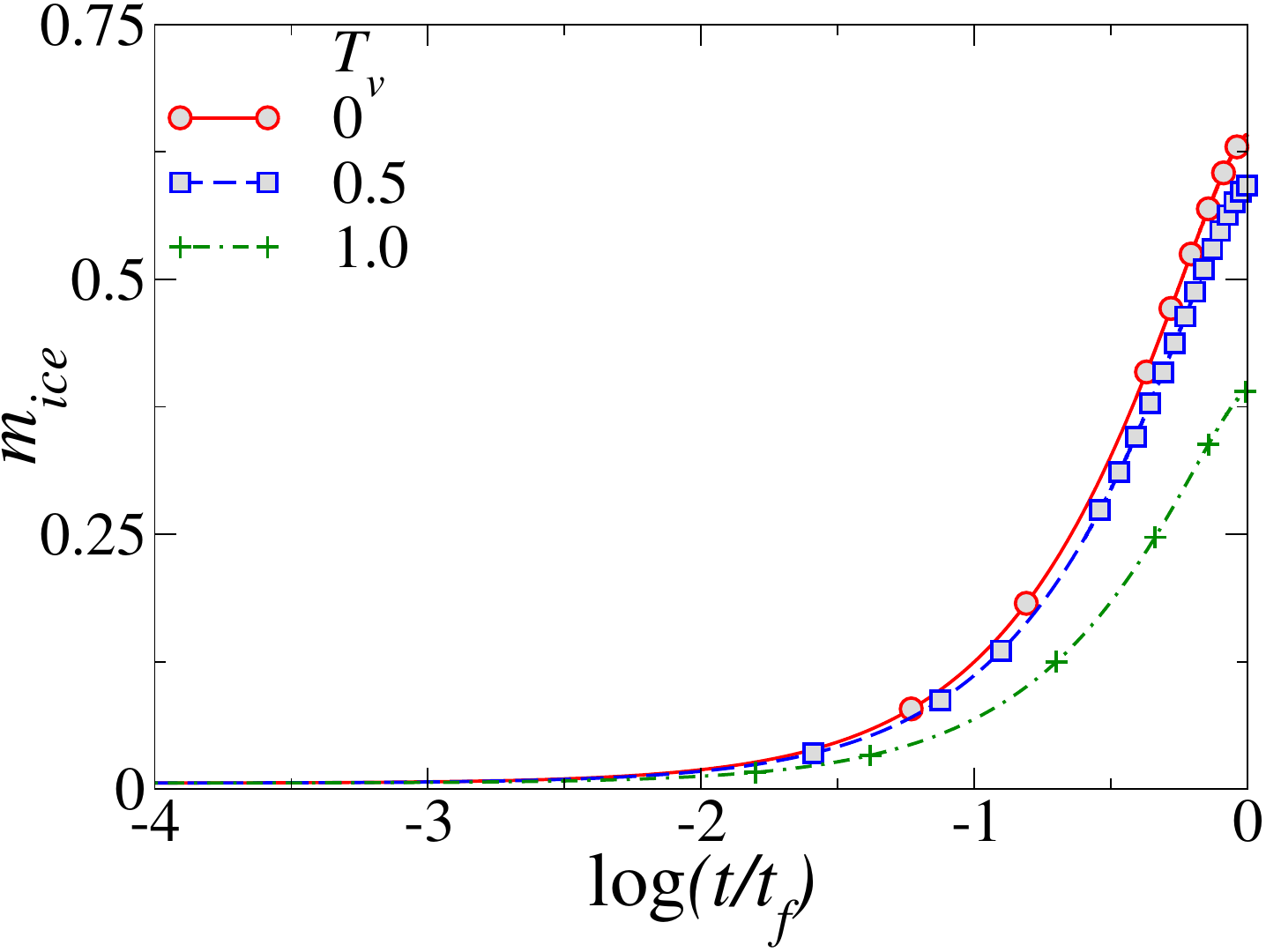}
\caption{Variation of the evaporative flux along the substrate for (a) $T_{v} = 0$ and (b) $T_{v} = 1$. Temporal variation of (c) the width of the frost halo $(d_{halo})$, (d) the left $(m_{cl})$ and right $(m_{cr})$ ends of the condensation halo over the substrate, and (e) the location of maximum condensation flux $(x_{halo})$ for different values of $T_{v}$. In panel (d), the solid and dashed lines represent the variation of $m_{cl}$ and $m_{cr}$, respectively. The values of the rest of the dimensionless parameters are $Ste = 1.22 \times 10^{-3}$, $A_{n} = 6.25$, $D_{g} = 2$, $D_{s} = 0.9$, $\Lambda_{S} = 3.89$, $\Lambda_{W} = 0.33$, $\chi = 0.01$, $V_{\rho,r} = 0.85$, $K = 8\times10^{-4}$, $Bi = 0.16$, $D_{w} = 15$, $\epsilon=0.2$, $D_{v} = 1.65 \times 10^{-6}$, $\Delta = 10^{-4}$, $\Psi = 0.94$ and $Pe_{v} = 1$.} The values of $t_{f}$ for $T_{v} = 0$, $0.5$ and $1$ are $t_f=5121$, 7919 and 19060, respectively.
\label{fig:Tg_Evaporation_Halo}
\end{figure}

Finally, we investigate the effect of the ambient temperature ($T_{v}$) on the freezing behaviour of the droplet. Here, $V_{\rho,r}= 0.85$ and the rest of the parameters are the same as those used to generate figure \ref{fig:RH_vapour} (`base' parameters). Here, we consider a thick substrate with low thermal conductivity (i.e. $D_w = 15$ and $\Lambda_W = 0.33$), resulting in less heat conduction through the substrate. It can be seen in figure \ref{fig:Tg_Evaporation_Halo}(a) that the evaporation is suppressed except for a region near the contact line where the evaporation flux is maximized. Due to the effect of latent heat, the local temperature decreases, favouring freezing near the contact line and leading to a concave shape of the liquid-ice interface. In contrast, when the ambient temperature is higher (e.g. for $T_v=1$, the gas temperature is equal to the melting point temperature), the convective heat transfer occurs from the ambient to the droplet, raising the temperature of the remaining liquid in the droplet. Besides delaying the freezing process, this results in the enhancement of vaporization (see figure \ref{fig:Tg_Evaporation_Halo}(e)) and leads to a more uniform evaporation flux along the liquid-gas interface, as shown in figure \ref{fig:Tg_Evaporation_Halo}(b). This, in turn, results in a nearly flat profile of the liquid-ice interface. Figure \ref{fig:Tg_Evaporation_Halo}(c) and (d) depict the temporal variation of the width of the frost halo $(d_{halo})$, and the left $(m_{cl})$ and right $(m_{cr})$ ends of the condensation halo over the substrate for different values of $T_v$. It can be seen that, as $T_v$ increases, the timescale of the halo formation decreases {even though the maximum width of the halo and the extent of the halo do not change considerably since the evaporation along the contact line does not change significantly for different values of $T_v$.

\section{Conclusions} \label{sec:Conc}
We have numerically investigated the freezing behaviour of a sessile liquid droplet placed on a cold, horizontal solid substrate using the lubrication approximation in the finite element method framework. We present a comprehensive model for freezing front dynamics, encompassing solid-liquid phase change, heat balance across the liquid-gas interface, heat flux between the cold substrate and ice, and mass conservation involving the liquid, solid, and gas phases. The current work mainly focuses on the frost halo formation, which has only been observed experimentally \citep{jung2012frost}. We found that evaporation has a distinct effect on the freezing process of a sessile droplet, in contrast to when evaporation is neglected \citep{Tembely2019,zadravzil2006droplet}. In the initial freezing stages, we observe a negative evaporation flux surrounding the droplet, indicating that vapor generated during freezing condenses on the substrate near the contact line, forming a frost halo. The accumulation of condensate triggers re-evaporation, causing a temporal shift of the frost halo region away from the contact line and ultimately, the halo disappears due to the diffusive nature of water vapor far from the droplet. We found that an increase in relative humidity extends the lifetime of the frost halo, as it significantly reduces evaporation, thereby prolonging the presence of net condensate on the substrate.

We also performed a parametric study to examine the effect of the thermal resistance of the substrate and ambient temperature on the freezing behaviour of the droplet. It is observed that greater liquid volatility results in an increased evaporation flux, and condensation occurs closer to the droplet as a higher concentration of vapor is present in the surrounding region of the droplet. Additionally, decreasing the thermal conductivity of the substrate extends the total freezing time and increases the amount of vaporized liquid. This, in turn, results in condensation, with the concentration reaching supersaturation. As expected, reducing the thermal resistance of the substrate, by increasing the thermal conductivity or reducing the substrate thickness, promotes freezing. We found another interesting behaviour - the droplet exhibits a concave freezing front for a low thermal conductivity substrate. This is because, in the case of a less-conductive substrate, the liquid-gas interface dissipates more heat through convection and evaporative cooling than the substrate itself, resulting in a lower temperature near the triple contact line than at the centre of the droplet. \\

\noindent{\bf Acknowledgement:} {K.C.S. thanks the Science \& Engineering Research Board and IIT Hyderabad, India for their financial supports through grants CRG/2020/000507 and IITH/CHE/F011/SOCH1, respectively. C.S.S. acknowledges IIT Ropar for financial support through ISIRD (Grant No. 9-388/2018/IITRPR/3335). We also acknowledge Vineet Kumar (IIT Ropar) for preliminary results on sessile droplet freezing in figure 1.} \\

\noindent
{\bf Declaration of interests:} The authors report no conflict of interest.

\appendix
\numberwithin{equation}{section}
\makeatletter
\newcommand{\section@cntformat}{Appendix \thesection:\ }
\makeatother

\section{Lubrication model} \label{sec:lub}

By employing the scalings shown in Eq. (\ref{eq:characteristic_scales}) and incorporating the lubrication approximation ($\epsilon \ll 1$), we obtain the following dimensionless governing equations and boundary conditions in the liquid, ice and gas phases. 

\subsubsection{Liquid phase}
The dimensionless governing equations in the liquid phase are given by
\begin{eqnarray} 
\partial^2_z u &=& \partial_x p,\label{eq:xmom_scaled}
\\ 
\partial_z p &=& 0,\label{eq:zmom_scaled}
\\
\partial_x u + \partial_z w &=& 0,\label{eq:cont_scaled}
\\
\partial^2_z T_l &=& 0.\label{eq:Tl_scaled}
\end{eqnarray}
The boundary conditions at the liquid-gas interface ($z=h(x,t)$) are
\begin{eqnarray} 
p  &=& - \kappa_{lg} \gamma_{lg} - \Pi, \label{eq:ph_scaled}
\\ 
\partial_z u  &=& 0, \label{eq:tstress_scaled}
\\ 
\partial_z T_l  &=& - \chi J_v + Bi \left(T_{v}-T_{l}\right), \label{eq:ebc_scaled}
\\ 
\partial_t h + u \partial_x h - w  &=& -D_v J_v. \label{eq:kin_scaled}
\end{eqnarray}
Here, $D_v = \rho_{ve}(T_m)/ \rho_l$ represents the density ratio, $\chi = \epsilon \rho_{ve}(T_m) U L_v H / (\lambda_l \Delta T)$ denotes the scaled latent heat of vaporization and $Bi = h_{tc}H/\lambda_l$ is the Biot number.

The boundary conditions at the liquid-ice interface (at $z=s(x,t)$) are  
\begin{equation} \label{eq:noslip_ls_scaled}
u  = 0 \; \; {\rm and} \;\; T_l = T_f,
\end{equation}
where $T_f=T_m$.

The dimensionless disjoining pressure is given by 
\begin{equation}
\Pi = \epsilon^{-2} A_{n} \left[ \left( \frac{\beta}{h-s} \right)^n - \left( \frac{\beta}{h-s} \right)^m \right],
\label{eq:dimensionless_disj_press}
\end{equation}
where $\beta$ is of the same order as the equilibrium precursor layer thickness \citep{pham2019imbibition} and $A_{n} = H A / \gamma_{lgo}$ denotes the dimensionless Hamaker constant. The interaction between the repulsive and attractive components of eq. (\ref{eq:dimensionless_disj_press}) dictates the value of the equilibrium contact angle $\theta_{eq}$ \citep{Schwartz1998,zadravzil2006droplet,espin2015droplet,pham2019imbibition,Tembely2019}. This can be approximated \citep{pham2019imbibition}
\begin{equation}
\theta_{eq} \approx \sqrt{\beta A_{n}}.
\end{equation}

The full mean curvatures at the liquid-gas and liquid-ice interfaces, retaining higher order contributions, are given by 
\begin{equation}
\kappa_{lg} = \frac{h_{xx}}{(1+\epsilon^2 h_x^2)^{\frac{3}{2}}} ~ {\rm and} ~
\kappa_{sl} = \frac{s_{xx}}{(1+\epsilon^2 s_x^2)^{\frac{3}{2}}}.
\end{equation}

\subsubsection{Solid (ice) phase}

Under the lubrication approximation ($\epsilon \ll 1$), the dimensionless energy conservation equations for the ice phase is given by
\begin{equation}
\partial^2_z T_s = 0.
\end{equation}
The energy balance at the liquid-ice interface ($z=s(x,t)$) gives
\begin{equation} \label{Stefan_scaled}
Ste \left( \Lambda_s \partial_z T_s - \partial_z T_l \right) = J_s,
\end{equation}
where $\Lambda_s = \lambda_s / \lambda_l$ is the thermal conductivity ratio and $Ste = \lambda_l \Delta T / ( \epsilon \rho_s U L_f H )$ denotes the Stefan number. At $z=s(x,t)$, the conservation of mass (eq. \ref{eq:mass_bc_solid_ND}) gives
\begin{equation} \label{eq:Js_scaled}
\partial_t s  - w = D_s J_s, ~ {\rm where} ~ J_s = \partial_t s.
\end{equation}
At the ice-solid interface ($z=0$), we impose
\begin{equation}
T_s = T_w.
\end{equation}

\subsubsection{Gas phase}

The dimensionless conservation equation for the vapor concentration becomes
\begin{equation}\label{eq:rho_v_scaled}
Pe_v({\partial_t\rho_v}+u_g\partial_x\rho_v+w_g\partial_z\rho_v) = \partial^2_z\rho_v +\epsilon^2\partial^2_x\rho_v.
\end{equation}
At the liquid-gas interface ($z=h(x,t)$), the conservation of mass reduces to
\begin{equation}\label{bc:1_mass_bal}
Pe_v J_v = - \partial_z \rho_v, 
\end{equation}
while the constitutive equation for the evaporation flux gives
\begin{equation}\label{bc:k}
K J_v = 1 - \rho_v  + \Delta p + \Psi \left (T_{l}-1 \right).
\end{equation}
The boundary condition far from the droplet ($z = D_g$) is given by
\begin{equation} \label{eq:vap_conc_hum_scaled}
\rho_v = V_{\rho,r}.
\end{equation}
The various dimensionless numbers appearing in eqs. (\ref{eq:rho_v_scaled} - \ref{bc:k}) are
\begin{equation}
\begin{gathered}
Pe_v = \frac{\epsilon U H}{D_m}, ~ K = \frac{\epsilon U}{\alpha } \left( \frac{2 \pi M}{R_g T_{m}} \right)^{1/2}, ~ \Delta = \frac{\mu_{l}ULM}{H^{2}\rho_{l}R_gT_m}, \\  \Psi = \frac{L_{v}M\Delta T}{R_{g}T_{m}^2}, ~ {\rm and} ~ V_{\rho,r} ={\rho_{vi} \over \rho_{ve}(T_m)}.
\end{gathered}
\end{equation}

\subsubsection{Solid substrate}

The scaled energy conservation equation for the solid substrate is given by
\begin{equation}
\partial^2_z T_w = 0.
\end{equation}
The boundary condition at the solid-ice interface ($z=0$) is
\begin{equation}
\Lambda_s \partial_z T_s = \Lambda_w \partial_z T_w,
\end{equation}
where $\Lambda_w = \lambda_w / \lambda_l$ denotes the thermal conductivity ratio. At the bottom of the substrate (at $z=-D_w$), we impose
\begin{equation}
T_w  = 0.
\end{equation}

\subsection{Evolution equations}

By integrating eqs. (\ref{eq:xmom_scaled}) and (\ref{eq:zmom_scaled}) with respect to $z$ and using eqs. (\ref{eq:tstress_scaled}), (\ref{eq:noslip_ls_scaled}) and (\ref{eq:ph_scaled}), we get 
\begin{equation}
u = \frac{\partial_x p}{2}   (z^2 - s^2)  - h \partial_x p (z-s),
\end{equation}
\begin{equation}
p = - \kappa_{lg} \gamma_{lg} - \Pi.
\end{equation}

By integrating eq. (\ref{eq:cont_scaled}) and using eq. (\ref{eq:kin_scaled}), we get the following evolution equation: 
\begin{equation}
\partial_t h - \partial_t s = - \partial_x q_l - D_v J_v - D_s J_s,
\end{equation}
where
\begin{equation}
q_l = \frac{\partial_x p}{2} \left (\frac{h^3}{3} - s^2 h + \frac{2s^3}{3} \right) - h \partial_x p  \left (\frac{h^2}{2}-s h  + \frac{s^2}{2} \right).
\end{equation}

Similarly, by integrating eq. (\ref{eq:Tl_scaled}) and using eqs. (\ref{eq:ebc_scaled}) and (\ref{eq:noslip_ls_scaled}), we get 
\begin{equation}
T_l = (-\chi J_v+Bi(T_{v}-T_{l}\big|_{h})) (z-s) + T_f.
\end{equation}
The temperature distribution in the ice phase is governed by
\begin{equation}
T_s = \frac{T_f}{D_w+s \Lambda_w/\Lambda_s} \left( D_w + z  \frac{\Lambda_w}{\Lambda_s} \right).
\end{equation}
Using the above expression along with eq. (\ref{eq:Js_scaled}) and introducing them into eq. (\ref{Stefan_scaled}), we get
\begin{equation}\label{eq:Freezing_rate}
\partial_t s = Ste \left(  \frac{\Lambda_w T_f}{D_w+s \Lambda_w/\Lambda_s} + \chi J_v - Bi(T_{v}-T_{l}\big|_{h}) \right).
\end{equation}

The temperature profile in the solid substrate is given by 
\begin{equation}
T_w = \frac{T_f}{D_w + s \; \Lambda_w/\Lambda_s}(z+D_w).
\end{equation}

\section{Grid independence test} \label{sec:grid}

%Figure 15
\begin{figure}[H]
\centering
 \includegraphics[width=0.7\textwidth]{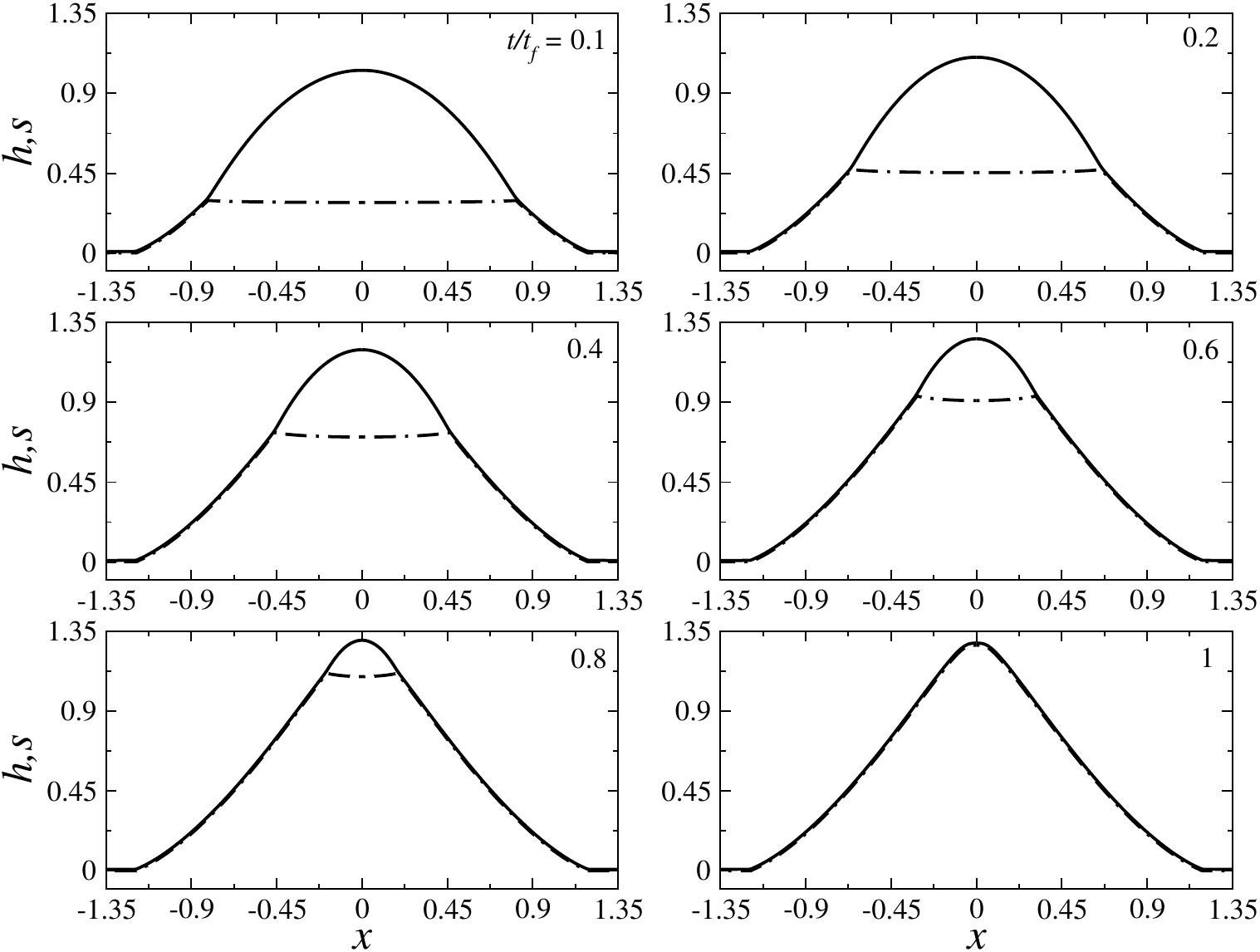}  
\caption{Evolution of the freezing front, $s$ (dot-dashed lines) and magnified shape of the upper part of the droplet, $h$ (solid line) placed on a cold substrate. The results obtained using 2001, 4801, and 8001 grid points were indistinguishable. The rest of the dimensionless parameters are $V_{\rho,r}= 0.85$, $\chi = 0.23$, $\Delta = 10^{-4}$, $\Psi = 0.14$ and $Pe_{v} = 1.0$, $D_{w} = 7.5$, $\epsilon=0.2$, $Ste = 1.7 \times 10^{-4}$, $T_{v} = 0.2$, $A_{n} = 6.25$, $D_{g} = 2$, $D_{s} = 0.9$, $\Lambda_{S} = 3.82$, $\Lambda_{W} = 191$, $K = 8\times10^{-4}$, $Bi = 2.29$ and $D_{v} = 4.85 \times 10^{-6}$. The values of the dimensionless total freezing time $(t_{f})$ of the droplet is 1220, respectively.}
\label{fig:Grid_Convergence}
\end{figure}

\section{Tip angle} \label{sec:tip}

%Figure 16
\begin{figure}[H]
\centering
\includegraphics[width=0.45\textwidth]{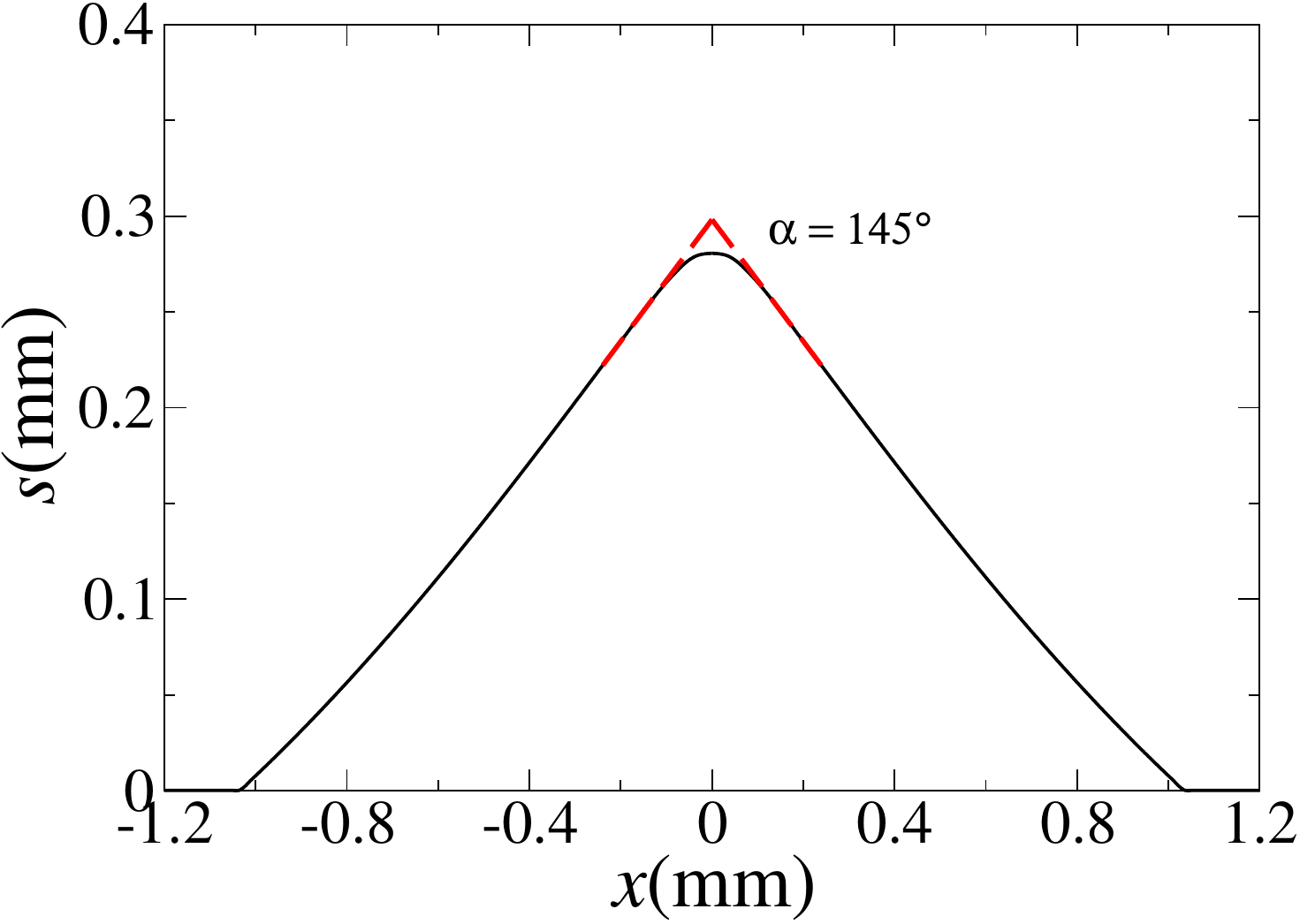}
\caption{Figure showing the estimation of the tip angle for a typical set of parameters ($Ste = 1.22 \times 10^{-3}$, $T_{v} = 1$, $A_{n} = 6.25$, $D_{g} = 2$, $D_{s} = 0.9$, $\Lambda_{S} = 3.89$, $\Lambda_{W} = 698$, $\chi = 0.01$, $K = 8\times10^{-4}$, $Bi = 0.16$, $D_{w} = 15$, $V_{\rho,r} = 0.70$, $\epsilon=0.2$, $D_{v} = 1.65 \times 10^{-6}$, $\Delta = 10^{-4}$, $\Psi = 0.94$ and $Pe_{v} = 1$).}
\label{fig:Tip_angle}
\end{figure}

\clearpage
%\bibliography{Bibliography}

\end{document}